\documentclass[11pt,a4paper]{article}
\usepackage{jheppub}
\usepackage{graphicx}
\usepackage{color}
\usepackage{latexsym}
\usepackage{slashed}
\usepackage{amsthm}
\usepackage{amssymb,mathtools}
\usepackage{bbm}
\usepackage{indentfirst} 
\definecolor{ao(english)}{rgb}{0.0, 0.5, 0.0}
\hypersetup{colorlinks=true,linkcolor=blue,urlcolor=blue,citecolor=ao(english)}
\usepackage{mathrsfs}
\usepackage[latin1,utf8]{inputenc}
\usepackage[makeroom]{cancel}

\newcommand{\be}{\begin{eqnarray}}
\newcommand{\ee}{\end{eqnarray}}
\newcommand{\bea}{\begin{eqnarray}}
\newcommand{\eea}{\end{eqnarray}}

\newcommand{\eq}[1]{(\ref{#1})}
\newcommand{\n}[1]{\label{#1}}

\def\cob{\color{blue}}

\newcommand{\au}[2]{#1.~#2}
\newcommand{\book}[5]{\emph{#1}, #2, #3, #4 (#5)}

\newcommand{\oarX}[1]{\href{http://arxiv.org/abs/#1}{{\ttfamily\cob arXiv:#1}}}
\newcommand{\arX}[1]{\href{http://arxiv.org/abs/#1}{{\ttfamily\cob arXiv:#1}}}
\newcommand{\doin}[6]{\href{http://dx.doi.org/#1}{{\cob {\it #2} {\bf #3 #4} (#6) #5}}}
\newcommand{\doinn}[5]{\href{http://dx.doi.org/#1}{{\cob {\it #2} {\bf #3} (#5) #4}}}
\newcommand{\doij}[5]{\href{http://dx.doi.org/#1}{{\cob {\it #2} {\bf #3} (#5) #4}}}
\newcommand{\ndoin}[6]{\href{#1}{{\cob {\it #2} {\bf #3 #4} (#6) #5}}}
\newcommand{\ndoinn}[5]{\href{#1}{{\cob {\it #2} {\bf #3} (#5) #4}}}
\newcommand{\procsinm}[5]{in \emph{#1}, #2 (eds.), #3, #4 (#5)}

\newcommand{\tia}[1]{\textit{#1},}

\numberwithin{equation}{section}

\newcommand{\al}{\alpha}
\newcommand\ga{\gamma}

\newcommand\de{\delta}
\newcommand\De{\Delta}

\newcommand\ka{\kappa}
\newcommand\la{\lambda}
\newcommand\La{\Lambda}

\newcommand\ph{\varphi} 
\newcommand\om{\omega}

\def\cL{\mathcal{L}}

\newcommand\na{\nabla}
\def\loc{{\rm loc}}
\newcommand{\rd}{\mathrm{d}}

\newtheoremstyle{sltheorem}
{}                
{}                
{\slshape}        
{}                
{\bfseries}       
{.}               
{ }               
{}                
\theoremstyle{sltheorem}
\newtheorem{theorem}{Statement}

\begin{document}


\title{Renormalizability of nonlocal quantum gravity coupled to matter}

\author[a]{Gianluca Calcagni,}
\emailAdd{g.calcagni@csic.es}
\affiliation[a]{Instituto de Estructura de la Materia, CSIC, Serrano 121, 28006 Madrid, Spain}

\author[b]{Breno L.\ Giacchini,}
\emailAdd{breno@sustech.edu.cn}

\author[b]{Leonardo Modesto,}
\emailAdd{lmodesto@sustech.edu.cn}

\author[b]{Tib\'{e}rio de Paula Netto,}
\emailAdd{tiberio@sustech.edu.cn}

\affiliation[b]{Department of Physics, Southern University of Science and Technology, Shenzhen, 518055, China}

\author[c]{Les{\l}aw Rachwa\l{}}
\emailAdd{grzerach@gmail.com}
\affiliation[c]{Departamento de Física -- Instituto de Ciências Exatas, Universidade Federal de Juiz de Fora, Campus Universitário, Rua José Lourenço Kelmer, s/n -- São Pedro 33036-900, Juiz de Fora, MG, Brazil}


\abstract{We extensively study the ultraviolet quantum properties of a nonlocal action for gravity nonminimally coupled to matter. The theory unifies matter and gravity in an action principle such that all the classical solutions of Einstein's theory coupled to matter are also solutions of the nonlocal theory. At the quantum level, we show that the theory is power-counting super-renormalizable in even dimensions and finite in odd dimensions. A simple extension of the model compatible with the above properties is finite also in even dimensions.}



\maketitle
\tableofcontents


\section{Introduction}

Nonlocal quantum gravity is a perturbative quantum field theory of the gravitational force where both the classical and quantum dynamics of the graviton are characterized by nonlocal form factors, operators with infinitely many derivatives \cite{Krasnikov:1987yj,Kuzmin:1989sp,Modesto:2011kw,Biswas:2011ar,Dona:2015tra}. Nonlocality is a classical fundamental feature instead of an emerging one at the quantum level. Matter fields can be introduced separately with a minimal coupling through the spacetime measure weight $\sqrt{|g|}$, where $g$ is the determinant of the metric $g_{\mu\nu}$. This theory has received a lot of attention in the last decade thanks to its rigor and conceptual simplicity (it is based on traditional techniques of quantum fields) \cite{BasiBeneito:2022wux} and its applications to black holes \cite{Buoninfante:2022ild} and cosmology \cite{CANTATA:2021ktz,Koshelev:2023elc}, with a promise of more phenomenology.

Recently, a new version of the theory has been proposed where gravity and matter are nonminimally coupled while keeping intact all the good ultraviolet (UV) properties of the original formulation \cite{Modesto:2021ief,Modesto:2021okr,Modesto:2021soh,Modesto:2022asj,Calcagni:2022tuz}. The theory unifies matter and gravity in an action principle strongly constrained according to four consistency requirements: (i) all the classical solutions of Einstein's theory coupled to matter are also solutions of the nonlocal theory; (ii) such solutions have the same stability properties at the linear and nonlinear level; (iii) the tree-level scattering amplitudes are the same as in Einstein's gravity; (iv) macrocausality is not violated. The advantages of this new version of the theory are that, on one hand, it delivers predictions testable in the very near future \cite{Calcagni:2022tuz} and, on the other hand, it accounts for matter fields in a natural way in contrast with the minimally coupled versions. In fact, while the nonminimally coupled version implements a property (recovery of covariant Einstein's equations as a sub-case of the full equations of motion) that shapes the nonminimal coupling in a rather rigid way and imposes the same type of form factors in all sectors, in the case of the minimally coupled versions matter fields are freely added by hand and their form factors are specified \emph{ad hoc}, independently of those in the gravitational sector. By unifying the gravitational and matter sectors in such a manner, one is widening the interesting but limited realm of quantum gravity to a theory of everything, thus moving towards the amplitude of scope of supergravity or string theory.

In this paper, we use the power counting of divergences in Feynman diagrams to show that nonminimally coupled nonlocal gravity is super-renormalizable (i.e., superficially divergent diagrams are finite in number \cite{Kaku:1993ym,PeSc,Buchbinder:2021wzv})\footnote{\label{foot1} 
This definition \cite{Kaku:1993ym,PeSc,Buchbinder:2021wzv} applies also to the case of gravity after noting that, when written in terms of the graviton $h_{\mu\nu}$, instead of a finite number of divergent diagrams we have a finite number of \textit{families} of divergent diagrams~\cite{Buchbinder:2021wzv}, which (owing to locality and covariance) result in a finite number of covariant divergent terms in the effective action.
} or finite (no divergent diagrams), thus fulfilling the expectations advanced in \cite{Modesto:2021ief}. After presenting the classical and quantum theory in section \ref{Sec.Act}, we introduce a simplified model where the only matter field is a real scalar. The most general scalar-tensor action of this model is discussed in section \ref{model gravity-scalar}, while we systematically develop the power counting in section \ref{Sec.PC}. The general renormalizability properties of the theory for any matter content are stated in section \ref{genpc}. Finiteness is discussed in section \ref{finite}, while conclusions are in section \ref{concl}. Appendices contain some accessory material.


\section{Overview of the theory}
\label{Sec.Act}

In this section, we shortly recall the theory proposed in~\cite{Modesto:2021ief,Modesto:2021okr,Modesto:2022asj} as a UV completion of the Standard Model of particle physics coupled to gravity. We also elaborate on the asymptotic limit of the action in the UV limit, and the general structure of vertices. Finally, we present the explicit form of the UV action in the case of gravity coupled to a scalar field, which is the fundamental element for the renormalizability considerations of the next sections.


\subsection{Classical theory}

The classical action $S$ of the theory is constructed as a nonlocal and nonminimal extension of a subsidiary local action $S_{\loc}$ as follows~\cite{Modesto:2021ief}:
\begin{align}
&
\n{action}
S [\Phi] = \int \rd^D x \sqrt{|g|} \left[ \cL_{\loc} + E_i F^{ij} (\hat{\De}) E_j \right]
,
\\
&
S_{\loc} = \int \rd^D x \sqrt{|g|} \, \cL_{\loc}
,
\n{action-localEq1}
\\
&
\cL_{\loc} = \frac{1}{\ka^2} R  + \cL_{\rm m} (g_{\mu\nu}, \ph, \psi, A^\mu)
,
\\
&
E_i (x) = \frac{\de S_{\loc}}{\de \Phi^i (x)}
,
\label{EiEq1}
\\
&
\De_{ij} (x,y) = \frac{\de E_{i} (x)}{\de \Phi^j(y)} =  \frac{\de^2 S_{\loc}}{\de \Phi^j(y) \de \Phi^i (x)} = \hat{\De}_{ij} \, \de^D (x,y)
,
\label{HessianEq1}
\\
&
2 \hat{\De}_{ik} \, F^{k} {}_j (\hat{\De}) = \left[ \frac{e^{{\rm H}(\hat{\De}_{\La_*} )} }{e^{{\rm H}(0)} } - 1 \right]_{ij}, \label{FF2}
\end{align}
where $D$ is the number of spacetime topological dimensions ($D=4$ in the physical case), $\ka^2 = 16 \pi G$, $G$ is Newton's constant, $\de^D(x,y) = \de^D (x-y)/\sqrt{|g(x)|}$ is the covariant delta function, $\Phi^i = (g_{\mu\nu}, \ph, \psi, A^\mu)$
is the set of all fields (metric, scalars, fermions, gauge fields), $F^{ij}$ is a symmetric tensorial entire function whose argument is the Hessian operator $\hat{\De}_{ij}$, and ${\rm H}( \hat{\De}_{\La_*} )$ is an entire analytic function whose argument is the dimensionless Hessian
\be
\label{DefStar}
( \hat{\De}_{\La_*}) _{ij} = \frac{\hat{\De}_{ij}}{\left( \La_* \right)^{[\hat{\De}_{ij}]}}\, .
\ee
In the above formula, $\La_*$ is the nonlocality scale with mass (energy) dimensionality. We use $\hbar = c = 1$ units and the notation $[X]$ indicates the dimensionality of the quantity $X$ in powers of mass units, i.e., $X$ scales as $(\text{mass})^{[X]}$. In particular, $[\La_*]=1$ and the dimensionality of the component $(i,j)$ of the Hessian is 
\be
[\hat{\De}_{ij}] = D - [\Phi^i] - [\Phi^j] .
\ee
A detailed dimensional analysis of the model is carried out in Appendix~\ref{AppA}.

The form factor $F(\hat{\De})$ is defined in terms of an entire function $\exp {\rm H}$ of the Hessian operator which does not have any poles or zeros in the whole complex plane at finite distance. This type of nonlocality is called weak in contrast with strong nonlocality where nonlocal operators have poles or zeros, such as inverse powers of $\hat\De$. Note that $F$ vanishes in the infrared (IR) local limit $\Lambda_*\to\infty$, where ${\rm H}(\hat{\De}_{\La_*})\to {\rm H}(0)$ and one recovers the underlying local theory. In (\ref{FF2}), ${\rm H}(\hat{\De}_{\La_*})$ is a special function called complementary exponential integral \cite[formula 6.2.3]{NIST}, an entire analytic function with asymptotic logarithmic behavior:
\be
{\rm H}(z)  =  \int_0^{p(z) } \rd w \, \frac{1 - e^{- w}}{w}
=  \gamma_{\rm E} + \Gamma[ 0, p(z) ] + \ln [ p(z) ] \, , 
\label{H}
\ee
where $\gamma_{\rm E}$ is the Euler--Mascheroni constant and $p(z)$ is a generic polynomial of degree $n+1$ in the variable $z$, namely 
\be
p(z) = a_0 + a_1 z + a_2 z + \dots + a_{n+1} z^{n+1} \, , \qquad a_i \in \mathbb{R} \, . 
\label{Poly}
\ee
Notice that the function ${\rm H}(z)$ is defined both for ${\rm Re} \, p(z) > 0$ and ${\rm Re} \, p(z) < 0$. Moreover, the following identity is true in the complex $z$ plane:
\be
e^{{\rm H}(z)} \equiv e^{\gamma_{\rm E} + \Gamma[ 0, p(z) ]} \,  p(z) \, ,\qquad e^{{\rm H}(0)} = a_0 \, e^{\gamma_{\rm E} + \Gamma(0, a_0)}\, .
\label{MCFormFactor}
\ee
Let us now study the UV limit of $\exp {\rm H}(z)$. For large $z$, the form factor (\ref{MCFormFactor}) simplifies to a polynomial,
\be
\hspace{-1cm} 
e^{{\rm H}(z)} \,\, & \simeq &  \,\, e^{\gamma_{\rm E}} p(z) \quad {\rm for} \quad z \gg 1 \, .
\label{UVlimit} 
\ee
More precisely,
\be
e^{{\rm H}(z)} &=& \left[ e^{{\rm H}(z)} - e^{\gamma_{\rm E}} p(z) \right] + e^{\gamma_{\rm E}} p(z)  \nonumber \\
&\simeq&   e^{\gamma_{\rm E}} p(z)\left\{ e^{\frac{e^{-p(z)} [p(z)-1]}{p(z)^2}} - 1 \right\} + e^{\gamma_{\rm E}} p(z) 
=   e^{\gamma_{\rm E}} p(z) + e^{\gamma_{\rm E}} e^{-p(z)}  +  \dots   , 
\label{corrInf}
\ee
where the second term in the last equality in (\ref{corrInf}) provides the leading correction to the asymptotic polynomial limit of the form factor $\exp {\rm H}(z)$. On the other hand, for small $z$ the analyticity of the form factor provides an IR expansion of the action in higher-derivative operators. 

To complete this short review on the classical theory, let us also recall that the nonlocal equations of motion are at least linear in $E_i$ \cite{Modesto:2021ief},
\be
\left[e^{{\rm H}(\hat{\De}_{\La_*})}\right]_{kj} \, E_j + O(E^2) = 0 \,,\label{LEOM}
\ee
which implies that all the solutions of the local equations of motion $E_i=0$ of the associated local system are also solutions of the nonlocal one. It is then not difficult to show that these solutions have the same stability properties as in the subsidiary local theory (e.g., Einstein's gravity with matter), so that a stable solution in the local theory is also stable in the nonlocal one \cite{Modesto:2021soh}.


\subsection{Unitarity}

Moving on to the quantum theory, the unitarity issue has already been tackled in general analytic or weakly nonlocal quantum field theories in \cite{Modesto:2021soh} at tree-level and in \cite{Briscese:2018oyx} at any perturbative order. Indeed, the tree-level unitarity of the theory is guaranteed by the fact that its tree-level scattering amplitudes are the same of the associated local system (this is due to the structure of the nonlocal equations of motion (\ref{LEOM})), or, equivalently, by the structure of the propagator \cite{Modesto:2011kw}, which is the same as in the local theory up to analytic nonlocal functions that do not include extra poles. Indeed, the structure of the equations of motion (\ref{LEOM}) implies that the tree-level scattering amplitudes are the same of the underling local theory if the background is an exact solution of the local theory itself \cite{Modesto:2021soh}. 

On the other hand, unitarity at all perturbative orders in the loop expansion can be checked via the Cutkosky rules adapting the same procedure of \cite{Briscese:2018oyx} to the action \eq{action}. The proof provided in \cite{Briscese:2018oyx} is valid for all weakly nonlocal field theories with or without gauge symmetry. A detailed analysis has been given in \cite{complexGhosts} for the case of gauge theories. The generalization to gravity, also discussed in \cite{complexGhosts}, is more tedious but equally straightforward once the Cutkosky rules are derived.


\subsection{Structure of nonlocal vertices}

In this section, we show that the nonlocal analytic theory (\ref{action}) has the same divergences of an asymptotic polynomial higher-derivative theory that will be defined at the end of this section. The latter statement requires to study the general analytic structure of the vertices in presence of matter. In other words, we here extend the result in \cite{Eran:1998pga}. 

Let us focus on the nonlocal operator
\be
\sqrt{|g|}\, E_i F_{i j} E_j \, , 
\ee
and consider a vertex with $N^\prime+N+N^{\prime \prime}$ external legs including the graviton and any matter field. 
Here with $N^\prime$ we indicate the legs of the vertex coming from the Taylor expansion of the first operator on the left, i.e. $\sqrt{|g|}\, E_i$, by $N^{\prime \prime}$ we mean the legs of the vertex resulting from the expansion of $E_j$ on the right, and by 
$N$ the number of legs of the Taylor expansion of the form factor 
\be
F_{ij}(\hat\Delta)=\sum_{r=0}^{+\infty} c_r (\hat\Delta^r)_{ij}\, . 
\label{EFFE}
\ee
Explicitly, the variation with respect to any field $\Phi_i$, namely 
\be\label{tusev}
\delta F_{ij}=\sum_r c_r \delta(\hat\Delta^r)_{ij}\,, 
\ee
 contains expressions like
\be
&&(\Delta_{0}^{p_1})_{ii_1}[{\cal I}^{(m_1)}]_{i_1j_2}(\Delta_0^{p_2})_{j_2i_2}[{\cal I}^{(m_2)}]_{i_2j_3}(\Delta_0^{p_3})_{j_3i_3}\dots (\Delta_0^{p_l})_{j_li_l}[{\cal I}^{(m_l)}]_{i_lj_{l+1}}(\Delta_0^{p_{l+1}})_{j_{l+1}j}\nonumber\\
&& 
\to \quad \left[\prod_{k=1}^l{\cal I}^{(m_k)}\right]_{i_1j_2i_2j_3\dots i_l j_{l+1}} \quad 
\prod_{\substack{n=1\\ ( j_1 \equiv i,  \, i_{l+1} \equiv j )}}^{l+1} \hat\Delta^{p_n}_{0,j_n i_n}(Q_n)\,,
\label{DeltaQ}
\ee
where $\mathcal{I}^{(m_i)}$ is the nondiagonal $m_i$-legged piece of the operator 
$\mathcal{I}=\hat\Delta-\hat\Delta_{\Phi_i^{(0)}}$ 
and $\Delta_0 \equiv \hat\Delta_{\Phi_i^{(0)}}$ is the Hessian operator evaluated on the backgrounds $\Phi_i^{(0)}$, namely,
\be
 g^{(0)}_{\mu\nu} = \eta_{\mu\nu} \,\, \mbox{and the matter fields are zero or constant}. 
 \ee
The exponents $m_i$ sum to the total number $N$ of perturbations coming from the expansion of (\ref{EFFE}), namely 
\be\label{tutwe}
m_1+m_2+\dots+m_l\leq N.
\ee 

By the arrow $\to$ in (\ref{DeltaQ}), we mean transformation to momentum space. As we will see below, each operator $\hat\Delta_0$ yields a factor $Q_n^2$, where $Q_n$ is the the sum of all the momenta appearing on the right of any given $\hat\Delta^{p_n}_{0,j_n i_n}(Q_n)$, as evident from the first line of (\ref{DeltaQ}). Since the legs of any vertex are finite in number in perturbation theory, so is the number of momenta in each $Q_n$. In fact, according to \eq{tutwe}, the number of field-dependent functions ${\cal I}^{(m_i)}$ is finite and equals $\ell$ because we expand $F$ to a finite order $N$ in the fields. From each ${\cal I}^{(m_i)}$, we get $m_i$ external legs that collect to $N$ consistently with \eq{tutwe} because we have a total finite number of insertions ${\cal I}^{(m_i)}$. In the last product in (\ref{DeltaQ}), we defined $j_1=i$ and $i_{l+1}=j$. Finally, the operator $\hat\Delta_{0,j_n i_n}(Q_n)$ is diagonal (see \cite{Ohta:2020bsc} where several examples of Hessians have been computed explicitly for gravity, fermions, scalar fields, and Abelian gauge fields). However, the eigenvalues depend on the kind of matter with respect to which we are differentiating. Hence, in short,
\be
\hat\Delta_{0,j_n i_n}(Q_n)= \left( C^{(2)}_{i_n} Q^2_n + C^{(0)}_{i_n} \right)  \, \de_{i_nj_n}\, ,
\label{DC}
\ee
where $C^{(2)}_{i_n}$ and $C^{(0)}_{i_n}$ are constants. As explicit examples, $Q_1= k_1+k_2 + \dots + k_{l +1}$, 
$Q_2 = k_2 + \dots + k_{l+1}$, $\dots$, $Q_{l+1} = k_{l+1}$, where $k_i$ is the sum of the momenta for the $m_i$ legs emanating from ${\cal I}^{(m_i)}$. Given the above expression (\ref{DC}), we can implement the same formula derived in \cite{Eran:1998pga} paying attention to replace $Q_i^2$ with $C^{(2)}_{i} Q^2_i + C^{(0)}_{i}$ for each fixed string of indices $i_1l_2i_2j_3\dots i_l j_{l+1}$ in (\ref{DeltaQ}). Therefore, for each $i_n$ and $j_n$ pair, we can focus on
\be
\hspace{-1cm}\sum_{r = l}^{\infty} c_r \sum_{\{ p_n \} } \prod_{n=1}^{l+1}\hat\Delta^{p_n}_{0,j_n i_n}(Q_n) 
& = & \sum_{r = l}^{\infty} c_r \sum_{\{ p_n \} } \prod_{n=1}^{l+1} \left( C^{(2)}_{i_n} Q^2_n + C^{(0)}_{i_n} \right)^{p_n} 
\nonumber \\
& = & \sum_{n=1}^{l +1} \frac{F( C^{(2)}_{i_n} Q^2_n + C^{(0)}_{i_n} ) }{\prod_{m \neq n} \left[ \left( C^{(2)}_{i_n} Q^2_n + C^{(0)}_{i_n} \right) - \left( C^{(2)}_{i_m} Q^2_m + C^{(0)}_{i_m}  \right) \right] }
\, ,
\label{EramLeo}
\ee
which depends on $i_1, \dots, i_{\ell +1}$.

Using now the identity
\be
\hspace{-1cm}2 \hat{\De}_{ik} \, F^{k} {}_j (\hat{\De}) = \left[ \frac{e^{{\rm H}(\hat{\De}_{\La_*} )} }{e^{{\rm H}(0)} } - 1 \right]_{ij}
\equiv 
\underbrace{\left[ \frac{e^{{\rm H}(\hat{\De}_{\La_*} )} - e^{\gamma_{\rm E}} p ( \hat{\De}_{\La_*})} {e^{{\rm H}(0)}} -1 \right]_{ij}}_{{\left( 2 \Delta F^{\rm conv} \right)}_{i j}} + \underbrace{e^{\gamma_{\rm E}} p( \hat{\De}_{\La_*})_{i j}}_{{\left( 2 \Delta F^{\rm poly} \right) }_{i j}} \, , 
\label{FF2bis}
\ee
and applying the general result (\ref{EramLeo}) to the first operator in square brackets, one can figure out that such operator does not contribute to one-loop divergences because the high-energy scaling of $F^{\rm conv}$ is at most $1/Q_n^2$ or even more decreasing. Notice that, in the vertices, $F^{\rm conv}$ is evaluated on the background $\Phi_i^{(0)}$ and, hence, it is a simple analytic function of $Q_n^2$. 

Therefore, in the rest of the paper we will focus on the asymptotic polynomial form factor $F^{\rm poly}$.


\subsection{Asymptotically polynomial regime}

Making use of the property (\ref{corrInf}) of the nonlocal form factor and given the above considerations, one can see that in the UV limit the theory (\ref{action}) reduces to a local higher-derivative theory whose action reads
\be
\n{UV-act}
S_{\rm UV} = \int \rd^D x \sqrt{|g|} \, \Big[ \cL_{\loc} + \alpha \sum_{k = 0}^{n} E_i \frac{(\hat{\De}^k_{\La_*})^{ij}}{( \La_* )^{[\hat{\De}_{ij}]}} E_j \Big],
\ee
where $\al = e^{\ga_{\rm E}} e^{-{\rm H}(0)}/2$. In particular, in $D=4$ dimensions the action (\ref{UV-act}) can be regarded as a generalization of Stelle's gravity \cite{Stelle:1976gc} in which both gravity and matter obey higher-derivative equations of motion; Stelle's gravity is recovered for $n=0$. In what follows, we shall discuss two examples, namely, the cases of pure gravity and, in the next section, of gravity coupled to a scalar field.

If matter fields are switched off, the UV action~\eq{UV-act} is equivalent to a polynomial-derivative gravity model, which is renormalizable in $D=4$ dimensions if $n=0$ \cite{Stelle:1976gc}, or super-renormalizable, if $n \geqslant 1$ \cite{Asorey:1996hz}. The result on power-counting renormalizability can be easily sketched in the case of pure gravity. To this end, let us omit coefficients and tensorial indices. 
Since we are interested in the UV regime, we focus on the operators with the maximum number of derivatives. The scaling of the graviton propagator in the large-$k^2$ limit is
\be
D_g(k) \propto \frac{1}{ k^{2n+4}}\,.
\label{gravitonScalarProp}
\ee
Although we have many different vertices, we concentrate on those with the maximum number of derivatives, which equals the one present in the propagators. Therefore, we can estimate the maximum degree of divergence of a general amplitude ${\cal A}^{(L)}$ with $L$ loops, $I$ internal lines and $V$ vertices [not to be confused with the potential $V(\ph)$ introduced later] as
\be
{\cal A}^{(L)} &\propto& \delta^4(K) \, \Lambda_*^{2 n (L-1)} \int (\rd^4 k)^L \left( \frac{1}{k^{2 n+4}} \right)^I \left( k^{2 n +4} \right)^{V} \nonumber \\
&=&
\delta^4(K) \, \Lambda_*^{2 n (L-1)} \int (\rd^4 k)^L \left( \frac{1}{k^{2 n +4}} \right)^{L-1} 
\nonumber \\
&=& \delta^4(K) \, \Lambda_*^{2 n (L-1)} \, \left( \Lambda_{\textsc{uv}}\right)^{\omega({\cal G})}  ,  \qquad 
\omega({\cal G}) = 4 - 2 n  (L - 1) \, ,
\label{PC}
\ee
where $\delta^4(K)$ symbolizes the Dirac delta for momentum conservation, we used the topological relation $V=I+1-L$, and $\Lambda_{\textsc{uv}}$ is a UV cutoff. This simple power counting shows that, for $n$ large enough, we have divergences only at one loop and the maximal divergences have dimension four. 

For pure gravity, in the background-field method, the structure of the counterterms is fully fixed by the number of derivatives therein, thanks to general covariance and the uniqueness of the Riemann tensor. Therefore, the theory is renormalizable if all the possible operators of dimension up to four are present in eq.\ (\ref{action}), being super-renormalizable if $n \geqslant 1$. However, the same does not apply when one also has a scalar or other matter fields, in which case one needs to consider in detail the structure of the vertices. This will be our task for the rest of the paper. 


\section{Scalar-tensor model: action}
\label{model gravity-scalar}

In order to study the renormalizability of the model~\eq{UV-act} in the presence of matter, in what follows we consider in full detail the case of gravity coupled to a scalar field. The starting point is the subsidiary local action of a scalar field \emph{minimally coupled}\footnote{As discussed in Appendix~\ref{AppB}, if the local subsidiary action $S_{\loc}$ is nonmimally coupled, then the resultant UV model based on $S_{\rm UV}$ in~\eq{UV-act} is nonrenormalizable. Notice, however, that this statement is not in conflict with the well-established results regarding the renormalizability of scalar fields coupled to gravity, which do require certain nonminimal structures (see, e.g.,~\cite{Buchbinder:2021wzv} for an introduction). In our case, the required nonminimal terms are generated in $S_{\rm UV}$ by the construction of the nonlocal model, even if they are not present in $S_{\loc}$ (see, e.g., Eq.~\eq{1x} below).} to gravity,
\be
\label{aclocal}
S_{\loc} = \int \rd^D x \sqrt{|g|} \left[ \frac{1}{\ka^2} \,R  - \frac12 (\na \ph)^2 - V(\ph) \right].
\ee
The ingredients to construct the UV action~\eq{UV-act} resulting from the nonlocal model~\eq{action} with fields $\Phi^i = (g_{\mu\nu}, \ph)$ are the extremals $E_i$ (i.e., the left-hand side of the local equations of motion $E_i=0$ of the associated local system),
\begin{align}
\label{EOMmunu}
 E^{\mu\nu} &\coloneqq \frac{\de S_{\loc}}{\de g_{\mu\nu}} =  \frac{1}{\ka^2} \left(
- R^{\mu \nu } + \frac{1}{2} g^{\mu \nu } R \right)  
+ \frac{1}{2} (\nabla^{\mu }\ph) (\nabla^{\nu }\ph) 
-  \frac{1}{4} g^{\mu \nu } (\nabla \ph)^2  
-  \frac{1}{2} g^{\mu \nu } V(\ph)\,,
\\
\label{EOMph}
 E_\ph &\coloneqq \frac{\de S_{\loc}}{\de \ph} = \Box \ph -  V'(\ph)\,,
\end{align}
and the Hessian $\hat{\De}_{ij}$, whose explicit expression can be found in Appendix \ref{App3}. Since the action~\eq{aclocal} describes a two-derivative system, the maximal number of derivatives in the Hessian components is two. Moreover, the terms containing two derivatives can only depend on the field $\ph$ through its derivatives (see eqs.~\eq{He11}-\eq{He22}), because the system is minimally coupled. This is a crucial element in the proof of the power-counting renormalizability of the general model~\eq{UV-act}, as we discuss in section \ref{Sec.PC}.

Before studying the most general scenario in which the UV action~\eq{UV-act} contains a form factor given by an arbitrary power $n$ of the Hessian, let us first present the model with $n = 0$. This example is not only the simplest one, but it is also of utmost importance because, as we show in section \ref{Sec.PC}, this is the part of the UV action~\eq{UV-act} that renormalizes in $D=4$ dimensions. In this case, since 
\be
\left(\hat{\Delta}_{\Lambda_{*}}^{n}\right)^{i}{}_{j} \, \bigg\vert_{n=0} = \, \delta^{i}_{j}\,,
\ee
where the identity in the space of fields reads
\be
\label{id}
\delta^{i}_{j} = 
\begin{pmatrix}
\de_{\mu\nu}^{\al\beta}  & 0 \\
0 & 1
\end{pmatrix} ,
\quad \quad \quad 
\de_{\mu\nu}^{\al\beta} \coloneqq \frac12 \left( \de_{\mu}^{\al} \de_{\nu}^{\beta} + \de_{\mu}^{\beta} \de_{\nu}^{\al} \right) ,
\ee
the UV action~\eq{UV-act} reduces to
\be
\label{pequena}
S_{\rm UV}^{n=0} = \int \rd^D x \sqrt{|g|} \left[ \cL_{\loc} + \frac{\al }{\La_*^D} E_{\mu\nu} E^{\mu\nu} + \frac{\al}{\La_*^{2}} E_\ph E_\ph\right].
\ee

It is worthwhile mentioning that a generalization of the model can generate more structures in the UV action. This is achieved if the summation over the repeated field-space indices $i,j$ in~\eq{UV-act} is performed with a nontrivial field-space metric $\mathscr{G}_{ij}$, instead of the identity~\eq{id}. Here we define $\mathscr{G}_{ij}$ such that the line element  $\rd s$ in the space of fields is
\be
\rd s^2 \equiv \int \rd^D x \sqrt{|g|} \, \mathscr{G} _{ij} \, \rd \Psi^i (x)  \rd \Psi^j (x)\,, \qquad \rd \Psi^i  = (\ka \, \rd g_{\mu\nu}, \rd \ph)\,.
\ee
In particular, this means that, in our conventions, the field space metric $\mathscr{G}_{ij}$ is dimensionless.
For a metric-scalar theory, $\mathscr{G}_{ij}$ has the general form
\be
\n{de-field-co}
\mathscr{G}_{ij} = 
\begin{pmatrix}
\mathscr{G}^{\mu\nu,\al\beta}  & \tilde\gamma_3\, g^{\mu\nu} \\
\tilde\gamma_4\, g^{\al\beta} & \tilde\gamma_5
\end{pmatrix} ,
\quad \quad \quad 
\mathscr{G}^{\mu\nu,\al\beta} = \tilde\gamma_1 \de^{\mu\nu,\al\beta} + \tilde\gamma_2 g^{\mu\nu} g^{\al\beta},
\ee
where, in general, $\tilde\gamma_i = \tilde\gamma_i (\ph)$. 
Here we assume that $\tilde\gamma_i$ are constants, otherwise the power-counting renormalization properties of the model would change.\footnote{In fact, if the coefficients $\tilde\gamma_i$ are functions of the scalar field $\ph$, then the field-space metric can generate vertices similar to the ones described in Appendix~\ref{AppB}.}
Moreover, we assume $\tilde\gamma_1,\tilde\gamma_5 \neq 0$, so that the propagator of the resultant model~\eq{UV-act} has a homogeneous behavior both in the gravitational and the scalar sector.

Accordingly, the contravariant field-space metric $\mathscr{G}^{ij}$ is defined as its inverse, 
\be
\n{inv_def}
\mathscr{G}_{ik} \mathscr{G}^{kj} = \de_i^j\,.
\ee
Using~\eq{de-field-co} and~\eq{inv_def}, we find 
\be
\n{de-field}
\mathscr{G}^{ij} = 
\begin{pmatrix}
\mathscr{G}_{\mu\nu,\al\beta}  & \ga_3\, g_{\mu\nu} \\
\ga_4\, g_{\al\beta} & \ga_5
\end{pmatrix} ,
\quad \quad \quad 
\mathscr{G}_{\mu\nu,\al\beta} = \ga_1 \de_{\mu\nu,\al\beta} + \ga_2 g_{\mu\nu} g_{\al\beta},
\ee
where 
\be
\label{OsGamas}
\hspace{-.8cm}\ga_1 = \frac{1}{\tilde\gamma_1},
\qquad
\ga_2 = \frac{   \tilde\gamma_3 \tilde\gamma_4 - \tilde\gamma_2 \tilde\gamma_5 }{\tilde\gamma_1 X },
\qquad
\ga_3 = - \frac{\tilde\gamma_3}{X},
\qquad
\ga_4 = - \frac{\tilde\gamma_4}{X},
\qquad
\ga_5 = \frac{\tilde\gamma_1 + D \tilde\gamma_2}{X}
\ee
and 
\be
\n{O_Famoso_Z}
X = (\tilde\gamma_1 + D \tilde\gamma_2)\tilde\gamma_5 - D \tilde\gamma_3 \tilde\gamma_4.
\ee 
The restrictions on the coefficients $\tilde\gamma_i$ for the existence of the inverse of the field-space metric are, therefore, $\tilde\gamma_1 \neq 0$ (which we also assumed as definition) and $X \neq 0$. Also, using~\eq{OsGamas} and~\eq{O_Famoso_Z} it is straightforward to verify that 
\be 
\label{ga1+Dg2}
\gamma_1 + D \gamma_2 = \frac{\tilde\gamma_5}{X} \neq 0\,,
\ee 
where we used that $\tilde\gamma_5 \neq 0$ by definition. 
Notice that, if the field-space metric is diagonal, i.e., $\tilde\gamma_{3}=\tilde\gamma_{4} = 0$, then we have $\ga_3 = \ga_4 =0$, $\ga_5 = 1/\tilde\gamma_5$ and
$$
\ga_2 = -\frac{\tilde\gamma_2}{\tilde\gamma_1 ( \tilde\gamma_1 + D \tilde\gamma_2 )}\,, 
$$
such that, in this case, $\mathscr{G}_{\mu\nu,\al\beta}  \mathscr{G}^{\mu\nu,\al\beta} = \delta^{\alpha\beta}_{\mu\nu}$, as expected for a block-diagonal matrix.

In the generalized model which uses the field-space metric~\eq{de-field} to contract the indices, we have
\be
E_i (\hat{\De}_{\La_*}^{n})^{i} {}_{j} E^j \Big\vert_{n=0} &=& E_i E^i = E_i \mathscr{G}^{ij} E_j\nonumber\\
& =& E^{\mu\nu} \mathscr{G}_{\mu\nu,\al\beta} E^{\al\beta}
+ \ga_3 E^{\mu\nu} g_{\mu\nu} E_\ph
+ \ga_4  E_\ph g_{\al\beta} E^{\al\beta}
+ \ga_5  E_\ph E_\ph\,
,
\ee
and for $n = 0$ the UV action becomes
\be
\n{act-UV-gen}
S_{\rm UV}^{n=0} = \int \rd^D x \sqrt{|g|} \, \Big[ \cL_{\loc} +  \alpha_1 E_{\mu\nu} E^{\mu\nu} 
+ \alpha_2 E_{\mu}^\mu E^{\nu}_{\nu} 
+ \alpha_3 E_{\mu}^\mu E_\ph
+ \alpha_4 E_\ph E_\ph  \Big]
,
\ee
where
\be
\label{OsAlfas}
\al_1 = \frac{\al }{\La_*^D} \ga_1,
\qquad
\al_2 = \frac{\al }{\La_*^D} \ga_2,
\qquad
\al_3 = \frac{\al}{\La_*^{\frac{D+2}{2}}} (\ga_3 + \ga_4),
\qquad
\al_4 = \frac{\al}{\La_*^{2}} \ga_5\,.
\ee
Notice that the UV action~\eq{pequena} is the particular case of~\eq{act-UV-gen} for $\ga_1 = \ga_5 =1$ and $\ga_2=0=\ga_3+\ga_4$. Thus, for the sake of generality, in the following we work with the action in~\eq{act-UV-gen}.

Performing the explicit calculation of each term in~\eq{act-UV-gen}, we get
\be
E_{\mu\nu} E^{\mu\nu} &=&
\frac{1}{\ka^4} \left[ R_{\mu\nu } R^{\mu\nu } 
+ \left(\frac{D-4}{4} \right) R^2 \right]\nonumber\\
&&
- \frac{1}{\ka^2} \, 
\bigg[
 \left( \frac{D-4}{4} \right)  R  \, (\na \ph)^2
+  R^{\mu\nu } (\nabla_{\mu }\ph) (\nabla_{\nu }\ph )
+ \left( \frac{D-2}{2} \right) V(\ph) R  
\bigg]\nonumber\\
&&
+ \frac{D}{16} \,  (\na \ph)^4
+ \left(\frac{D-2}{4} \right) V(\ph) (\na \ph)^2
+ \frac{D}{4}   V^2(\ph)\,,
\ee
\be
E_\mu^\mu E_\nu^\nu &=&
\frac{1}{\ka^4} \left (\frac{D-2}{2} \right)^2  R^2 
- \frac{1}{\ka^2}  \left[  \left(\frac{D-2}{2} \right)^2   R \, (\na \ph)^2  +  \frac{D(D-2)}{2} \,   V(\ph) R   \right] 
\nonumber\\
&&
+ \left( \frac{D-2}{4} \right)^2  (\na \ph)^4+ \frac{D(D-2)}{4}  \, V(\ph) (\na \ph)^2+ \frac{D^2}{4}  V^2(\ph)   \,,
\ee
\be
E_\mu^\mu E_\ph &=&
 \frac{1}{\ka^2} \left[  \left(\frac{D-2}{2} \right)  R \square \ph    -\left(\frac{D-2}{2} \right) V'(\ph) R \right] 
- \left( \frac{D-2}{4} \right)   (\na \ph)^2 \Box \ph\nonumber\\ 
&&+ \left(\frac{D-2}{4} \right) V'(\ph) (\na \ph)^2
-  \frac{D}{2}  V(\ph) \square \ph  
+ \frac{D}{2}  V(\ph) V'(\ph) \,,
\ee
\be
E_\ph E_\ph &=& (\square \ph )^2 
- 2 V'(\ph) \square \ph  
+ V'^2(\ph)\,,
\ee
where we used the simplified notations $(\nabla \ph)^2 = (\nabla_{\mu }\ph) (\nabla^{\mu } \ph)$ and $(\na \ph)^4 = [(\nabla_{\mu }\ph) (\nabla^{\mu } \ph)]^2$.

Therefore, by using integration by parts and discarding boundary terms, it is possible to cast~\eq{act-UV-gen} in the form
\be
S_{\rm UV}^{n=0} &= & \int \rd^D x \sqrt{|g|} \, \big[ 
 a_1 R_{\mu\nu\alpha\beta }R^{\mu\nu\alpha\beta }  + 
a_2 R_{\mu\nu}R^{\mu\nu} + a_3 R^2 + a_4 (\ph ) R\nonumber\\
&&+ b_1 ( \Box \ph )^2 
+  b_2 (\nabla \ph)^2 \Box \ph +  b_3 (\nabla \ph)^4
+  b_4 (\ph ) (\nabla \ph)^2 
+  b_5 (\ph )\nonumber\\
&&
+  c_1 R   (\nabla \ph)^2 
+  c_2 R^{\mu\nu} ( \nabla_\mu \ph
) ( \nabla_\nu \ph )
+ c_3 R \Box \ph \big],       
\label{1x}
\ee
where
\begin{subequations}
\be
\label{OsAs1}
a_1&=& 0\,,\\
\label{OsAs2}
a_2&=& \frac{1}{\ka^4} \, \al_1\,,\\
\label{OsAs3}
a_3&=& \frac{1}{\ka^4} \left[ \al_1 \left( \frac{D-4}{4} \right) + \al_2 \left (\frac{D-2}{2} \right)^2 \right]\,,\\
\label{OsAs4}
a_4 (\ph) &=& - \frac{1}{\ka^2} \left[  \al_1  \left( \frac{D-2}{2} \right)   V(\ph) + \al_2  \frac{D(D-2)}{2} \,   V(\ph)
+ \al_3 \left(\frac{D-2}{2} \right) V'(\ph) 
- 1 \right],\\
\label{OsBes1}
b_1&=& \al_4\,,\\
\label{OsBes2}
b_2&=& - \al_3 \left( \frac{D-2}{4} \right),\\
b_3 &=& \al_1 \frac{D}{16} + \al_2  \left( \frac{D-2}{4} \right)^2,\\
\label{OsBes4}
b_4 (\ph) &=& \al_1 \left( \frac{D-2}{4} \right)  \, V(\ph) + \al_2  \frac{D(D-2)}{4}  \, V(\ph)
+ \al_3 { \left(\frac{3D-2}{4} \right) V'(\ph)} 
+ 2 \al_4 { V''(\ph)}- \frac{1}{2}\,,\nonumber\\
&&\label{OsBes5}\\
b_5 (\ph) &=& \al_1\frac{D}{4} \,  V^2(\ph) + \al_2 \frac{D^2}{4}  V^2(\ph) 
+ \al_3  \frac{D}{2}  V(\ph) V'(\ph) 
+ \al_4 V'^2(\ph)- V(\ph)\,,\\
\label{OsCes1}
c_1&=& - \frac{1}{\ka^2} \left[  \al_1\left( \frac{D-4}{4} \right) + \al_2 \left(\frac{D-2}{2} \right)^2 \right],\\
c_2&=& - \frac{1}{\ka^2} \, \al_1\,,\\
c_3&=& \frac{1}{\ka^2} \, \al_3 \left(\frac{D-2}{2} \right).\label{OsCes3}
\ee
\end{subequations}
It is important to mention that an action of the form~\eq{1x}, with coefficients depending on $\ph$, represents the most general fourth-derivative system of a scalar coupled to gravity. In fact, as shown in~\cite{Elizalde:1994nz}, any other structure not explicitly written in~\eq{1x} can differ from that action only by total derivatives. In our case, however, only the coefficients of the terms with zero and two derivatives are functions of the scalar field, i.e., $a_4(\ph)$, $b_4(\ph)$ and $b_5(\ph)$, while all the others are constant. This difference with respect to \cite{Elizalde:1994nz} is simply due to the fact that, while the goal of \cite{Elizalde:1994nz} was to consider the most general four-derivative scalar-tensor action, our model is constructed from the recipe above and, in this way, only some coefficients happen to depend on $\ph$.

A comment regarding the coefficient $a_1$ is also in order. Since the dependence of the extremals $E_i$~\eq{EiEq1} on the metric curvature is only via $R_{\mu\nu}$ and $R$, the term quadratic in the Riemann tensor appears in~\eq{1x} with a null coefficient. In $D = 4$, however, the absence of this term in the action is not so critical, because its renormalization is equivalent to the one of the Gauss--Bonnet term, $\cL_{\rm GB} = R_{\mu\nu\al\beta}^2 - 4 R_{\mu\nu}^2 + R^2$, which is topological and does not affect the equations of motion, similarly to the omitted superficial terms.

Finally, with the action in the form~\eq{1x}, we can discuss one of the main differences between the models~\eq{pequena} and~\eq{act-UV-gen}. In the former, we have $\al_3 = 0$, which yields $b_2 = c_3 = 0$ in~\eq{1x}; this means that the fourth-derivative terms with an odd number of fields $\ph$, namely, $(\nabla \ph)^2 \Box \ph $ and $ R \Box \ph$, are not present in~\eq{pequena}. As a result, if the scalar potential $V(\ph)$ is an even function, the action~\eq{pequena} is also even in $\ph$. This feature is not {exclusive to} the action~\eq{pequena}; indeed, it happens if the field-space metric~\eq{de-field-co} has $\tilde\gamma_3 + \tilde\gamma_4 =0$ (in particular, this is true for a diagonal metric).\footnote{The same property regarding the parity of the UV action on $\ph$ is also true for the model with $n > 0$ based on a diagonal field-space metric; see section \ref{SubSecD=4}.} Nevertheless, as we discuss in the following section, the specific form of the field-space metric does not affect the power-counting renormalizability of the model.


\section{Scalar-tensor model: power counting}
\label{Sec.PC}

In this section, we investigate the structure of divergences of the quantum effective action associated with the general model~\eqref{UV-act} with fields $\Phi^i = (g_{\mu\nu}, \ph)$, presented in the previous section. The power-counting analysis based on Feynman diagrams in flat spacetime is efficient in this case since, in the framework of the background-field method using dimensional regularization (assumed here), divergences are covariant and, moreover, there are theorems which guarantee that they are local (see, e.g.,~\cite{Lavrov:2019nuz} and references therein for a detailed discussion on the gauge-invariant renormalization of quantum gravity, and~\cite{tHooft:1974toh,Deser:1974cz,gosa,vandeVen:1991gw} for explicit calculations). These results, guaranteed to hold here because the UV limit of the theory is local, significantly constrain the form of the counterterms and, by evaluating the number of derivatives acting in the external lines of a given diagram, one can start to classify the possible structure of divergences associated.

The main elements to employ in the power-counting analysis are the scaling of the propagators and the structure of the interaction vertices. In what concerns the former, we recall that for the model~\eqref{UV-act} with local subsidiary action \eqref{aclocal} the propagators of the gravitational and scalar sectors are homogeneous and scale like $1/k^{2n+4}$. The propagators of the Faddeev--Popov ghosts can have this same behavior by introducing an appropriate weight operator, in a similar way as done in~\cite{Modesto:2011kw} (see also the discussion in~\cite{Lavrov:2019nuz}). 

Regarding the interaction vertices, the number of scalar legs in a given vertex is limited  by the form of the potential $V(\ph)$ and the power $n$ of the Hessian in the UV action~\eqref{UV-act}. On the other hand, the number of gravitational legs is unrestricted, as they are originated from the expansion of the nonlocal action in terms of the metric fluctuation around Minkowski spacetime, through $g_{\mu\nu} = \eta_{\mu\nu} + h_{\mu\nu}$. Even though this expansion produces an infinite number of vertices, the number of derivatives in such vertices is always bounded by $2n+4$; in other words, these vertices can have $0,2,4,\ldots, 2n+4$ derivatives distributed among matter, gravity and ghost legs, depending on the vertex.


\subsection{Case \texorpdfstring{$n=0$}{n=0}, \texorpdfstring{$D=4$}{D=4}}
\label{Sec.PCn=0}

It is instructive to begin with the simple case $n=0$ in $D=4$ dimensions, based on the action~\eqref{act-UV-gen}. In the following sections, we generalize the discussion for the case of arbitrary $n$ and higher spacetime dimensions. 
When $n=0$ and $D=4$, the propagators scale homogeneously as $1/k^4$ and the number of derivatives in the vertices are either 4, 2 or 0, which can be distributed among matter, gravity and ghost legs. For instance, pure-gravity and graviton-scalar vertices with 4 derivatives are originated from the terms in the action~\eq{act-UV-gen} with the coefficients $a_1$, $a_2$, $a_3$, $b_1$, $b_2$, $b_3$, $c_1$, $c_2$ and $c_3$, vertices with 2 derivatives come from the terms with $a_4$ and $b_4$ and, finally, vertices without derivative come from the term with $b_5$.

Taking all these results into consideration, the superficial degree of divergence $\om(\mathcal{G})$ of a given diagram $\mathcal{G}$ with a number $L$ of loops, $I$ internal lines, $V_{2N}$ vertices with $2N$ derivatives ($N=2,1,0$), and $d$ derivatives in the external lines can be expressed in the simple formula,
\be
\om(\mathcal{G}) = 4L - 4 I + 4 V_4 + 2 V_2 - d\, ,
\ee
regardless of the exact types of internal lines or vertices in $\mathcal{G}$. The central argument is that each loop contributes a factor $k^4$, each internal line a factor $k^{-4}$, while the contribution of vertices depends on the number of derivatives they have. 
Using the topological relation
\be
L = I - V_4 - V_2 - V_0 + 1\,, 
\ee
we obtain
\be\label{om53}
\boxed{
\om(\mathcal{G}) = 4 - 2 V_2 - 4 V_0 - d\,.
}
\ee
The formula~\eq{om53} means that:
\begin{itemize}
\item[(i)] All diagrams with more than 1 vertex without derivative ($V_0>1$) are finite.
\item[(ii)] All diagrams with more than 2 vertices with 2 derivatives ($V_2>2$) are finite.
\item[(iii)] The number of vertices with 4 derivatives does not affect the superficial degree of divergence, since~\eq{om53} does not depend on $V_4$.
\item[(iv)] Logarithmically divergent diagrams ($\om = 0$) can be classified by the number $d$ of derivatives in the external lines: 
\\
(a) If $d=4$, the diagrams must have $V_0=V_2=0$; 
\\
(b) If $d=2$, the diagrams must have $V_0=0$ and $V_2=1$; 
\\
(c) If $d=0$, the diagrams must have either $V_0=0$, $V_2=2$  or  $V_0=1$, $V_2=0$. 
\\
In all these situations, $V_4$ is arbitrary.
\end{itemize}

Note that, for diagrams containing only gravity lines, the estimate \eq{om53} of the superficial degree of divergence reproduces the well-known result of pure Stelle gravity \cite{Stelle:1976gc}. In fact, as mentioned before, the theory~\eqref{act-UV-gen} in $D=4$ dimensions can be regarded as Stelle gravity coupled to a fourth-derivative scalar field; thus, they coincide when the scalar field is switched off.

Before considering each of the cases in item (iv) separately, an observation concerning the vertices with four derivatives is in order. Since all the coefficients $a_1$, $a_2$, $a_3$, $b_1$, $b_2$, $b_3$, $c_1$, $c_2$ and $c_3$ of the terms in the action~\eqref{act-UV-gen} that generate such vertices do not depend on $\ph$, the only dependence these terms can have on the scalar field is through its derivative, $\nabla \ph$. Thus, it is immediate to see that \emph{there will always be at least one derivative associated to each scalar line in the vertices with four derivatives}.\footnote{This result is true even upon integration by parts in the classical action, since vertices are defined as higher-order functional derivatives and are invariant under this operation.} In particular, an external scalar line originated from a vertex of this type will always contribute to the number $d$ of derivatives in the external lines. 
 The situation is analogous to the one considered in \cite{Stelle:1976gc} in the context of four-derivative gravity, in which there is a linkage between derivatives and the lines corresponding to scalar and ghost fields. Therefore, although formula~\eq{om53} does not explicitly depend on $V_4$, the increase on the number of external scalar lines originated from these vertices can only reduce $\om(\mathcal{G})$, thus improving the convergence of the loop integrals.

In order to keep track of the powers of $\ph$ that can appear in the counterterms, let us assume that the potential has the form of a monomial,
\be
\label{potencial}
V(\ph) = \la \ph^\ell .
\ee
We can now continue the analysis of the structure of divergences, expanding the cases (a)-(c) of the above item (iv):
\begin{itemize}
\item[(a)] 
The divergences involve terms with $d=4$ derivatives. In what concerns the pure gravity sector, they correspond to the curvature-squared terms in~\eqref{1x}. Regarding the scalar field, notice that since $V_0=V_2=0$, all vertices in such diagrams have four derivatives and, according to the above observation, there will always be at least one derivative acting upon each scalar line. Therefore, the divergences can be at most quartic in $\nabla\ph$ or quadratic in $\Box\ph$, corresponding to the structures in~\eqref{1x} with coefficients $b_{1,2,3}$ and $c_{1,2,3}$ (apart from topological and boundary terms).

Depending on the form of the field-space metric, less diverging structures can occur. For instance, the divergences  with four derivatives and an odd number of $\ph$, namely, proportional to $\sqrt{\vert g\vert} R \Box \ph $ and $\sqrt{\vert g\vert} (\nabla \ph)^2 \Box \ph $, can only be generated if the field-space metric is such that $\tilde\gamma_3 + \tilde\gamma_4 \neq 0$. To prove this statement, notice that the coefficients $b_2$ and $c_3$ of these terms in the action~\eqref{1x} are proportional to $\al_3$ ---which, in turn, vanishes if and only if $\tilde\gamma_3 + \tilde\gamma_4 = 0$ (see eqs.~\eq{OsGamas} and~\eq{OsAlfas}). Hence, if $\al_3 = 0$ the model has no four-derivative vertices with an odd number of scalar legs. Now, the number $\mathcal{E}_{\ph}$ of external scalar lines in any diagram is given by
\be 
\label{Ext}
\mathcal{E}_{\ph} = \sum_{k=1}^{V_0+V_2+V_4} \nu_{k}  - 2 I_\ph,
\ee
where $\nu_{k}$ is the number of scalar legs in the $k$-th vertex and $I_\ph$ is the number of internal scalar lines in the diagram. Since $V_0=V_2=0$ for the case under consideration, it follows that, if $\nu_k$ is even for all $k$ (as it is if $\al_3 = 0$), then $\mathcal{E}_{\ph}$ is also even, proving that divergences with an odd number of $\ph$ can only occur if $\al_3 \neq 0$.

\item[(b)] In the case of $d=2$, the only possible forms for the logarithmic divergences are the two-derivative operators 
\be
\label{2derst}
\quad \sqrt{\vert g\vert} \ph^k R\,,  \qquad \sqrt{\vert g\vert} \ph^k ( \nabla \ph)^2,
\ee
with $k \in \mathbb{N}$. Moreover, they are 
originated from diagrams containing an arbitrary number of vertices with 4 derivatives and only one vertex with 2 derivatives, which is related to the terms $ \sqrt{\vert g\vert} a_4(\ph ) R$ and $ \sqrt{\vert g\vert}  b_4(\ph ) ( \nabla \ph )^2$ (see eq.~\eqref{1x}), namely, to the structures
\be
&&(\al_1 + 4 \al_2) \sqrt{\vert g\vert} \ph^\ell R , \qquad 
\al_3 \sqrt{\vert g\vert} \ph^{\ell-1} R\,, \nonumber\\
&&(\al_1 + 4 \al_2) \sqrt{\vert g\vert} \ph^\ell ( \nabla \ph )^2  , \qquad 
\al_3 \sqrt{\vert g\vert} \ph^{\ell-1} ( \nabla \ph )^2  , \qquad 
\al_4 \sqrt{\vert g\vert} \ph^{\ell-2} ( \nabla \ph )^2  , \nonumber\\
&&\sqrt{\vert g\vert} ( \nabla \ph )^2,\label{VDD}
\ee
where we used~\eq{potencial} and reintroduced the coefficients $\al_i$ (see eqs.~\eq{OsAs4} and~\eq{OsBes4}) to keep track of how they are originated, depending on the metric on the space of fields. Since these terms can have up to $\ell+2$ scalar legs ---of which at most $\ell$ are without derivatives (and since all the scalars attached to four-derivative vertices carry derivatives), the maximal number of external lines without derivatives is $\ell$, whence $k \leqslant \ell$. This result guarantees that the counterterms related to the divergences with $d=2$ are in a finite number. Yet, it is possible to develop a more detailed analysis of the possible values for $k$ (of course, according to power-counting arguments), as we show in what follows.

To this end, recall that the action~\eqref{1x} generates vertices with four derivatives and an odd number of scalar fields if, and only if, $\al_3 \neq 0$. In particular, in this circumstance there are graviton-scalar vertices with only one scalar, and it is possible to have logarithmically diverging diagrams formed by connecting scalar legs of a vertex originated from~\eq{VDD} to internal graviton or scalar lines in vertices with four derivatives and only one $\ph$. Since $\al_1 + 4 \al_2 \neq 0$ (see eq.~\eq{ga1+Dg2}), the term with maximal power $k=\ell$ will always be present and the outcome is that, if  $\al_3 \neq 0$, the divergences have the form~\eq{2derst} with $k\in\lbrace \ell, \ell-1 , \ell-2,\ldots , 0 \rbrace$. On the other hand, if $\al_3=0$, then we have $k \in \lbrace \ell, \ell-2, \ell-4, \ldots, 0 \rbrace$.

\item[(c)] Finally, the divergences with $d=0$ can only take the general form 
\be 
\label{phi^m}
\sqrt{\vert g\vert} \ph^k. 
\ee
There are two classes of diagrams that contribute to this type of divergences, namely, the ones with $V_0=0$, $V_2=2$ and the ones with $V_0=1$ and $V_2=0$; in both cases, $V_4$ is arbitrary. In what follows we show that $k \leqslant 2\ell$, ensuring that these counterterms are in limited number. 

First, let us consider the case $V_2=2$, which is very similar to the case (b) just discussed above. Again, each of these two-derivative vertices have up to $\ell+2$ scalar legs, of which at most $\ell$ are without derivatives. Since in these divergent diagrams all the derivatives must be in the internal lines, the maximal number of external lines without derivatives is $\ell V_2 = 2\ell$, whence $k \leqslant 2\ell$.

Second, for the case $V_0=1$, the vertices without derivatives are originated from the terms in the action (see eqs.~\eqref{1x} and~\eq{OsBes5})
\be
\label{VnD}
(\al_1 + 4\al_2) \sqrt{\vert g\vert}  \ph^{2\ell}, \qquad  \al_3 \sqrt{\vert g\vert} \ph^{2\ell-1}, \qquad \al_4 \sqrt{\vert g\vert} \ph^{2\ell-2}, \qquad \sqrt{\vert g\vert} \ph^{\ell}.
\ee
Since the maximal number of external lines without derivatives is $2\ell$, it follows that $k \leqslant 2\ell$ also for this type of diagram.

The exact possible values of $k$, by power-counting arguments, can be deduced by a reasoning similar to the one used in item (b). To summarize the results, if $\al_3 \neq 0$ there will be divergences of the type~\eq{phi^m}  with $k\in\lbrace 2\ell, 2\ell-1 , 2\ell-2,\ldots, 0 \rbrace$, whereas $k\in\lbrace 2\ell, 2\ell-2,\ldots, 0 \rbrace \cup\lbrace \ell, \ell-2, \ell-4, \ldots, 0 \rbrace$ if $\al_3 = 0$. The former case is immediate; as for the latter, recall that there is no four-derivative vertex with and odd number of scalars if $\al_3=0$. In this case, any logarithmically diverging diagram formed by an arbitrary number of four-derivative vertices and only one vertex with no derivative coming from the first and third terms in~\eq{VnD} must have an even number of external scalar legs (apply, e.g., eq.~\eq{Ext}). Therefore, the first and third terms in~\eq{VnD} generate divergences of the type~\eq{phi^m} with $k\in\lbrace 2\ell, 2\ell-2,\ldots, 0 \rbrace$, while the fourth term (which comes from $\cL_{\loc}$ in~\eq{action}) will generate divergences with $k\in\lbrace \ell, \ell-2, \ell-4, \ldots \rbrace$. The same consideration applies, \textit{mutatis mutandis}, to the diagrams with two two-derivative vertices originated from~\eq{VDD}. In particular, if $\ell$ is even and $\al_3 = 0$, all the divergences of this type have an even number of scalars ---this is indeed expected, since diagrams with an odd number of external scalar legs can only occur if the model has vertices with an odd number of scalars; see eq.~\eq{Ext}. 
\end{itemize}
Note that, in the case of pure gravity, there would still be diagrams falling into the above three categories, corresponding to the renormalization of the fourth-derivative terms, the Einstein--Hilbert term, and the cosmological constant.

The conclusion is that the counterterms are in finite number, regardless of the choice for the field-space metric~\eq{de-field-co}, and the model is power-counting renormalizable. Moreover, powers of $\ph$ higher than the ones already present in \eq{act-UV-gen} do not occur in the counterterms and a general model defined by the action~\eqref{1x} in $D=4$ dimensions with independent couplings
is multiplicatively renormalizable (up to boundary and topological terms) if the potential has the form $V(\ph) = \sum_{k=0}^\ell \la_k \ph^k$ for a certain~$\ell$. In the case of the specific model considered here, the coefficients $a_{1,2,3,4}$, $b_{1,2,3,4,5}$ and $c_{1,2,3}$  of the UV action~\eqref{1x} are not independent (they depend on only a few free parameters $\alpha_{1,2,3,4}$; see~\eqref{act-UV-gen}). Therefore, although the model is power-counting renormalizable, it may not be multiplicatively renormalizable because 
the number of divergences is bigger than the one of free parameters in the action. This problem can be solved in the models with $n>0$, which, as we show in section \ref{SubSecD=4}, are power-counting super-renormalizable and can be made finite by the introduction of appropriate killer operators (see the discussion in section \ref{finite}). Alternatively, another possible solution might be to apply the technique used for nonlocal theories in~\cite{Tomboulis:1997gg,Modesto:2011kw,Modesto:2014lga}, as discussed in section \ref{secnew}.

One of the central arguments in the proof of the power-counting renormalizability is the fact that the terms in the action that generate vertices with four derivatives cannot yield external scalar legs without derivatives. This hypothesis is true \emph{provided that the local subsidiary action does not have nonminimal terms}. However, if nonminimal terms are present in $\cL_{\loc}$, the hypothesis is violated and the theory is \emph{not} power-counting renormalizable; see Appendix~\ref{AppB}. The theory studied in \cite{Modesto:2022asj,Calcagni:2022tuz} does not have any such nonminimal terms in $\cL_{\loc}$; at the same time, the full nonlocal action of the model generates the nonminimal terms required for renormalization, as it can be seen from Eq.~\eq{1x} and the above discussion.


\subsection{General case}
\label{Sec.PC-gen}

We now consider the general theory with UV-limit action~\eq{UV-act}, i.e., the case with $n\geqslant0$ and $D$ arbitrary. It is useful to define
\be
N \coloneqq n +2\,,
\ee
such that the operators in the action have either 0, 2, $4, \ldots, 2N$ derivatives, and no term with more than $2N$ derivatives. 
Note that, as in the case of section \ref{Sec.PCn=0}, the terms with maximal number of derivatives $2N$ can only depend on the scalar field through its derivatives $\nabla \ph$. This happens because\footnote{For a more explicit proof of this statement, see the discussion related to eq.~\eq{commax} below.} the terms with maximal number of derivatives in the extremals \eq{EOMmunu} and \eq{EOMph} and in the Hessian~\eq{HssianApp} can depend only on $\nabla \ph$, and also because we assumed that the metric in the space of fields~\eq{de-field-co} does not depend on $\ph$.

In a $D$-dimensional spacetime, the superficial degree of divergence of a diagram $\mathcal{G}$ is given by
\be
\om(\mathcal{G}) = DL - 2N I +  \sum_{k = 0}^N 2 k \, V_{2k} - d\,,
\ee
where $V_{2k}$ is the number of vertices with $2k$ derivatives. 
Using the topological relation
\be
L = I -  \sum_{k = 0}^N \, V_{2k} + 1\,, 
\ee
we obtain
\be
\n{pc-gen-fi}
\boxed{
\om(\mathcal{G}) = D - (2N - D) ( L - 1 ) -  \sum_{k = 1}^N  2 k  \, V_{2(N-k)} - d\, .
}
\ee
This means that:
\begin{itemize}
\item[{(I)}] A requirement for power-counting super-renormalizability is
\be
\n{con1}
2 N > D\,,\qquad \Longrightarrow\qquad n>\frac{D-4}{2}\,.
\ee
\item[{(II)}] One-loop divergences are always present. Indeed for $L=1$, we get
\be
\om (\mathcal{G}^\textrm{1-loop}) \, = \, D \, - \,  \sum_{k = 1}^N  2 k  \, V_{2(N-k)} \, - \, d\, .
\ee
Now, if eq.~\eq{con1} holds true,  we can rewrite the above equation as 
\be
\label{59}
\om (\mathcal{G}^\textrm{1-loop}) = D -  \sum_{k = 1}^{\lfloor \frac{D}{2} \rfloor} 2 k  \, V_{2(n-k)} \, -  \sum_{k = \lfloor\frac{D}{2}\rfloor+1}^N 2 k  \, V_{2(N-k)} - d\,,
\ee
where $\lfloor x \rfloor$ is the floor function, i.e., 
\be
\Big\lfloor \frac{D}{2} \Big\rfloor
= 
\left\{ 
\begin{array}{l l}
\dfrac{D}{2} \, ,  &  \text{if } D \text{ is even},\\
&\\
\dfrac{D-1}{2}  \, , &  \text{if } D \text{ is odd}.\\
\end{array} \right . \,
\ee
Thus, every diagram with at least one vertex with $k \in \{ {\lfloor\frac{D}{2}\rfloor+1}, {\lfloor\frac{D}{2}\rfloor+2}, \ldots, N -1 , N \} $ is already superficially finite ($\om <0$). The only diagrams that have logarithmic divergences are those such that the number of vertices $V_{2N}$ is arbitrary and
\be
\label{61}
d = D -  \sum_{k = 1}^{\lfloor \frac{D}{2} \rfloor}  2k  \, V_{2(N-k)}\,.
\ee
Therefore, the one-loop logarithmic divergences have up to $D$ derivatives and they are in a finite number because, as before, all scalar legs in the vertices with $2N$ derivatives carry at least one derivative.

\item[{(III)}] Since higher loops improve the convergence of the diagram when $2N > D$, in this case, the set of possible vertices in divergent diagrams is smaller at each order of the loop expansion. For instance, diagrams with $L=2$ with any number of vertices $V_{2(N-\lfloor \frac{D}{2} \rfloor)}$ are already superficially finite (there may be still divergent subdiagrams). Moreover, if $N > D$, then only the one-loop divergences remain.

\item[{(IV)}] Finally, if $2N = D$, the term $(2N - D) ( L - 1 )$ in~\eq{pc-gen-fi} vanishes and the model can be strictly renormalizable by power counting, generalizing the $n=0$ and $D=4$ result of section \ref{Sec.PCn=0}. In this case, we have
\be\label{stric}
\om (\mathcal{G}) = D -  \sum_{k = 1}^{{\lfloor \frac{D}{2} \rfloor}}  2k  \, V_{2(N-k)}  - d\,,
\ee
even for multi-loop diagrams (compare with eq.\ \eq{61}). Hence, the number of counterterms at any loop order is finite also in this case, and the logarithmic divergences must have a total number of derivatives given by
\be
d = D -  \sum_{k = 1}^{{\lfloor \frac{D}{2} \rfloor}}  2 k  \, V_{2(N-k)}\,.
\ee

\end{itemize}

\subsection{Case \texorpdfstring{$n>0$}{n>0}, \texorpdfstring{$D=4$}{D=4}}
\label{SubSecD=4}

Particularizing the above discussion for $D=4$ and $n>0$, we notice that only the diagrams containing vertices with $2N$, $2N-2$ and $2N-4$ derivatives can generate logarithmic divergences. Power-counting super-renormalizability can be achieved if
\be
N > 2\,.
\ee 
In this case,  the relation~\eq{61} satisfied by the one-loop logarithmically divergent diagrams reads
\be
 d = 4 - 2 V_{2N-2} - 4 V_{2N-4}\,,
\ee
with $V_{2N}$ arbitrary. 
As mentioned in item {(III)}, if $N>2$, diagrams with higher loops cannot generate other types of divergences, thus making it sufficient to perform the analysis of these one-loop diagrams. Therefore, once more, one can classify the possible forms of the counterterms by their number of derivatives $d$:
\begin{itemize}
\item[{A)}] $d=4$ with an arbitrary number $V_{2N}$ of $2N$-derivative vertices and  $V_{2N-2} =  V_{2N-4} = 0$. In this case, the counterterms have four derivatives and all the dependence on the scalar field is through $\nabla\ph$ or $\Box \ph$.
 They have the same structure as the terms proportional to $a_{2,3}$, $b_{1,2,3}$ and $c_{1,2,3}$ in~\eq{1x}, namely,
\be
\hspace{-.5cm}&&\sqrt{|g|}  R_{\mu\nu}R^{\mu\nu} , \qquad 
\sqrt{|g|}  R^2 , \qquad
\sqrt{|g|}  ( \Box \ph )^2 , \qquad 
\sqrt{|g|}  (\nabla \ph)^2 \Box \ph\,, 
\nonumber \\
\hspace{-.5cm}&&\sqrt{|g|}  (\nabla \ph)^4 , \qquad 
\sqrt{|g|}   R   (\nabla \ph)^2 , \qquad 
\sqrt{|g|}  R^{\mu\nu} ( \nabla_\mu \ph) ( \nabla_\nu \ph ) , \qquad 
\sqrt{|g|}  R \Box \ph \, .\label{Ozum23}
\ee 
The term proportional to $a_1$, i.e., the Riemann-squared term, might also be present. However, in $ D = 4$, the renormalization of $R_{\mu\nu\al\beta}^2$ is equivalent to the renormalization of the topological Gauss--Bonnet term~\cite{tHooft:1974toh}; which can be included in the action~\eq{action} without changing its equations of motion~\eq{LEOM} (the same applies to the superficial terms with four derivatives, such as $\Box R$ and others).

\item[{B)}] $d=2$, with $V_{2N}$ arbitrary, $V_{2N-2} = 1$ and $V_{2N-4} = 0$. The general structure of the divergences are
\be
\label{A4B4}
\sqrt{\vert g\vert} A_4 (\ph ) R \qquad {\rm and} \qquad
\sqrt{\vert g\vert} B_4 (\ph ) ( \nabla \ph )^2 .
\ee

\item[{C)}] $d = 0$; again, in this case we have two possibilities: $V_{2N-2} = 2$, $V_{2N-4} = 0$ or $V_{2N-2} = 0$, $V_{2N-4} = 1$. In both cases $V_{2N}$ is arbitrary. The counterterms have the form
\be
\label{B5}
\sqrt{\vert g\vert} B_5 (\ph)\,.
\ee 
\end{itemize}

The structure of the counterterms~\eq{A4B4} and~\eq{B5} is similar to eqs.~\eq{2derst} and~\eq{phi^m} of the case $n=0$, as the former are the generalization of the latter ones. Like we did in section \ref{Sec.PCn=0}, in order to identify the bounds on the form of $A_4 (\ph)$ and $B_{4,5} (\ph)$ it is necessary to check the exact structure of the higher-derivative sector of the action~\eqref{UV-act}, namely,
\be
\label{ac}
\sum_{k = 0}^{N-2} E^j (\hat{\De}^k)_{ij} E^j .
\ee
After inspecting the structure of the Hessian (see Appendix~\ref{App3}), a moment's reflection shows that $(\hat{\De}^{N-2})_{ij}$ can be written as a combination of terms of the type (in schematic form omitting the indices)
\be
\label{68}
\na^{2\ell_1}  
\left[ \left( \na \ph \right)\na \right]  ^{\ell_2}  \,
\left( \na^2 \ph \right) ^{\ell_3}  \,
\left[ \left( \na \ph \right) \left( \na \ph \right) \right] ^{\ell_4}  \,
{\cal R}^{\ell_5}  \,
V^{\ell_6}(\ph) \,
V^{\prime\ell_7}(\ph) \,
V^{\prime\prime\ell_8}(\ph)\,  ,
\ee
constrained by 
\be 
\sum_{i=1}^8 \ell_i = N-2\, . 
\ee 
We use the notation ${\cal R}^{\ell}$ for a product of a number $\ell$ of curvature tensors.

The terms in~\eqref{ac} with the highest number of derivatives $2N$ are originated from the terms~\eqref{68} with $\ell_6=\ell_7=\ell_8 = 0$ and the two-derivative terms coming from the extremals \eqref{EOMmunu} and~\eqref{EOMph}. They have the form
\be
\label{commax}
\sqrt{|g|} (\na^p {\cal R}^{q}) \prod_{k=1}^{2N-2} (\na^k \ph)^{r_k}
,
\qquad
p + 2q + \sum_{k=1}^{2N-2} k \, r_k = 2N\, . 
\ee
Hence, $\ph$ in such terms always occurs with at least one derivative, which means that each external scalar leg originated from these vertices contributes at least one derivative to the number $d$ of derivatives in the external lines.

On the other hand, the terms in eq.~\eqref{ac} with $2N-2$ derivatives may contain structures without derivatives acting on $\ph$. Let us assume, again, a potential of the form $V(\ph) \propto \ph^\ell$. Then, the terms of this type \emph{with maximal number of scalars without derivative} have the form
\be
\label{V2n-2}
\sqrt{|g|} V(\ph) (\na^p {\cal R}^{q}) \prod_{k=1}^{2N-2} (\na^k \ph)^{r_k}
,
\qquad 
p + 2q + \sum_{k=1}^{2N-2} k \, r_k = 2N-2\, .
\ee
Likewise, the terms with $2N-4$ derivatives and maximal number of scalars without derivative are of the type
\be
\label{V2n-4}
\sqrt{|g|} V^2(\ph) (\na^p {\cal R}^{q}) \prod_{k=1}^{2N-4} (\na^k \ph)^{r_k}
,
\qquad
p + 2q + \sum_{k=1}^{2N-4} k \, r_k = 2N-4\, .
\ee

As discussed above, the analysis of these three types of terms~\eq{commax}--\eq{V2n-4} is sufficient for studying the possible counterterms in~\eq{A4B4} and~\eq{B5}. Let us start with the latter. Since diagrams formed exclusively by vertices with $2N$ derivatives always carry at least one derivative in each external scalar leg, these diagrams do not contribute to $B_5 (\ph)$. The logarithmically diverging diagrams with $V_{2N}$ arbitrary and $V_{2N-2}=2$, on the other hand, can have at most $2\ell$ external scalar legs without derivatives (since each vertex contributes with at most $\ell$ external scalar legs of this type; see~\eq{V2n-2}). The same happens with the logarithmically diverging diagrams with $V_{2N}$ arbitrary and $V_{2N-4}=1$. Therefore, $B_5 (\ph)$ has the general form
\be
\label{B5Vl}
B_5 (\ph) = \sum_{k=0}^{2\ell} \la_k \ph^k .
\ee

Similarly, $A_4 (\ph)$ and $B_4 (\ph)$ in~\eq{A4B4} are bounded by the terms in the action with $2N-2$ derivatives. So, 
for $V(\ph) \propto \ph^\ell$ we have
\be
\label{A4B4Vl}
A_4(\ph) = \sum_{k=0}^{\ell} \bar{\la}_{k} \, \ph^k , \qquad  B_4 (\ph) = \sum_{k=0}^{\ell} \tilde{\la}_{k} \, \ph^k .  
\ee
This means that \emph{there are no counterterms of these forms with powers of $\ph$ higher than those present in the classical action}. 

Last but not least, we remark that if the metric~\eq{de-field-co} in the space of fields is diagonal, i.e., $\tilde\gamma_3 = \tilde\gamma_4 = 0$, then the action~\eq{UV-act} does not contain terms with maximal number of derivatives and an odd number of $\ph$. This can be proved by noticing that only the off-diagonal components ($\hat{\De}_{12}$ and $\hat{\De}_{21}$) of the Hessian~\eq{HssianApp} contain such terms, whereas the terms in $E^{\mu\nu}$ with derivatives are even in $\ph$, and those in $E_\ph$ are odd; see eqs.~\eq{EOMmunu} and~\eq{EOMph}. Therefore, if  $\tilde\gamma_3 = \tilde\gamma_4 = 0$, the quantity $E^i(\hat{\De}^{k})_{ij}E^j$ (for any $k$) cannot contain terms with maximal number of derivatives and an odd number of $\ph$. If, in addition, the potential $V(\ph)$ is even, all the terms in the UV action are even in $\ph$, which means that all the diagrams have an even number of external scalar lines. In particular, if  $\tilde\gamma_3 = \tilde\gamma_4 = 0$, then the divergences of the type $\sqrt{|g|} ( \nabla \ph)^2 \Box \ph $ and $\sqrt{|g|} R \Box \ph$ cannot be generated, which generalizes a similar statement made in section \ref{Sec.PCn=0} regarding the case $n=0$.


\section{General renormalizability statements}\label{genpc}

Based on the above outcomes, in this section we state general renormalizability properties that apply to the theory (\ref{action}) with any content of matter. In particular, we can make two statements providing necessary conditions for super-renormalizability, namely, to have a finite number of divergences (excluding possible divergent subdiagrams and taking on board the caveat for gravity mentioned in footnote \ref{foot1}).
\begin{theorem} 
In order to have only a finite number of superficially divergent diagrams, in all operators having the higher number of derivatives (the same number as in the kinetic operator), any matter field must carry at least one derivative.
\end{theorem}
Also, in the power-counting analysis we had to assume that the potential be polynomial, and it turned out that, in general, all its monomial were renormalized. Therefore, one might be tempted to state a second necessary condition:
\begin{theorem}[\textbf{too strong}]
In order to have only a finite number of superficially divergent diagrams, all operators must contain a finite number of matter fields. In particular, the potential for the scalar field has to be polynomial, it cannot be an analytic nonpolynomial function (like the Starobinsky potential) because it will produce an infinite number of counter-terms.
\end{theorem}
However, this statement turns out to be too strong and can be relaxed. Although the analysis reported in the previous sections about the number of divergences is correct, it provides an overestimation of their number. As we will show with several explicit examples, 
\begin{theorem}
Once (i) the requirement {\bf 1} is secured, (ii) the number of derivatives in the vertices is less than in the propagator, and (iii) the theory has divergences only at one loop, then the number of divergent one-loop diagrams is finite, regardless of the type of potential.
\end{theorem}
In other words, theories satisfying the condition {\bf 1} and having a sufficiently high number of derivatives in the kinetic term are super-renormalizable, independently of whether the scalar potential is a polynomial or a nontrivial analytic function with infinitely many terms. 

In sections \ref{secex1}--\ref{secex3}, we illustrate these features in three examples of scalar fields living on Minkowski spacetime.


\subsection{Two-derivative scalar field theory with polynomial potential}\label{secex1}

As a first simple example, one can consider a two-derivative scalar field theory in $D=4$ dimensions with the polynomial potential
\be
V(\ph) = \sum_{n=3}^N c_n V^{(n)}(\ph)\,, 
\label{GV}
\ee
in which any term in the above sum is a monomial $V^{(n)} \propto \ph^n$. The complete action reads:
\be
S_{\partial 2} = \int d^4 x \left[ -\frac{1}{2} \partial_\mu \ph \, \partial^\mu \ph - \sum_{n=3}^N c_n V^{(n)}(\ph) 
\right] . 
\label{Sd0}
\ee

At first, one could be induced to claim that the number of one-loop divergent diagrams is larger than $N$ and infinite if $N=+\infty$, thus confirming statement {\bf 2}. Indeed, if we focus on the bubble diagram, each term of the sum (\ref{GV}) gives the following divergent diagram:
\be
\int \rd^4 k  \, \langle V^{(n)}(\ph) V^{(n)}(\ph) \rangle \propto \frac{1}{\varepsilon} V^{(n-2)}(\ph) V^{(n-2)}(\ph) \,,
\ee
where $\langle\cdot\rangle_r$ denote the Feynman contraction of two fields in a vertex with two fields in the other, and we used dimensional regularization and $\varepsilon=(D-4)/2$. Moreover, for $n \neq m$, we can have other divergences involving different terms in the sum (\ref{GV}), 
\be
\int \rd^4 k  \, \langle V^{(n)}(\ph) V^{(m)}(\ph) \rangle \propto \frac{1}{\varepsilon} V^{(n-2)}(\ph) V^{(m-2)}(\ph)\,, \qquad n \neq m  \, .
\ee
For each term $n$ in the sum (\ref{GV}), there are no other bubble divergences in $D=4$ dimensions. In fact, the diagrams with three external legs are convergent because we have three propagators and no derivatives in the vertices, 
\be
\hspace{-.5cm}\int \rd^4 k  \, \langle V^{(n)}(\ph) V^{(n)}(\ph) V^{(n)}(\ph) \rangle \propto 
\int \rd^4 k \, V^{(n-2)}(\ph) V^{(n-2)}(\ph) V^{(n-2)}(\ph) \left( \frac{1}{k^2} \right)^3 < \infty 
 \, . 
\ee
Finally, the tadpole divergence reads
\be
\int \rd^4 k  \, \langle V^{(n)}(\ph) \rangle \propto \frac{1}{\varepsilon} \Box V^{(n-2)}(\ph) \, . 
\ee
Therefore, if we increase the number $N$ of monomials $V^{(n)}$, we get more and more divergences at one loop. In particular, if we consider $N=\infty$, which is the case of an analytic potential such as, for example, the one in Starobinsky's theory \cite{Starobinsky:1980te,Vilenkin:1985md,Maeda:1987xf}, then one is led to conclude that the theory has an infinite number of divergences. 

However, the last statement is not correct. We show this by evaluating the one-loop quantum effective action built on the Hessian of the theory in the background-field method. We introduce the background field $\Phi$ and the perturbation $\phi$,
\be
\ph = \Phi + \phi\,. 
\label{BFM}
\ee
The Hessian of the theory is
\be
\hat\Delta_{\partial 2} = \left.\frac{ \delta^2 S }{\delta \phi \, \delta \phi}\right|_{\phi=0} = \Box + V^{\prime \prime}(\Phi) \, , 
\ee
so that the one-loop quantum effective action is (notice that we here consider the whole potential, not each of its constituent monomials)
\be
 \Gamma^{(1)}_{\partial 2} & = & \frac{i}{2}\,  {\rm Tr} \, {\rm ln} \,\hat\Delta_{\partial 2}
 = \frac{i}{2}\,  {\rm Tr} \, {\rm ln} \, \left[ \Box \left( 1 +  V^{\prime \prime} \frac{1}{\Box} \right) \right] 
= \frac{i}{2}\,  {\rm Tr} \, {\rm ln} \, \Box 
+ 
\frac{i}{2}\,  {\rm Tr} \, {\rm ln} \, \left( 1 +  V^{\prime \prime} \frac{1}{\Box} \right) \nonumber \\
& 
= & \frac{i}{2}\,  {\rm Tr} \, {\rm ln} \, \Box 
+ 
\frac{i}{2}\,  {\rm Tr}  \,  V^{\prime \prime} \frac{1}{\Box} 
- \frac{i}{2} \frac{1}{2} {\rm Tr}  \,  \left( V^{\prime \prime} \frac{1}{\Box} \, V^{\prime \prime} \frac{1}{\Box} \right) 
+ \frac{i}{2} \frac{1}{3} {\rm Tr}  \,  \left( V^{\prime \prime} \frac{1}{\Box} \, V^{\prime \prime} \frac{1}{\Box} 
\, V^{\prime \prime} \frac{1}{\Box} \right) 
+ \dots \, ,\nonumber\\
\label{E1L}
\ee
where the dots stand for higher-order terms in the Taylor expansion of the logarithm. Since this example is in the absence of gravity, the first trace is just a constant. Then, only the second and third terms in (\ref{E1L}) are divergent, while the forth is convergent because proportional to $\int \rd^4k \, k^{-6}$. Hence, we do not have an infinite number of divergences but only the following two divergent contributions to the quantum effective action:
\be
 \Gamma^{(1) {\rm div}}_\ph = \frac{1}{\varepsilon} \beta_1 \int \rd^4 x \, \Box \, V^{\prime \prime} 
 + \frac{1}{\varepsilon} \beta_2  \int \rd^4 x \, V^{\prime \prime} \, V^{\prime \prime} \, . 
 \label{TDiv}
\ee 

Therefore, contrary to what claimed in statement {\bf 2}, a more careful analysis shows that we do not have an infinite number of 
divergences at one-loop. The explicit computation of the effective action based on the Hessian of the theory provides only the two divergences in (\ref{TDiv}). 

Notice that the model (\ref{Sd0}) is nonrenormalizable because $(V^{\prime\prime})^2$ is not present in the classical action unless $N\leqslant 4$. In fact, for general $n$, $(V^{\prime\prime})^2 \propto (V^{n-2})^2$ is a polynomial in $\ph$ of degree higher than the potential present in the classical action. The higher monomial in $(V^{\prime\prime})^2$ and in the potential 
$V$ are $\ph^{2n - 4}$ and $\ph^{n}$ respectively. Only if $N \leqslant 4$ is the theory renormalizable, since all the monomials of $(V^{\prime\prime})^2$ are already present in the classical action (we do not consider the boundary term). Nevertheless, the main concern in this section was not about the renormalizability problem, but on the number of divergences at one loop. Therefore, according to the above example, we realize that, in general, statement {\bf 2} is too strong and model-dependent. 

 
\subsection{Toy model for Stelle gravity}\label{secex2}

Let us now go to four derivatives and consider a scalar toy model for Stelle's theory of gravity \cite{Stelle:1976gc}. The action is defined following the recipe (\ref{action}) with form factor $F=1$ and reads\footnote{According to the recipe (\ref{action}) with $F=1$, the toy model in this section is inspired by Stelle's gravitational theory for a particular choice of the front coefficients for the operators $R^2$ and ${\rm Ric}^2$, namely, the quadratic part of the action is $G_{\mu\nu} G^{\mu \nu}$.}
\be
S_{\partial 4} = \int \rd^4 x \left\{ \frac{1}{2}  \ph \, \Box \ph - V(\ph) 
+ \lambda \left[ \Box \ph - V^{\prime }(\ph) \right] \left[ \Box \ph - V^{\prime }(\ph) \right]
\right\} , 
\label{Sd4}
\ee
where we can take the potential to be again (\ref{GV}). The coupling $\lambda$ has dimensionality $[\lambda] = -2$, while $[ V ] =  4$, $[ V^\prime ] =  3$, $[ V^{\prime \prime} ] =  2$, $[ V^{\prime \prime \prime} ] =  2$, and $[\ph] = 1$. In order to isolate the vertices form the kinetic term, we rewrite the action as
\be
S_{\partial 4} = \int \rd^4 x \left[ \frac{1}{2}  \ph \, \Box \ph + \lambda \ph \Box^2 \ph - V(\ph) 
- 2  \lambda  (\Box \ph ) V^{\prime }(\ph) 
 + \lambda  V^{\prime }(\ph) V^{\prime }(\ph) 
\right] .
\label{Sd4B}
\ee
If we Taylor expand at the second order in the perturbation $\phi$ defined in (\ref{BFM}), we get
\be
S^{(2)}_{\partial 4} = \frac{1}{2} \int \rd^4 x \, \phi \, \left(\left.\frac{\delta^2 S_{\partial 4}}{\delta \phi \delta \phi}\right|_{\phi=0} \right) \phi 
=  \frac{1}{2} \int \rd^4 x \, \phi \, \hat\Delta_{\partial 4} \, \phi \,,
\ee
where the Hessian is
\be
\hat\Delta_{\partial 4} &=& 
 \Box -  V^{\prime \prime}(\Phi) + 2 \lambda \Box^2 
- 4 \lambda \left[ V^{\prime \prime}(\Phi) \Box + \frac{1}{2} V^{\prime \prime \prime}(\Phi) (\Box \Phi)
\right] \nonumber\\
&&+ 2 \lambda V^{\prime}(\Phi) V^{\prime \prime \prime}(\Phi) 
+ 2 \lambda V^{\prime \prime}(\Phi) V^{\prime \prime}(\Phi) \, .
\ee
The quantum effective action is
\be 
 \Gamma^{(1)}_{\partial 4} & = & \frac{i}{2}\,  {\rm Tr} \, {\rm ln} \, \hat\Delta_{\partial 4}
 = \frac{i}{2}\,  {\rm Tr} \, {\rm ln} \, (\Box + 2 \lambda \Box^2)
 \nonumber  \\
 &&+ \frac{i}{2} {\rm Tr} \, {\rm ln} 
\left\{  1 +  \left[ 
 V^{\prime \prime}(\Phi) - 4 \lambda \left( V^{\prime \prime}(\Phi) \Box + \frac{1}{2} V^{\prime \prime \prime}(\Phi) (\Box \Phi)
\right) 
+ 2 \lambda V^{\prime}(\Phi) V^{\prime \prime \prime}(\Phi)\right.\right.\nonumber\\ 
&&\qquad\left.\left.+ 2 \lambda V^{\prime \prime}(\Phi) V^{\prime \prime}(\Phi) \right] \frac{1}{\Box + 2 \lambda \Box^2} 
  \right\} \nonumber 
  \ee
  \be
 & = & \frac{i}{2}\,  {\rm Tr} \, {\rm ln} \, (\Box + 2 \lambda \Box^2) \nonumber \\
&& + \frac{i}{2} {\rm Tr} \, {\rm ln} 
\left\{  1 +  \left[ 
 V^{\prime \prime}(\Phi) \, \frac{1}{\Box + 2 \lambda \Box^2} 
- 4 \lambda  V^{\prime \prime}(\Phi) \Box \, \frac{1}{\Box + 2 \lambda \Box^2} \right. \right. \nonumber \\
&&\qquad - 4 \lambda \frac{1}{2} V^{\prime \prime \prime}(\Phi) (\Box \Phi) \, \frac{1}{\Box + 2 \lambda \Box^2} 
+ 2 \lambda V^{\prime}(\Phi) V^{\prime \prime \prime}(\Phi) \, \frac{1}{\Box + 2 \lambda \Box^2} \nonumber\\
&& \left. \left. \qquad+ 2 \lambda V^{\prime \prime}(\Phi) V^{\prime \prime}(\Phi)  \frac{1}{\Box + 2 \lambda \Box^2} 
\right]   \right\}.
\label{QEAd4}
\ee
The first trace is a constant, so that we can focus on the second and expand the logarithm in Taylor's series,
\be
 \Gamma^{(1)}_{\partial 4}  & = & 
  \frac{i}{2} {\rm Tr} \, 
   \left[ 
 V^{\prime \prime}(\Phi) \frac{1}{\Box + 2 \lambda \Box^2} 
 - 4 \lambda  V^{\prime \prime}(\Phi) \Box \frac{1}{\Box + 2 \lambda \Box^2} 
 \right. 
 \nonumber \\ 
 && \qquad
 -4 \lambda  \frac{1}{2} V^{\prime \prime \prime}(\Phi) (\Box \Phi) \frac{1}{\Box + 2 \lambda \Box^2} 
+ 2 \lambda V^{\prime}(\Phi) V^{\prime \prime \prime}(\Phi) \frac{1}{\Box + 2 \lambda \Box^2} 
\nonumber \\
&& 
\left. \qquad+ 2 \lambda V^{\prime \prime}(\Phi) V^{\prime \prime}(\Phi)  \frac{1}{\Box + 2 \lambda \Box^2} \right] 
\nonumber \\
&& 
- \frac{1}{2} \frac{i}{2} 
{\rm Tr} \, \left[ 
16 \lambda^2  V^{\prime \prime}(\Phi) \Box \frac{1}{\Box + 2  \lambda \Box^2} \, 
   V^{\prime \prime}(\Phi) \Box \frac{1}{\Box + 2 \lambda \Box^2} 
\right] + \dots \, , 
\label{QEAd4B}
\ee
where the ellipsis stand for convergent contributions. 
Since the theory is only four derivatives, in order to figure out the kind of UV divergent contributions to the quantum effective action it is sufficient to expand the propagator for large $\Box$ and retain only the leading term:
\be
 \frac{1}{2 \lambda \Box^2 \left( 1 + \frac{1}{2 \lambda \Box}  \right) }  =  
 \frac{1}{2 \lambda \Box^2} \left( 1 - \frac{1}{2 \lambda \Box}  \right)  + O \left( \frac{ 1 }{\Box^4} \right) .
 \ee
Replacing the above expansion in (\ref{QEAd4B}), we get
\be
 \Gamma^{(1)}_{\partial 4}  & = & 
  \frac{i}{2} {\rm Tr} \, 
   \Big[ 
\underbrace{ V^{\prime \prime}(\Phi) \frac{1}{\Box + 2 \lambda \Box^2} }_{\frac{1}{\varepsilon} \frac{1}{\lambda} V^{\prime \prime}(\Phi)}
- \underbrace{4 \lambda  V^{\prime \prime}(\Phi) \cancel{\Box} \frac{1}{2 \lambda \Box^{\cancel{2}} } \left( 1 - \frac{1}{2 \lambda \Box}  \right) }_{ \frac{1}{\varepsilon} \Box V^{\prime \prime}(\Phi)  + \frac{1}{\varepsilon} \frac{1}{\lambda}  V^{\prime \prime}(\Phi)  }
 \nonumber \\ 
 && \qquad
- \underbrace{ 4 \lambda  \frac{1}{2} V^{\prime \prime \prime}(\Phi) (\Box \Phi) \frac{1}{\Box + 2 \lambda \Box^2} }_{ \frac{1}{\varepsilon} V^{\prime \prime \prime}(\Phi) (\Box \Phi)}
+ \underbrace{2 \lambda V^{\prime}(\Phi) V^{\prime \prime \prime}(\Phi) \frac{1}{\Box + 2 \lambda \Box^2} }_{\frac{1}{\varepsilon} V^{\prime}(\Phi) V^{\prime \prime \prime}(\Phi)}\nonumber\\
&&\qquad
+ \underbrace{
2 \lambda V^{\prime \prime}(\Phi) V^{\prime \prime}(\Phi)  \frac{1}{\Box + 2 \lambda \Box^2} }_{\frac{1}{\varepsilon} V^{\prime \prime}(\Phi) V^{\prime \prime}(\Phi)}
\Big]
\nonumber \\
&& 
- \frac{1}{2} \frac{i}{2} 
{\rm Tr} \, \Big[ 
\underbrace{16 \lambda^2  V^{\prime \prime}(\Phi) \Box \frac{1}{\Box + 2  \lambda \Box^2} \, 
   V^{\prime \prime}(\Phi) \Box \frac{1}{\Box + 2 \lambda \Box^2} }_{\frac{1}{\varepsilon} V^{\prime \prime}(\Phi) V^{\prime \prime}(\Phi)}
\Big] + \dots \,.
\label{QEAd4C}
\ee
Denoting as $c_i$ the coefficients, the above explicit computation provides the following result for the divergent part of the effective action:
\be
\Gamma^{(1)}_{\partial 4}  & = & \frac{1}{\varepsilon} \int \rd^4 x \,
 \left[
  \frac{1}{\lambda}  c_1 V^{\prime \prime} 
+ c_2 \Box V^{\prime \prime} 
+ c_3 V^{\prime \prime \prime} \Box \Phi 
+ c_4 V^{\prime} \, V^{\prime \prime \prime} 
+ c_5 V^{\prime\prime } \, V^{\prime \prime } 
\right]  .
\label{GSD}
\ee
Again, we have a finite number of divergences. Moreover, if the potential is a polynomial all the above terms (\ref{GSD}) are already present in the action (\ref{Sd4B}) and the theory is renormalizable. On the other hand, if the potential is a nonlinear analytic function such as Starobinsky's potential, the theory is not renormalizable because we have to redefine an infinite number of coefficients. This issue can be avoided in super-renormalizable theories, as we will show in the next subsection. 


\subsection{Higher derivatives toy model}\label{secex3}

Let us now consider a model that includes derivatives in the vertices:
\be
S_{\partial \gamma} & = & \int \rd^4 x \left\{\frac{1}{2} \ph \, \Box \ph - V(\ph) 
+ \lambda \left[ \Box \ph - V^\prime(\ph) \right]  \frac{\Box^\gamma}{\Lambda^{2 \gamma} } \left[ \Box \ph - V^\prime(\ph) \right] \right\} \nonumber \\
& = & \int \rd^4 x \left[\frac{1}{2} \ph \Box \ph + \lambda \, \ph \frac{\Box^{\gamma + 2} }{\Lambda^{2 \gamma} } \ph 
- V(\ph) - 2 \lambda \, (\Box \ph)  \frac{\Box^{\gamma } }{\Lambda^{2 \gamma} } V^\prime(\ph) 
+ \lambda \, V^\prime( \ph )  \frac{\Box^{\gamma } }{\Lambda^{2 \gamma} } V^\prime(\ph) 
\right] \, , \nonumber\\
\label{STM}
\ee
where $\gamma \in \mathbb{N}$. 
The first two operators in (\ref{STM}) define the higher derivative free theory, while the others are interaction vertices. With the decomposition (\ref{BFM}), the Hessian is
\be
S^{(2)}_{\partial \gamma} &=& \frac{1}{2} \int \rd^4 x \, \phi \left(\left.\frac{\delta^2 S_{\partial \gamma}}{\delta \phi \delta \phi}\right|_{\phi=0} \right) \phi 
= \frac{1}{2} \int \rd^4 x \, \phi \, \hat\Delta_{\partial \gamma} \, \phi \, , \nonumber \\
\hat\Delta_{\partial \gamma} &=& \left[ \Box  + 2\lambda \,  \frac{ \Box^{\gamma +2}}{\Lambda^{2 \gamma}} 
- V^{\prime \prime}(\Phi) - 4 \lambda \, V^{\prime \prime}(\Phi)  \, \frac{ \Box^{\gamma +1}}{\Lambda^{2 \gamma}} 
- 2 \lambda \, V^{\prime \prime \prime} \left( \frac{ \Box^{\gamma +1}}{\Lambda^{2 \gamma}} \Phi \right)\right.\nonumber\\
&&\left.\qquad - 2 \lambda \, V^{\prime \prime \prime} \left( \frac{ \Box^{\gamma}}{\Lambda^{2 \gamma}} V^\prime(\Phi) \right)
\right]\delta^4(x - y) \nonumber \\
&&
+ 2 \lambda V^{\prime \prime} (\Phi) \frac{\Box^{\gamma}}{2 \gamma}
\left[  V^{\prime \prime} (\Phi) \delta^4(x-y) \right] .
\label{Hdn}
\ee
The quantum effective action for the model (\ref{STM}) is
\be 
 \Gamma^{(1)}_{\partial \gamma} & = & \frac{i}{2}\,  {\rm Tr} \, {\rm ln} \, \hat\Delta_{\partial \gamma}
 = \frac{i}{2}\,  {\rm Tr} \, {\rm ln} \, \left( \Box + 2 \lambda \frac{\Box^{\gamma + 2}}{\Lambda^{2 \gamma}} \right)
 \nonumber  \\
&& +\frac{i}{2} {\rm Tr} \, {\rm ln} 
\left(\vphantom{\frac{1}{ \Box + 2 \lambda \frac{\Box^{\gamma + 2}}{\Lambda^{2 \gamma}} } }  1 +  \left[ 
 - V^{\prime \prime}(\Phi) - 4 \lambda \, V^{\prime \prime}(\Phi)  \, \frac{ \Box^{\gamma +1}}{\Lambda^{2 \gamma}} 
- 2 \lambda \, V^{\prime \prime \prime} \left( \frac{ \Box^{\gamma +1}}{\Lambda^{2 \gamma}} \Phi \right)\right.\right.\nonumber\\
	&&\qquad\left.- 2 \lambda \, V^{\prime \prime \prime} \left( \frac{ \Box^{\gamma}}{\Lambda^{2 \gamma}} V^\prime(\Phi) \right)
  \right] \frac{1}{ \Box + 2 \lambda \frac{\Box^{\gamma + 2}}{\Lambda^{2 \gamma}} }  \nonumber \\
 & &\qquad +
  \!\left\{
  2 \lambda V^{\prime \prime}(\Phi) V^{\prime \prime}(\Phi) \frac{\Box^{\gamma }}{\Lambda^{2 \gamma}}
  + 2 \lambda V^{\prime \prime}(\Phi) \left[ \partial V^{\prime \prime}(\Phi) \right] \frac{\Box^{\gamma -1 } \partial}{\Lambda^{2 \gamma}}\right.\nonumber\\
	&&\qquad\left.\left.
  + 2 \lambda V^{\prime \prime}(\Phi) \left[ \Box V^{\prime \prime}(\Phi) \right] \frac{\Box^{\gamma -1 } }{\Lambda^{2 \gamma}} + \dots \right\}   \frac{1}{ \Box + 2 \lambda \frac{\Box^{\gamma + 2}}{\Lambda^{2 \gamma}} } 
  \right) ,
\ee
where the dots are other derivative terms of the product $V^{\prime \prime} (\Phi) \delta^4(x-y)$ in (\ref{Hdn}) that do not contribute to the divergent part of the quantum effective action. Expanding the logarithm at second order and omitting convergent terms, we end up with
\be 
\hspace{-1cm}
 \Gamma^{(1)}_{\partial \gamma} & = & \frac{i}{2}\,  {\rm Tr} \, {\rm ln} \, \left( \Box + 2 \lambda \frac{\Box^{\gamma + 2}}{\Lambda^{2 \gamma}} \right)
 \nonumber  \\
& & +\frac{i}{2} {\rm Tr} \, 
\left\{  \left[ 
 - V^{\prime \prime}(\Phi) - 4 \lambda \, V^{\prime \prime}(\Phi)  \, \frac{ \Box^{\gamma +1}}{\Lambda^{2 \gamma}} 
- 2 \lambda \, V^{\prime \prime \prime} \left( \frac{ \Box^{\gamma +1}}{\Lambda^{2 \gamma}} \Phi \right)
- 2 \lambda \, V^{\prime \prime \prime} \left( \frac{ \Box^{\gamma}}{\Lambda^{2 \gamma}} V^\prime(\Phi) \right)
  \right. 
  \right. \nonumber \\
  & & +
  \! 
  2 \lambda V^{\prime \prime}(\Phi) V^{\prime \prime}(\Phi) \frac{\Box^{\gamma }}{\Lambda^{2 \gamma}}
  + 2 \lambda V^{\prime \prime}(\Phi) \left( \partial V^{\prime \prime}(\Phi) \right) \frac{\Box^{\gamma -1 } \partial}{\Lambda^{2 \gamma}}\nonumber\\
	&&\left.\left.
  + 2 \lambda V^{\prime \prime}(\Phi) \left( \Box V^{\prime \prime}(\Phi) \right) \frac{\Box^{\gamma -1 } }{\Lambda^{2 \gamma}} \right] 
  \frac{1}{ \Box + 2 \lambda \frac{\Box^{\gamma + 2}}{\Lambda^{2 \gamma}} } 
  \right\} \nonumber \\
  &  & - \frac{1}{2} \frac{i}{2} {\rm Tr} 
  \left[ 16 \lambda^2 \, V^{\prime \prime}(\Phi)  \, \frac{ \Box^{\gamma +1}}{\Lambda^{2 \gamma}} 
  \frac{1}{ \Box + 2 \lambda \frac{\Box^{\gamma + 2}}{\Lambda^{2 \gamma}} } \, V^{\prime \prime}(\Phi)  \, \frac{ \Box^{\gamma +1}}{\Lambda^{2 \gamma}} 
  \frac{1}{ \Box + 2 \lambda \frac{\Box^{\gamma + 2}}{\Lambda^{2 \gamma}} } \right] + \dots \,.
  \label{traceM}
  \ee
Distributing the propagator at the right side of each vertex, the only divergent contributions are the second in the second line of (\ref{traceM}), the first in the third line, and the one in the last line:
\be 
\hspace{-1cm}
 \Gamma^{(1)}_{\partial \gamma} & = & \frac{i}{2} {\rm Tr} \, 
 \left[ 
 - 4 \lambda \, V^{\prime \prime}(\Phi)  \, \frac{ \Box^{\gamma +1}}{\Lambda^{2 \gamma}} \frac{1}{ \Box + 2 \lambda \frac{\Box^{\gamma + 2}}{\Lambda^{2 \gamma}} }
  +  2 \lambda V^{\prime \prime}(\Phi) V^{\prime \prime}(\Phi) \frac{\Box^{\gamma }}{\Lambda^{2 \gamma}}
  \frac{1}{ \Box + 2 \lambda \frac{\Box^{\gamma + 2}}{\Lambda^{2 \gamma}} } \right]
   \nonumber \\
  & &  -\frac{1}{2} \frac{i}{2} {\rm Tr} 
  \left[ 16 \lambda^2 \, V^{\prime \prime}(\Phi)  \, \frac{ \Box^{\gamma +1}}{\Lambda^{2 \gamma}} 
  \frac{1}{ \Box + 2 \lambda \frac{\Box^{\gamma + 2}}{\Lambda^{2 \gamma}} } \, V^{\prime \prime}(\Phi)  \, \frac{ \Box^{\gamma +1}}{\Lambda^{2 \gamma}} 
  \frac{1}{ \Box + 2 \lambda \frac{\Box^{\gamma + 2}}{\Lambda^{2 \gamma}} } \right] \nonumber \\
  & = & \frac{1}{\varepsilon} \int \rd^4 x 
  \left[ b_1 \Box V^{\prime \prime}(\Phi) + b_2 V^{\prime \prime}(\Phi) V^{\prime \prime}(\Phi) \right] \, ,
  \label{twoOP}
\ee
where $b_1$ and $b_2$ are two dimensionless constants. Contrary to the toy model (\ref{Sd4}) for Stelle's theory, the two operators (\ref{twoOP}) are not present in the classical action. However, the model (\ref{STM}) can be made super-renormalizable if we add the operators (\ref{twoOP}) into (\ref{STM}). Indeed, such operators do not change the above derivation because they cannot give rise to extra divergences. 

Moreover, the potential does not need to be polynomial, but can also be a generic analytic function of the scalar field. In fact, for this toy example the renormalization procedure involves only a finite number of functional derivatives of the potential and not all the Taylor's coefficients of $V(\ph)$. In other words, one has to renormalize a function or the finite number of functions in (\ref{twoOP}). This is neither new nor unconventional, since it is what we actually have in gravity where we do not renormalize each coefficient in the Taylor's expansion of $R^2$ or ${\rm Ric}^2$ in powers of the graviton $h_{\mu\nu}$, but we simply renormalize multiplicatively such operators as single quantities when seen as functions of $h_{\mu\nu}$. Hence, there is not conceptual difference among $R^2$ or ${\rm Ric}^2$ and $V^{\prime  \prime}$, $V^{\prime \prime} \, V^{\prime  \prime}$ or $V^{\prime} \, V^{\prime \prime \prime}$, $\dots$, in the multiplicative renormalization, as evident from the comparison
\be
(Z_{V^{\prime \prime}} -1) V^{\prime \prime} \, , \quad 
[Z_{(V^{\prime \prime})^2} -1] V^{\prime \prime} \, V^{\prime \prime}  \, , \quad \dots
\quad \mbox{similar to} \quad 
(Z_R -1) R \, , \quad (Z_{\rm Ric} -1) {\rm Ric}^2
\, . \nonumber\\
\ee


\section{Finite quantum gravity}\label{finite}

Having established under which conditions the theory is super-renormalizable and thus has only a finite number of divergent diagrams (excluding divergent subdiagrams, if any), we are now ready to investigate the possibility of removing all divergences at all loop orders and ending up with a finite quantum theory. In the following, we consider the versions of the theory which are one-loop super-renormalizable, i.e., that have divergences only at the one-loop level.


\subsection{Finiteness in odd dimensions}

If the topological dimension $D$ is odd, then the theory is not just one-loop super-renormalizable but actually finite. In fact, since the form factor is asymptotically polynomial, the UV behavior of one-loop integrals has the general structure
\be\label{Ikn}
\int \frac{\rd^D k}{(2 \pi)^D} \frac{(k^2)^r}{(k^2 + {\rm C})^s} =-\frac{i}{2^D\pi^{\frac{D}{2}}} \frac{\Gamma\left(r+\frac{D}{2}\right) \Gamma\left(s-r-\frac{D}{2}\right)}{\Gamma\left(\frac{D}{2}\right)\Gamma(s)}\,{\rm C}^{\frac{D}{2}+r-s}\,,
\ee
where ${\rm C}$ depends on the external momenta. The integral \eq{Ikn} is convergent in odd dimensions because the gamma function 
\be
\Gamma \bigg(\underbrace{s - r - \frac{D}{2}}_{\textrm{semi-integer}}\bigg)
\label{gammaDiv}
\ee
has no poles if $r$ and $s$ are both integer, which is the present case. As a consequence, we do not have one-loop divergences and, due to the absence of divergences also at $L>1$, the theory is finite.  


\subsection{Killers}

When $D$ is even, one-loop divergences persist and we have to remove them modifying the theory. For this purpose, we have to include in the bare action other operators that do not get renormalized (i.e., we do not get quantum divergences proportional to such operators) and do not contribute to the propagator on Minkowski background. Such operators, introduced for the first time in \cite{Modesto:2014lga,Modesto:2015lna} (see also \cite{Modesto:2017sdr}), are named killers because they ``kill'' all the quantum divergences, i.e., they make all $\beta$-functions zero. Killers must be at least cubic in the extremals $E_i$ (this request ensures the stability properties reviewed in section \ref{Sec.Act}; see also the next subsection) and have the same number of derivatives of the quadratic operators in $E_i$ present in (\ref{UV-act}). Examples of killers quartic in the extremals are
\be
&& b_1 E_i E^i \, \Box^{n-2} E_j E^j \, , \qquad b_2 E_i E_j \, \Box^{n-2} E^i E^j \, , \qquad 
b_3 E_i E^i \, \Box^{n-3} \, \nabla_\mu E_j \nabla^\mu E^j \, , \nonumber\\
&& b_4 E_i E_j \Box^{n-2} \, \nabla_\mu E^i \nabla^\mu E^j \, ,\qquad b_5 E_i E^i \, \Box^{n-4} \, \nabla_\mu E_j \Box \nabla^\mu E^j \, ,\nonumber\\
&& b_6 E_i E_j \Box^{n-4} \, \nabla_\mu E^i \Box \nabla^\mu E^j \, , \qquad\dots \, , 
\label{exterminators} 
\ee
where $\Box$ must respect all the symmetries of the theory. The operators in (\ref{exterminators}) are quartic in the extremals in order to contribute linearly in the front coefficients $b_i$ to the quadratic (in $E_i$) divergent part of the action, that is to say, to the $\beta$-functions. Notice that increasing the integer $n$ makes the number of potential killers increase, although it does not necessarily imply that the number of independent divergent terms also increases.


\subsection{Mimetic killers}

The operators introduced in the previous subsection are sufficient to make the purely gravitational theory finite, but they are not enough if we also have matter field. Indeed, at the classical level we would like to preserve the nonlocal equations of motion (\ref{LEOM}), which ensures the perturbative stability of the solutions of Einstein's theory and the Standard Model of particle physics. However, to achieve finiteness, we will need not only operators at least cubic in the extremals preserving (\ref{LEOM}), but also, as we will see below, operators quadratic in the extremals that spoil such classical property. In order to solve this problem, we generalize the argument of the entire function ${\rm H}$, replacing the polynomial \eq{Poly} inside the Hessian operator with a more general one including other curvature invariants, namely (in $D$ dimensions)
\be
p(\hat{\De}_{\La_*}) = \hat\De \left[ \tilde a_{n+1} \hat{\De}^n + \tilde a_{n} \hat{\De}^{n-1} + \dots + \tilde a_1 + \left(  \sum_r c_r \mathcal{O}_r  \right) \Box^{n-2} \right],\qquad n\geqslant 2\,,
\label{PolyMode}
\ee
where $\tilde a_n=a_n/\Lambda_*^{[\hat\De]n}$ and we set $a_0=0$. In dimension $D$, $n$ must be greater than two because of super-renormalizability. Notice that, in order to safeguard (\ref{LEOM}) and the stability properties of the theory, only the first operator $\hat\Delta$ in the polynomial (\ref{PolyMode}) needs to be the Hessian of the local underlying theory. All the others can simply be gauge-invariant d'Alembertian operators in curved spacetime. To achieve finiteness, the operators $\mathcal{O}_r$ must include at least all possible one-loop counterterms (at most four-derivative) that are listed in (\ref{Ozum23}), (\ref{A4B4}), and (\ref{B5}) for the $D=4$ case. We name these operators $\mathcal{O}_r$ {\em mimetic} killers since, contrary to the killers \eq{exterminators} that appear explicitly in the Lagrangian, they mimetize inside the form factor.

When we replace the polynomial (\ref{PolyMode}) in the leading contributions to the Lagrangian (\ref{action}) in the UV regime [large $z$ in (\ref{corrInf})], we get
\be
\mathcal{L}_{\rm UV} &=& E_i \, e^{\gamma_{\rm E}} (\hat\Delta^{-1} \,\,  p)_{i j} E_j\nonumber\\
&=& e^{\gamma_{\rm E}} \, 
E_i   \left[ \tilde a_{n+1} \hat{\De}^n + \tilde a_{n} \hat{\De}^{n-1} + \dots + \tilde a_1\mathbbm{1} + \left(  \sum_r c_r \mathcal{O}_r  \right) \mathbbm{1}\Box^{n-2}\right]_{ij}\!\! E_j 
\, ,
\label{Mimetic1}
\ee
where in the last two terms the indices $ij$ are attached to $\mathbbm{1}_{ij}=\de_{ij}$ and the $\Box$ operator should preserve all the symmetries of the theory and be compatible with any chosen metric background. Let us now focus on the last operators in (\ref{Mimetic1}). For the sake of simplicity, we can assume the sum on the indices $i,j$ to be proportional to the identity in the space of fields (in general, we can introduce a field-dependent and nondiagonal metric such as (\ref{de-field})). Then, we get
\be
  E_i \left(  \sum_r c_r \mathcal{O}_r  \right) \delta_{i j} \Box^{n-2} E_j & = & \left(  \sum_r c_r \mathcal{O}_r  \right) E_i \Box^{n-2} E_i \nonumber \\
 & \sim &
\left(   \sum_r c_r
 \left[ (\nabla^2 \Phi)^2,(\nabla \Phi)^4,(\nabla \Phi)^2,\dots \right]_r \right)\nonumber\\
&&\times  \left[ (\Box \Phi + \dots) \Box^{n-2} (\Box \Phi + \dots ) \right] 
, 
\label{Mimetic2}
\ee
where $\Phi$ can be any field including the graviton and in the first square brackets we schematically indicate a linear combination of different operators for each of the ${\cal O}_r$ terms. In order to prove finiteness, we have to consider the second-order variation of the operator (\ref{Mimetic2}). The other leading operator in the UV regime is the first term in (\ref{Mimetic1}), that we can compactly write as
\be 
\tilde a_{n+1} \, E_i  (\hat\Delta^n)_{ij} E_j \sim \tilde a_{n+1} [(\partial^2 \Phi) + \dots] \Box^n [(\partial^2 \Phi) + \dots ] \, ,
\label{KinNoKin}
\ee
where we omitted terms without derivatives and $\partial^2 \Phi$ means fields with two derivatives, regardless of how these are contracted. Notice that, contrary to matter, only the graviton field $h_{\mu\nu}$ can appear in any self-interaction power and without derivatives attached to its legs. The latter is a property of our theory, in agreement with the general renormalizability statement {\bf 1}.

If we Taylor expand the second operator $E_i \Box^{n-2} E_i$ in (\ref{Mimetic2}) with respect to $\Phi_i$, the one-loop integrals can produce divergences proportional to $c_r \mathcal{O}_r$. Consider a generic diagram with only one vertex. In $D=4$ dimensions and in dimensional regularization, schematically (see \cite{Julve:1978xn} for a similar notation)
\be
c_r \mathcal{O}_r \!  \int \rd^4 k  \, \langle \delta E_i \, k^{2n-4} \,  \delta E_i \rangle 
&=& c_r \mathcal{O}_r  \! \int \rd^4 k \, k^2 \, k^{2n - 4}  \, k^2 \langle \Phi_i \Phi_i \rangle_r  \nonumber \\
&\sim&  c_r \mathcal{O}_r  \! \int \rd^4 k \, k^2 \, k^2 k^{2n - 4} \frac{1}{k^{2n + 4}  } \nonumber \\
& = & c_r \mathcal{O}_r  \! \int \frac{\rd^4 k}{k^{4}  }
= c_r \mathcal{O}_r  \frac{1}{\varepsilon} 
\label{O-EE}
\, ,
\ee
where, like in section \ref{secex1}, the angular brackets $\langle\cdot\rangle_r$ denote the Feynman contraction (i.e., the propagator) of two fields, in this case coming form the expansion of the two operators $\delta E$ and with an $r$-dependence signalled with the subscript $r$:
\be\label{lPPr}
\langle \Phi_i \Phi_j\rangle_r = b_r  \frac{\delta_{ij}}{k^{2n+4}}\sim \frac{1}{k^{2n+4}}\,,
\ee
where $b_r$ is an $r$-dependent coefficient. Here and in the following formul\ae, the symbol $\sim$ means omission of this coefficient $b_r$ and of the Kronecker delta $\delta_{ij}$ (here $\de_{ii}=1$ but it is nontrivial in the case of mixed field products with $i\neq j$). Notice that, in order to be logarithmically divergent, in the above integral each of the two expansions $\delta E_i$ should contain at least two derivatives acting on the contracted field, otherwise the integral is convergent. Other divergences can be originated by the contraction of a field in $\delta E_i$ with another field in $\mathcal{O}_r$, 
\be
c_r E_i  \!  \int \rd^4 k  \, \langle \delta \mathcal{O}_r^{(2,2)}  \, k^{2n-4} \,  \delta E_i \rangle 
& = & c_r E_i   \left( \partial^2 \Phi_i\dots\Phi_j\right) \! \int \rd^4 k \, k^2 \, k^{2n - 4}  \, k^2 \langle \Phi_i \Phi_i \rangle_r \nonumber \\
& \sim & c_r E_i   \left( \partial^2 \Phi_i\dots\Phi_j \right) \! \int \rd^4 k \, k^2 \, k^{2n - 4}  \, k^2 \frac{1}{k^{2 n +4}}  
\nonumber \\
& = & c_r E_i   \left( \partial^2 \Phi_i\dots\Phi_j \right) \! \int \rd^4 k   \frac{1}{k^{4}}   \nonumber \\
&=& c_r E_i   \left( \partial^2 \Phi_i\dots\Phi_j \right) \frac{1}{\varepsilon} 
\, , 
\label{E-OE}
\ee
where we assumed that ${\cal O}_r$ contains {\em four} derivatives, but at most two can act on a single field. This is a characteristic of the counterterms in (\ref{Ozum23})--(\ref{B5}) with (\ref{B5Vl}) and (\ref{A4B4Vl}). However, if ${\cal O}_r$ happens to have, for example, four derivatives on a field and zero on the others, we can get up to two more counterterms in the action (proportional to $1/\varepsilon$),
\be
  c_r \left( \partial^2 E_i   \right) \Phi_i \, , \quad c_r \left( \partial E_i   \right) \left(\partial \Phi_i \right) \, , 
\ee
which are equivalent to (\ref{E-OE}) up to total derivatives. Indeed, we also get the following extra divergences besides (\ref{E-OE}):
\be
c_r E_i  \!  \int \rd^4 k  \, \langle \delta \mathcal{O}_r^{(4,0)}  \, k^{2n-4} \,  \delta E_i \rangle 
& = & c_r E_i \left(\partial^0 \Phi_i\dots\Phi_j \right)\! \int \rd^4 k \, k^4 \, k^{2n - 4}\, k^2 \langle \Phi_i \Phi_i \rangle \nonumber \\
& = & c_r E_i \left(\partial^0 \Phi_i\dots\Phi_j \right)\! \int \rd^4 k \, k^4 \, k^{2n - 4}\, k^2 \frac{1}{k^{2 n +4}}  
\nonumber \\
& = & c_r E_i   \left( \partial^0 \Phi_i\dots\Phi_j \right) \! \int  \rd^4 k \,  k^2 \frac{1}{k^{4}}\nonumber \\
&\sim& c_r \left( \partial^2 E_i   \right) \Phi_ i \frac{1}{\varepsilon} 
+ c_r \left( \partial E_i   \right) \left(\partial \Phi_i \right) \frac{1}{\varepsilon} 
\, ,
\label{E-OE-S}
\ee
where by $\left( \partial^0 \Phi_i\dots\Phi_j \right)$ we mean no derivatives on any field (which can be more than one), but the operator $\mathcal{O}_r$ studied in (\ref{E-OE-S}) has in total four derivatives, namely it is one of the operators in (\ref{Ozum23}) after integration by parts and discarding boundary terms. The boundary terms will be consider later in this section.

On the other hand, if $\mathcal{O}_r$ has a total number of {\em two} derivatives (see the second operator in (\ref{A4B4})), then the most divergent integral reads
\be
c_r E_i  \!  \int \rd^4 k  \, \langle \delta \mathcal{O}_r  \, k^{2n-4} \,  \delta E_i \rangle 
=c_r E_i   \left( \partial^0 \Phi_i\dots\Phi_j \right) \! \int \rd^4 k   \frac{1}{k^{4}}  = c_r E_i   \left( \partial^0 \Phi_i\dots\Phi_j \right) \frac{1}{\varepsilon} 
. 
\ee

Conversely, no divergences result from the contraction of two fields in the operator $\mathcal{O}_r$, because $\mathcal{O}_r$ contains at most four derivatives, while the denominator contains $2n +4$ derivatives. Indeed,
\be
c_r \,  E_i \Box^{n-2} E_i  \!  \int \rd^4 k  \,  
\langle \delta \mathcal{O}_r \rangle 
& = & c_r E_i \Box^{n-2} E_i  \,  
  \! \int \rd^4 k \, k^2 \, k^2  \langle \Phi_j \Phi_j \rangle_r 
\nonumber \\
& \sim & c_r \, E_i \Box^{n-2} E_i    \! \int \rd^4 k \, k^2 \, k^2  \frac{1}{k^{2 n + 4}} \nonumber \\
& = & c_r \, E_i \Box^{n-2} E_i  
  \! \int \rd^4 k \,  \frac{1}{k^{2 n }} < + \infty 
\,.
\ee
Moreover, divergences can only be linearly proportional to the coefficients $c_r$ because one-loop integrals with a number of vertices $\geqslant 2$ are convergent if no more than two derivatives act on any single field in the operator $O_r$. For two vertices,
\be
&&  c_r \, c_s \, \mathcal{O}_r \, E_j  \!  \int \rd^4 k  \, \langle \delta \mathcal{O}_s  \, k^{2n-4} \,  \delta E_i \rangle 
 \, \langle \delta E_i \, k^{2n-4} \,  \delta E_j \rangle_r \nonumber \\
 &&\hspace{1.8cm} =  c_r \, c_s \, \mathcal{O}_r \, E_j  \, \mathcal{O}^\prime_{s, k} \!  \int \rd^4 k  \, k^2 \, k^2 \, 
\langle \Phi_k \, k^{2n-4} \, \Phi_i \rangle_s 
 \, k^2 \, k^2 \, \langle  \Phi_i \, k^{2n-4} \,   \Phi_j \rangle_r
 \nonumber \\
 &&\hspace{1.8cm} \sim   c_r \, c_s  \, \mathcal{O}_r \, E_j  \, \mathcal{O}^\prime_{s, k} \!  \int \rd^4 k  \, {{k^2}} \, {{k^2}} \, 
 k^{ {2n} -4} \frac{1}{k^{{2n} + {{4}}} } 
 \, {k^2 \, k^2} \, k^{ {{2n}} - 4} \, \frac{1}{k^{ {{2n}} + {4} } } \nonumber \\
 &&\hspace{1.8cm} 
 =c_r \, c_s  \, \mathcal{O}_r \, E_j \, \mathcal{O}^\prime_{s, k} \!  \int \frac{\rd^4 k}{k^8} 
< + \infty \, . 
 \ee
 The above integral is convergent also if, in the operator $\mathcal{O}_r$, four derivatives act on a  single field, since $\int \rd^4k/k^6 < +\infty$. Similarly, performing contractions only between the tensors $E_i$ in the two vertices, we obtain
\be
&&  c_r \, c_s \, \mathcal{O}_r   \, \mathcal{O}_s \!  \int \rd^4 k  \, \langle \delta E_i  \, k^{2n-4} \,  \delta E_j \rangle_r 
 \, \langle \delta E_j \, k^{2n-4} \,  \delta E_i \rangle_s \nonumber \\
&&\hspace{1.8cm} =  c_r \, c_s \, \mathcal{O}_r   \, \mathcal{O}_s  \!  \int \rd^4 k  \, k^2 \, k^2 \, 
\langle \Phi_i \, k^{2n-4} \, \Phi_j \rangle_r 
 \, k^2 \, k^2 \, \langle  \Phi_j \, k^{2n-4} \,   \Phi_i \rangle_s
 \nonumber \\
&&\hspace{1.8cm}\sim c_r \, c_s \, \mathcal{O}_r   \, \mathcal{O}_s  \!  \int \rd^4 k  \, {{k^2}} \, {{k^2}} \, 
 k^{2n-4} \frac{1}{k^{2n + {{4}}} } 
 \, {k^2 \, k^2} \, k^{ {{2n}} - 4} \, \frac{1}{k^{ {{2n}} + {4} } } \nonumber \\
 &&\hspace{1.8cm}   = c_r \, c_s \, \mathcal{O}_r   \, \mathcal{O}_s \!  \int \frac{\rd^4 k}{k^8} 
< + \infty \, . 
\ee

Finally, we consider the contraction of a mimetic killer with the operator $E_i \Box^n E_i$ in (\ref{KinNoKin}). For the case of two vertices, the divergent one-loop integral is
\be
&&  c_r \, \mathcal{O}_r  \!  \int \rd^4 k  \, \langle \delta E_i  \, k^{2n-4} \,  \delta E_j \rangle_r 
 \, \langle \delta E_j \, k^{2n} \,  \delta E_i \rangle \nonumber \\
&& \hspace{1.8cm}
 \sim c_r \, \mathcal{O}_r  \!  \int \rd^4 k  \, k^2 \, \frac{1}{k^{2n+4}} \, k^{2n - 4} \, k^2 \, 
 {{k^2 }} \, \frac{1}{k^{{{2n} }+ {{4}}}} \, {k^{{2n}}} \, {{k^2 }}\nonumber \\
 &&\hspace{1.8cm} 
  = c_r \,  \mathcal{O}_r  \! \int \frac{\rd^4 k}{k^{4}  }
= c_r \, \mathcal{O}_r  \frac{1}{\varepsilon} \, , 
 \label{IntStronzo}
 \ee
which is the same kind of divergence of \eq{O-EE}. Notice that, in order to be divergent, the variation $\delta E_i$ must provide a $k^2$ contribution in the internal loop momentum to the integral (\ref{IntStronzo}). Therefore, only $\Box \ph$ and $R_{\mu\nu}$ can give a divergent contribution (see the equations of motion for the scalar field as an example, but the result is general because $E_i$ are the equations of motion of a local two-derivative subsidiary theory). In both cases, the resummation of all the divergences for an arbitrary large number of external legs will reconstruct the measure weight $\sqrt{|g|}$. This resummation of divergences is more evident when the background-field method is properly implemented. Let us expand on this issue. 

The path integral for the full theory, including gravity and matter, the gauge-fixing operators, and the BRST-ghosts action reads
\be
&& \hspace{-0.5cm}
Z[ \Phi_i ] = \int \!  
\prod_{i} \mathcal{D}\Phi_i \prod_{j} \mathcal{D} c_j \,  
 e^{i \left(  S +  S_{\rm gf} + S_{\rm gh}   \right)} \,  ,
\label{QG}
 \ee
 where $i$ labels all the fields in the theory and $j$ all the BRST ghost fields. 
 
As we saw above in this section, in the background-field method one splits the fields in a background plus a fluctuation,
\be
\Phi_i \,\, \rightarrow \,\, \Phi_i + \phi_i . 
\ee
Hence, at one loop we can evaluate the functional integral explicitly and express the partition function as a product of functional determinants: 
  \be
  Z[\Phi_k] &=& e^{i S[{\Phi_k}]}   \left\{   \left. {\rm Det}\left[ 
   \frac{\delta^2 (S[\Phi_k +  \phi_k ] + S_{\rm gf}[\Phi_k + \phi_k ]) }{\delta \phi_i \, \delta \phi_j } \right\vert_{\phi_l=0} \right] \right\}^{-\frac{1}{2}} \nonumber\\
&&\times ({\rm Det} \, M ) \, \, 
({\rm Det} \, C )^{\frac{1}{2}}  
\,\, \prod_q ({\rm Det} \, M^{\rm YM}_q)
\, ,
 \ee
where $M$ and $C$ are the quadratic operators for the BRST ghosts of the gravitational sector, while  $M^{\rm YM}_q$ is the quadratic operator for the BRST ghosts of the gauge fields. 
Moreover, by $S[{\Phi_k}]$ we mean the classical action of the theory for the background fields $\Phi_k$.

Therefore, in order to get the one-loop effective action, we need to expand the action including the gauge-fixing term to the second order in the quantum fluctuation $\phi_i$. The result in $D=4$ dimensions is
\be
\delta^2 S = \frac{1}{2} \int \rd^4 x \sqrt{|g|} \, \phi_i \, \De^{ij} \, \phi_j \, , \qquad  \De^{ij}= \frac{\delta^2 (S + S_{\rm gf}) }{ \delta \phi_i  \, \delta \phi_j}\Bigg|_{\phi_l=0}\,. \label{Hij}
\ee
Following \cite{Asorey:1996hz}, we can recast the minimal Hessian in (\ref{Hij}) in the compact and general form
\be
\De^{ij} &=& 
\delta^{i j} \Box^{n+2} 
+ \underbrace{V^{i,j , i_1  \dots i_{2 n +2}} }_{\partial^2 \Phi_i}
\nabla_{i_1}  \cdots \nabla_{i_{2 n+2 }} 
\nonumber 
\\
&& 
+\underbrace{ W^{i j , i_1  \dots i_{2 n + 1}} }_{ \partial (\partial^2 \Phi_i) }
\nabla_{i_1} \cdots \nabla_{i_{2 n+1}}  
+ \underbrace{U^{i,j, i_1 \dots i_{2 n }} }_{(\partial^2 \Phi_i ) (\partial^2 \Phi_j) }
\nabla_{i_1}  \cdots \nabla_{i_{2 n}} +O(\nabla^{2n-1})  \nonumber \\
  &=& 
   \Box^{n+2} \left[ 
\delta^{i j}
+ \underbrace{V^{i,j , i_1  \dots i_{2 n +2}} }_{\partial^2 \Phi_i}
\nabla_{i_1}  \cdots \nabla_{i_{2 n+2 }} \, \frac{1}{ \Box^{n+2} }+\underbrace{ W^{i j , i_1  \dots i_{2 n + 1}} }_{ \partial (\partial^2 \Phi_i) }
\nabla_{i_1} \cdots \nabla_{i_{2 n+1}}  \,  \frac{1}{ \Box^{n+2} }
\right.
\nonumber 
\\
&& 
\hspace{1cm}
\left. 
+ \underbrace{U^{i,j, i_1 \dots i_{2 n }} }_{(\partial^2 \Phi_i ) (\partial^2 \Phi_j) }
\nabla_{i_1}  \cdots \nabla_{i_{2 n}} \,   \frac{1}{ \Box^{n+2} } +O(\nabla^{2n-1})  \,  \frac{1}{ \Box^{n+2} }
\right]\! . 
\label{acca}
\ee
When the background-field method is implemented and the second functional derivative taken, the mimetic killers contribute only to the operator $U$, if in the mimetic killers at most $2n-2$ derivatives act on the quantum field $\phi$ and $2$ on any other field. Mimetic killers can also contribute to the matrix $V$ if we include boundary terms or, equivalently, operators in which three or four derivatives act on a single field in some $\mathcal{O}_r$. These operators will be studied later in more detail.

Therefore, the contribution of the operator $E_i \Box^{n -2 }E_i$ in the Taylor expansion of (\ref{Mimetic2}) to the second-order variation giving rise to a divergence is proportional to
\be
\left[ (\nabla^2 \Phi)^2 + \dots \right] \phi \nabla^{2n} \phi \, , 
\label{Kvar}
\ee
which gives a contribution to $U$ according to the definition (\ref{acca}) (fourth term after the first equality). If a boundary term among the $\mathcal{O}_r$ is added to the list of mimetic killers (\ref{Mimetic2}), the variation (\ref{Kvar}) turns into
\be
\left(\nabla^4 \Phi + \dots \right) \phi \nabla^{2n} \phi \, .
\label{Kvar2}
\ee
In addition, we can vary the boundary operator $\mathcal{O}_r$, which can now provide also terms with three or four derivatives on a single field after explicit evaluation of the total derivative. Therefore, we get contributions also to the matrix $V$, 
\be
\left[ (\nabla^0 \Phi) (\nabla^4 \phi) + \dots \right] (\nabla^2 \Phi) \nabla^{2n-2} \phi \, . 
\label{Kvar3}
\ee
After integration by parts, we get
\be
\left[ (\nabla^0 \Phi) \phi + \dots \right] (\nabla^2 \Phi) \nabla^4 \nabla^{2n-2} \phi
= \left[ (\nabla^0 \Phi) (\nabla^2 \Phi) + \dots \right] \phi 
 \nabla^{2n+2} \phi \, .
\label{Kvar4}
\ee
The latter operator can contribute linearly in its front coefficient to any divergence with four derivatives, but it will contribute quadratically in the front coefficient only to operators having up two derivatives on a single field and two fields with two derivatives each, namely,
\be
 \frac{1}{\varepsilon} c_r^2 (\nabla^2 \Phi ) (\nabla^2 \Phi ) \, .
\label{KvarM}
\ee
However, the above divergences have two derivatives on a single field and, thus, they contribute to the $\beta$-functions that can be made to vanish by a proper choice of other constants $c_r,$ according to (\ref{O-EE}). 

Coming back to the term $\langle \delta E_j \, k^{2n} \,  \delta E_i \rangle$ in the first line of (\ref{IntStronzo}), it can only be originated by a nonminimal term in the Hessian in (\ref{acca}), so that
\be
\De^{ij} &=& 
\delta^{i j} \Box^{n+2} +  \underbrace{V^{(0)i,j , i_1  \dots i_{2 n +4}} }_{\partial^0 \Phi_i}
\nabla_{i_1}  \cdots \nabla_{i_{2 n+4 }} 
+ \underbrace{V^{i,j , i_1  \dots i_{2 n +2}} }_{\partial^2 \Phi_i}
\nabla_{i_1}  \cdots \nabla_{i_{2 n+2 }} 
\nonumber 
\\
&& 
+\underbrace{ W^{i j , i_1  \dots i_{2 n + 1}} }_{ \partial (\partial^2 \Phi_i) }
\nabla_{i_1} \cdots \nabla_{i_{2 n+1}}  
+ \underbrace{U^{i,j, i_1 \dots i_{2 n }} }_{(\partial^2 \Phi_i ) (\partial^2 \Phi_j) }
\nabla_{i_1}  \cdots \nabla_{i_{2 n}} +O(\nabla^{2n-1}) \, .
\label{HStronzo}
\ee
Indeed, the term proportional to $V^{(0)}$ has $2n+4$ derivatives like the inverse propagator and, when we take the trace of the Hessian, among all the divergences we also find one proportional to
\be\label{blabla}
{\rm Tr}  \left[ 
 V^{i,j , i_1  \dots i_{2 n +2} } 
\nabla_{i_1}  \cdots \nabla_{i_{2 n+2 } } \, \frac{1}{\Box^{ n + 2} } \,\, 
V^{(0)i,j , i_1  \dots i_{2 n +4}} 
\nabla_{i_1}  \cdots \nabla_{i_{2 n+4 }} \, \frac{1}{\Box^{ n + 2} } 
\right] ,
\ee
which is exactly (\ref{IntStronzo}). However, by a proper gauge choice we can always turn $V^{(0)}=0$ even in the presence of matter and, \emph{a posteriori}, we see that we can avoid (\ref{IntStronzo}) and such an infinite number of divergent contributions to the quantum effective action.

Finally, for completeness the one-loop effective action reads:
\be
\Gamma^{(1)}[\Phi_i] &=& - i  \ln Z[\Phi_i]\nonumber\\
& = &
 S[\Phi_i]  
+ \frac{i}{2}   \ln {\rm Det}
(\hat{\Delta}) - i  \ln {\rm Det}({M})
 - \frac{i}{2} \ln
{\rm Det}({C}) - i \sum_q \ln {\rm Det}(M^{\rm YM}_q) \, , \nonumber\\\label{Gamma1}
\ee
which can be obtained employing the universal trace formul\ae\ of Barvinsky and Vilkovisky \cite{Barvinsky:1985an}. 

Therefore, if we introduce a sufficient number of mimetic killers (\ref{Mimetic2}) in the theory in order to get extra divergences proportional to the operators in (\ref{Ozum23}) and (\ref{A4B4}), but linearly proportional to the front coefficients $c_r$, then 
 we can always solve the vanishing equations for all the $\beta$-functions. Such solutions for real coefficients $c_r$ always exist and are real since the system of equations is linear. In this way, all $\beta$-functions are zero and we achieve UV finiteness. 

Finally, we comment about boundary terms. In order to achieve conformal invariance at the quantum level, we need to cancel all divergences that can potentially contribute to the trace anomaly (see section \ref{sec.ci} for more details). This issue can be addressed by including 
 other operators $\mathcal{O}_r$ to the sum in (\ref{PolyMode}) that display fields with also three or four derivatives acting on them. Indeed, any four-derivative boundary term can be expanded into several operators having from one to four derivatives on a single field. 
Hence, we have mimetic killers with three or four derivatives on a single field in $\mathcal{O}_r$, but still a total number of four derivatives, i.e.,
\be
{\mathcal O}_r : \qquad  (\nabla^3 \Phi) (\nabla \Phi)(\dots)  \, , \quad (\nabla^4 \Phi) (\dots) \,
\label{3and4}
\ee
where by dots we mean other fields (if any) without derivatives.
This in turn will be equivalent to remove divergences proportional to  total derivatives after integration by parts. In other words, we can extend the list in (\ref{Ozum23}) to all possible operators having also three or four derivatives on a single field. 
According to the simplified analysis in (\ref{O-EE}), we get divergent contributions proportional to (\ref{3and4}) and linear in the front coefficient $c_r$. 

Following the example (\ref{E-OE-S}), other divergences can be originated by the contraction of a field in $\delta E_i$ with another field in $\mathcal{O}_r$, 
\be
\hspace{-.5cm}c_r E_i  \!  \int \rd^4 k  \, \langle \delta \mathcal{O}_r  \, k^{2n-4} \,  \delta E_i \rangle 
& = & c_r E_i   \left( \partial^0 \Phi_i \right) \! \int \rd^4 k \, k^4 \, k^{2n - 4}  \, k^2 \langle \Phi_i \Phi_i \rangle \nonumber \\
& = & c_r E_i   \left( \partial^0 \Phi_i \right) \! \int \rd^4 k \, k^4 \, k^{2n - 4}  \, k^2 \frac{1}{k^{2 n +4}}  
\nonumber \\
& = & c_r E_i   \left( \partial^0 \Phi_i \right) \! \int  \rd^4 k \,  k^2 \frac{1}{k^{4}}\nonumber \\
&\sim& c_r \left( \partial^2 E_i   \right) \Phi_ i \frac{1}{\varepsilon} 
+ c_r \left( \partial E_i   \right) \left(\partial \Phi_i \right) \frac{1}{\varepsilon} 
+ c_r  E_i    \left(\partial^2 \Phi_i \right) \frac{1}{\varepsilon}
\, .
\label{E-OE-S2}
\ee
The first two operators in (\ref{E-OE-S2}) also contribute to counterterms with three and four derivatives on a single field, but still linearly in the front coefficient $c_r$. Hence, they are not a problem to achieve full finiteness. 

However, divergences resulting from the mimetic killers that include (\ref{3and4}) can also contribute quadratically in $c_r$ to one-loop integrals.
For two vertices,
\be
&&  c_r \, c_s \, E_i \, E_j  \!  \int \rd^4 k  \, \langle \delta \mathcal{O}_r  \, k^{2n-4} \,  \delta E_i \rangle 
 \, \langle \delta \mathcal{O}_r \, k^{2n-4} \,  \delta E_j \rangle 
 \nonumber \\
&&\hspace{1.8cm} =   c_r \, c_s \, E_i \, E_j   \!  \int \rd^4 k  \, k^4 \, k^2 \, 
\langle \Phi_r \, k^{2n-4} \, \Phi_i \rangle 
 \, k^4 \, k^2 \, \langle  \Phi_s \, k^{2n-4} \,   \Phi_j \rangle 
 \nonumber \\
&&\hspace{1.8cm} =   c_r \, c_s  \, \, E_i \, E_j    \!  \int \rd^4 k  \, {\cancel{k^4}} \, {{k^2}} \, 
 k^{ \cancel{2n} -4} \frac{1}{k^{ \cancel{2n} + {\cancel{4}}} } 
 \, {\cancel{k^4} \, k^2} \, k^{ {\cancel{2n}} - 4} \, \frac{1}{k^{ {\cancel{2n}} + \cancel{4} } } \nonumber \\
 &&\hspace{1.8cm} =   c_r \, c_s  \, \, E_i \, E_j    \!  \int \frac{\rd^4 k  }{k^4} 
 =   c_r \, c_s  \, \, E_i \, E_j   \frac{1}{\varepsilon} 
 \label{quad4}
 \, . 
 \ee
The above divergences will contribute to the $\beta$-functions of the counterterms having only at most two derivatives on a single field. Such $\beta$-functions can be made to vanish thanks to other contributions that are certainly linear in the front coefficients according to (\ref{O-EE}). Hence, the divergences (\ref{quad4}) do not spoil the finiteness. On the other hand, if three derivatives act on a single field, integrals similar to (\ref{quad4}) and quadratic in $c_r$ are convergent. 

Let us also remark that, as long as it is finite, the number $N$ of divergences in the theory does not modify the killing procedure. Given $N$ operators $X_i$ associated with quantum divergences (e.g., $X_1=R_{\mu\nu}R^{\mu\nu}$, $X_2=R^2$, $X_3 = (\Box\ph)^2$, $X_4 = R \Box \ph$, and so on), one can construct $N$ killers of the form $s_iX_i \Box^{n-2} R^2$, where $s_i$ are coupling constants. All these operators contribute linearly to the respective beta functions, so that one must solve a system of $N$ linear equations in the $N$ variables $s_i$. For the above operators, one can show that such a system always admits a nontrivial set of solutions.


\subsection{Multiplicative renormalizability in even dimensions}\label{secnew}

Power counting does just what its name says: it counts the number of divergences in a given theory. However, it does not establish whether these divergences can actually be reabsorbed in counterterms reflecting the original structure of the bare action. If this happens, the theory is said to be multiplicatively renormalizable \cite{Buchbinder:2021wzv}. In other words, given a bare Lagrangian made of $n$ operators ${\cal O}_i$,
\be
{\cal L}=\sum_{i=1}^n {\cal O}_i\,,
\ee
the theory is multiplicatively renormalizable if the counterterms are of the form
\be
{\cal L}_{\rm ct}=\sum_{i=1}^n c_i {\cal O}_i\,,
\ee
for some coefficients $c_i$. In previous sections, we have seen that nonminimally coupled nonlocal quantum gravity is at least power-counting super-renormalizable, so that the number of superficial divergences is finite. However, due to the very rigid structure of the action as the sum of the square of the extremals $E_i$, the coefficients in front of $O({\cal R}^2)$ operators in the bare action are not mutually independent, which may lead to the conclusion that the theory has fewer free parameters in the classical action than divergences and, therefore, is not multiplicatively renormalizable. 

Let us illustrate the above issue in a higher-derivative nonunitary gravitational model with curvature square sector $G_{\mu\nu} G^{\mu\nu}$, which mimicks the theory $E_{\mu\nu} F(\Delta) E^{\mu\nu}$ for $F=1$ and in the absence of matter. The $GG$ model is special in the same sense of $EFE$ because the operators $R_{\mu\nu} R^{\mu\nu}$ and $R^2$ do not have arbitrary coefficients like in Stelle gravity. The full Lagrangian in $D=4$ dimensions is
\be
{\cal L} = 
\frac{1}{16 \pi G} (R + 2 \La) + \al G_{\mu\nu} G^{\mu\nu} = \frac{1}{16 \pi G} (R + 2 \La) + \al R_{\mu\nu} R^{\mu\nu}.
\label{modGR}
\ee
This model is a sort of overconstrained Stelle gravity with only one independent coefficient $\al$ in the quadratic sector and, thus, the masses of the ghost and the scalar are not independent. In order to better understand this, recall that Stelle's Lagrangian is \cite{Stelle:1976gc}
\be
\mathcal{L}_{\rm Stelle} = \frac{1}{16 \pi G} (R + 2 \La) + \al R_{\mu\nu} R^{\mu\nu} + \sigma R^2,
\label{stelle}
\ee
and the graviton propagator for Stelle gravity \eq{stelle} and the model \eq{modGR} (when $\sigma=0$) on Minkowski spacetime ($\La=0$) and with the minimal gauge choice is
\be
 \mathcal{O}^{-1}(k) & = & - \frac{P^{(2)}}{k^2[ (16 \pi G)^{-1} - \alpha k^2 ] } + \frac{P^{(0)}}{2 k^2[ (16 \pi G)^{-1} + k^2 (3 \sigma + 2 \alpha)] } \Big|_{\sigma =0} \nonumber \\
 & = & - \frac{P^{(2)}}{k^2[ (16 \pi G)^{-1} - \alpha k^2 ] } + \frac{P^{(0)}}{2 k^2[ (16 \pi G)^{-1}+ 2 \alpha k^2] } \, .
 \label{propa}
\ee
Moreover, the theory \eq{modGR} has the same power-counting divergences of Stelle gravity because vertices and kinetic terms have the same number of derivatives, if $\alpha\neq 0$. 

At the quantum level, Stelle theory \eq{stelle} is renormalizable and we get divergences proportional to the operators already present in the classical action (up to topological and superficial terms that we do not discuss here). Focusing on the higher-derivative part only, a power-counting analysis shows that the counterterms have the following form:
\be
\hspace{-0.7cm}
{\cal L}_{\rm ct} = - {\cal L}_{\rm div}  =  
\frac{1}{ 32 \pi^2  \varepsilon} \left(\beta_\alpha R_{\mu\nu} R^{\mu\nu} + \beta_\sigma R^2 \right) 
=
(Z_\alpha - 1) \alpha R_{\mu\nu} R^{\mu\nu} +(Z_\sigma - 1) \sigma R^2,
\label{RCcouS}
\ee
where $Z_{\alpha,\sigma}$ are the renormalization constants for the coupling parameters and $\beta_{\alpha, \sigma}$ are the beta functions for the two operators we are investigating. Note that the last equality holds only if the theory is multiplicatively renormalizable. The renormalized Lagrangian is given by 
\be\label{LrenS}
{\cal L}_{\rm ren} = {\cal L}(\mu) + {\cal L}_{\rm ct}\,,
\ee
where the classical Lagrangian ${\cal L}(\mu)=\mathcal{L}_{\rm Stelle}(\mu)$ depends on the renormalization scale $\mu$ because the coupling constants $\alpha$ and $\sigma$ both depend on the energy scale $\mu$. Now we introduce the bare parameters $\alpha_{\rm B}$ and $\sigma_{\rm B}$ in order to show that the quadratic part ${\cal L}_{\rm ren}^{\rm quad}$ of the Lagrangian ${\cal L}_{\rm ren}$ takes on the appearance of the classical Lagrangian \eq{stelle}, namely 
\be
{\cal L}_{\rm B}^{\rm quad} = \al_{\rm B} R_{\mu\nu} R^{\mu\nu} +\sigma_{\rm B}  R^2 \, . 
\label{bareS}
\ee
Indeed, comparing \eq{LrenS} with \eq{bareS} and ignoring the running of the Newton's and cosmological constants because it does not affect the issue considered here \cite{Avramidi:1985ki}, we get
\be
\alpha Z_\alpha = \alpha_{\rm B} \, , \quad \sigma Z_\sigma = \sigma_{\rm B}\,,
\ee
where the last equation is absent for the model \eq{modGR}. However, for the model \eq{modGR} we meet a problem. At one loop, we get a divergent contribution $\propto R^2$ because $\beta_\sigma\neq 0$ \cite{Stelle:1976gc,Buchbinder:2021wzv}, while in the classical action this operator is absent. Therefore, we cannot redefine any coupling to absorb such a divergence without avoiding the contradiction
\be
0=\sigma_{\rm B}\coloneqq (Z_\sigma - 1) \sigma = \frac{\beta_\sigma}{ 32 \pi^2  \varepsilon}\neq 0\,.
\ee

Therefore, the last equality in \eq{RCcouS} does not hold for the local $G_{\mu\nu} G^{\mu\nu}$ model \eq{modGR} and the latter is not multiplicatively renormalizable; this is exactly what happens also for the nonlocal model \eq{pequena} and its more general version \eq{1x}. In the latter case, since the coefficients $\al_1$ and $\al_2$ in \eq{OsAlfas} are independent when the general internal-space metric \eq{de-field-co} is adopted, the two structures $E_{\mu\nu} E^{\mu\nu}$ and $ E^\mu_\mu E^\nu_\nu$ are able to reabsorb the two covariant divergences proportional to $R_{\mu\nu}R^{\mu\nu}$ and $R^2$ (in $D=4$ dimensions) in a pure-gravity scenario with no matter fields. Therefore, in pure gravity the general metric \eq{de-field-co} fixes the problem since we have as many operators with independent coefficients as the number of counterterms. 

However, as soon as we switch on matter fields, we have fewer free parameters in the classical action than divergences and the model is not multiplicatively (super-)renormal\-izable. In particular, in $D=4$ dimensions the divergences repeat the form of the action~\eq{1x}; we have in this case 11 types of divergences (not considering topological and superficial terms) and only 4 free parameters given in \eq{OsAlfas}, so that it is not possible to satisfy all the renormalization group equations. In $D>4$ dimensions, the situation is even worse because new divergences pop up and their number increases with the dimension; in particular, the purely gravitational divergences with the maximum number of derivatives are not in the quadratic form $R_{\mu\nu}R^{\mu\nu}$ and $R^2$ (e.g., in $D=6$ dimensions, they are cubic in the curvatures). 

There are three ways to solve this problem. The first and most intuitive way is to consider the finite version of the theory, discussed above in this section~\ref{finite}. Indeed, if the theory is finite ($D$ odd, or $D$ even plus killer operators), there are no divergences at any loop order and we do not need to invoke any multiplicative renormalizability.

The second option is a partial application of what is spelled out above: One cancels only a subset of divergences by means of killers in order to make the number of free couplings equal to the number of divergences. Afterwards, one applies the multiplicative renormalization scheme.

The third way, first discussed in \cite{Tomboulis:1997gg,Modesto:2011kw,Modesto:2014lga}, is to take advantage of renormalization-group (RG) invariance and impose a particular initial condition on the RG flow. In fact, the definition of multiplicative renormalizability is somehow incomplete because, in the first place, it does not consider the quantum effective action, which is the actual physical quantity we have to use to compute the quantum equations of motion and the quantum tree-level scattering amplitudes; and, second, it does not take into account the initial conditions of the RG equations.  About the former point, it deserves to be stressed that, in general, the running of the coupling constants is only an indication of how the quantum action looks like, but it is not fully reliable. This problem is already evident in the case of the cosmological constant, where the running does not correspond to any logarithmic contribution to the quantum effective action~\cite{Shapiro:2009dh}.  

The definition of renormalizability is not simply based on introducing all possible operators consistent with the power counting, but on having a theory in which divergences correspond to the introduction of a finite number of counterterms at the quantum level and, therefore, of a finite number of parameters to be measured experimentally. What is crucial for renormalizability is that the number of divergences be finite. Only in this case is the quantum effective action uniquely specified by a finite number of new extra parameters (with respect to the classical theory), although it might not contain all possible operators as required by the definition of multiplicative renormalizability. On the other hand, if the number of counterterms is infinite, then we have to make an infinite number of measurements and the theory is nonpredictive. Consider, for instance, the extreme case of a finite theory devoid of divergences. In this case, the quantum effective action in some models may contain an infinite number of parameters, but they are all referable to the parameters present in the classical theory.

To extend the discussion in \cite{Modesto:2014lga} and fully illustrate the RG-based procedure avoiding the problem posed by multiplicative renormalizability, let us go back to Stelle gravity. Here we limit the discussion to $D=4$ dimensions, but we expect the scheme below to work also in higher dimensions. The theory \eq{stelle} is power-counting renormalizable and the one-loop RG equations for $\alpha$ and $\sigma$ at high energies are approximately~\cite{Fradkin:1981iu,Avramidi:1985ki}
\be
\frac{\rd \alpha}{\rd t} \simeq \beta_\alpha  \, ,  \qquad 
\frac{\rd \sigma}{\rd t} \simeq \beta_\sigma   \, , \qquad 
 t \coloneqq \frac{1}{(4\pi)^2}\ln\frac{\mu}{\mu_0} \,,\label{RGab}
\ee
equivalent to
\be
\al(\mu)\simeq\al(\mu_0)
  + \frac{\beta_\alpha}{(4\pi)^2} \ln \frac{\mu}{\mu_0}\,,\qquad\sigma(\mu)\simeq\sigma(\mu_0)
  + \frac{\beta_\sigma}{(4\pi)^2} \ln \frac{\mu}{\mu_0}\,,\label{RGab2}
\ee
where $\mu_0$ is an initial condition for the RG evolution. These equations also hold for the model \eq{modGR} as soon as we recognize that we can select a renormalization scale $\mu$ such that $\sigma=0$ (but \emph{with different beta functions}, since vertices and propagator are affected by the absence of the $R^2$ contribution). Such a value can be seen as a particular choice of the initial condition for the second equation in (\ref{RGab2}) made in order to keep the classical spectrum unchanged at the quantum level, independently of whether it contains ghosts or not. In fact, while perturbative unitarity (or its lack) is proved via the Cutkosky cutting rules derived at this particular scale \cite{Briscese:2018oyx}, RG invariance ensures that the spectrum is not affected by the running of the couplings. \emph{RG invariance} states that the quantum effective action $\Gamma$ has the same form at any RG scale $\mu$, at any loop order:
\be\label{RGI}
\Gamma(\mu) = \Gamma(\mu_0)\,.
\ee
Using \eq{RGab2}, it is easy to check \eq{RGI} at one loop in Stelle gravity, focusing on the quadratic part of the one-loop effective Lagrangian:
\bea
{\cal L}^{(1)\,{\rm quad}}(\mu) &=&  {\cal L}^{\rm quad}(\mu) + {\cal L}_{\rm ct} + {\cal L}_{\rm div} + {\cal L}_{\rm finite}(\mu)={\cal L}^{\rm quad}(\mu) + {\cal L}_{\rm finite}(\mu)\nonumber\\
&=& \al(\mu) R_{\mu\nu} R^{\mu\nu} + \sigma(\mu) R^2 \nonumber\\
&&+\frac{\beta_\alpha}{32 \pi^2} R_{\mu\nu} \ln \frac{- \Box}{\mu^2}  R^{\mu\nu}  + \frac{\beta_\sigma}{32 \pi^2} R \ln \frac{- \Box}{\mu^2} R \nonumber \\
&=& \left[\al(\mu_0)
  + \frac{\beta_\alpha}{16 \pi^2} \ln \frac{\mu}{\mu_0}\right] R_{\mu\nu} R^{\mu\nu} + \left[\sigma(\mu_0) + \frac{\beta_\sigma}{16 \pi^2} \ln \frac{\mu}{\mu_0}\right] R^2  \nonumber \\
 && +\frac{\beta_\alpha}{32 \pi^2} R_{\mu\nu} \ln \frac{- \Box}{\mu^2}  R^{\mu\nu} 
 + \frac{\beta_\sigma}{32 \pi^2} R \ln \frac{- \Box}{\mu^2} R \nonumber  \\
 &=& \al(\mu_0) R_{\mu\nu} R^{\mu\nu}  + \sigma(\mu_0) R^2\nonumber\\
&&  +  \frac{\beta_\alpha}{32 \pi^2} R_{\mu\nu} \ln \frac{- \Box}{\mu_0^2}  R^{\mu\nu} + \frac{\beta_\sigma(\mu_0)}{32 \pi^2} R \ln \frac{- \Box}{\mu_0^2} R \nonumber \\
 & = &{\cal L}^{(1)\,{\rm quad}}(\mu_0) \, . \label{QEA2}
\eea
Hence, if one derives the spectrum (or checks perturbative unitarity) at a scale $\mu$, its properties are preserved at any energy scale. 
From here, it follows that the case of the $GG$ model \eq{modGR} is obtained simply by a choice of initial RG scale. In fact, consider the RG-invariant scale 
\be\label{mumup}
\bar\mu \coloneqq \mu\, e^{-\frac{16\pi^2}{\beta_\sigma}\sigma(\mu)}\stackrel{\textrm{\tiny \eq{RGab2}}}{=} \mu_0\, e^{-\frac{16\pi^2}{\beta_\sigma}\sigma(\mu_0)}.
\ee
The invariance is manifest on the right-hand side since $\mu_0$ is a constant. If one assumes the initial condition $\sigma(\mu_0)=0$ in the RG flow, then
\be
\bar\mu=\mu_0\,.
\ee
In other words, the choice $\sigma(\mu_0)=0$ is equivalent to a particular selection of the RG-invariant scale.

As a double-check of our findings, we now show that the very same quantum effective action \eq{QEA2} with $\sigma(\mu_0)=0$ can be \emph{directly} obtained starting from the $G_{\mu\nu} G^{\mu\nu}$ model \eq{modGR}. Again, we concentrate on the quadratic part. Let us first generalize \eq{LrenS} as
\be\label{Lreg}
{\cal L}_{\rm reg} = {\cal L}(\mu) + {\cal L}_{\rm ct}\,.
\ee
This definition of a regularized one-loop Lagrangian is valid for any theory but only for renormalizable ones we can say that this Lagrangian is renormalized, ${\cal L}_{\rm reg}={\cal L}_{\rm ren}$. The one-loop quantum effective action coming from \eq{Lreg} is augmented by the divergent operators (which cancel the counterterms by construction) and a finite part. The quadratic part of the Lagrangian reads
\be
{\cal L}^{(1)\,{\rm quad}}_{\rm reg}(\mu_0) &=& {\cal L}_{\rm reg}^{\rm quad} + {\cal L}_{\rm div} + {\cal L}_{\rm finite}(\mu_0)={\cal L}^{\rm quad}(\mu_0) + {\cal L}_{\rm finite}(\mu_0) \nonumber \\
&=& \al(\mu_0) R_{\mu\nu} R^{\mu\nu} +\frac{\beta_\alpha}{32 \pi^2} R_{\mu\nu} \ln\frac{-\Box}{\mu_0^2}R^{\mu\nu} + \frac{\beta_{R^2}}{32 \pi^2} R \ln \frac{- \Box}{\mu_0^2} R\nonumber\\ 
&=& {\cal L}^{(1)\,{\rm quad}}(\mu_0)\big|_{\sigma(\mu_0)=0} \,, 
\ee
identical with the last equality in \eq{QEA2} with $\sigma(\mu_0)=0$, where $\beta_{R^2}=\beta_\sigma|_{\sigma=0}$.

To summarize, the model (\ref{modGR}) is singular in the space of parameters because the coefficient of the local operator $R^2$ vanishes. On the other hand, if the coefficients $\alpha$ and $\sigma$ are both nonzero, it is clear that the theory is renormalizable in a conventional manner. In the classical theory, $\alpha$ and $\sigma$ must have a fixed value corresponding to what is measured experimentally at a specific energy scale. The only theoretical assumption is that the theory must have a particular spectrum (or to be unitary, if this is the case). Hence, if the action is nonsingular in the space of parameters and $\alpha\neq0\neq \sigma$, then their classical values can only be the initial conditions of the RG equations that define the quantum effective action. In other words, the statement that the parameters in the classical action must be general in order to renormalize the theory should be replaced by the fact that they must have a specific value, which is the observed one. The spectrum (or perturbative unitarity, in the case of the nonlocal theory) and the Cutkosky rules are calculated at the specific scale $\mu_0$ in which we define the classical theory \cite{Briscese:2018oyx}, but they are secured at any scale by RG invariance. Indeed, the spectrum (perturbative unitarity) cannot be affected by perturbative quantum corrections. 

Let us move to super-renormalizable local higher-derivative and nonlocal theories, in which only a finite number of counterterms must be added. Following the previous example, we can reintroduce all such terms into the classical action simply assuming that the initial condition of the RG equations are the ones in which the coupling constants take specific constant values including zero. Exactly like for local quadratic gravity, the class of theories (\ref{action}) has singular points in the modular space of couplings because some of them are zero in the classical action on which unitarity and other properties are based on. The case where some of the coupling constants are nonzero but have specific values (for instance, a fixed value for the ratio of two or more couplings) is nothing new. Indeed, in any quantum field theory (including standard QED and QCD) the classical value for the couplings corresponds to the asymptotic value of the RG equations (in QED, for example, the asymptotic value for the fine structure constant $e^2/4 \pi$ is $1/137$). The fact that this value in the IR is nonzero can be explained by electron mass threshold phenomena, not by initial RG conditions (there is no non-Gaussian IR fixed point in QED) \cite{Eichhorn:2017muy}. However, the asymptotic conditions to be used to get the IR values of couplings are not very different from initial conditions. Aside from this analogy, the main difference with respect to higher-derivative and nonlocal gravitational or gauge theories is the crucial role played by the coupling constants in the unitarity issue. One can appreciate this in as simple an example as the six-derivative kinetic operator $\Box(1+\tau \Box^2)$, where the relative coefficient $\tau$ is important to determine the presence or absence of real ghosts (ghosts with real mass square) versus complex ghosts (ghosts with complex mass square). 

Similarly, in nonlocal theories the structure of the nonlocal terms is crucial to avoid extra degrees of freedom including ghosts. For instance, in the case of the minimal purely gravitational theory 
\be
{\cal L} = \frac{1}{\kappa^2} \left[ R + G_{\mu\nu} {\cal F}(\Box) R^{\mu\nu} \right], 
\label{PureG}
\ee
the coefficients in front of $R^2_{\mu\nu}$ and $R^2$ are $1$ and $-1/2$, consistently with ghost freedom for one suitable form factor ${\cal F}$ common for both terms. Such coefficients define the initial condition of the RG equations and they are exactly the analog of $1/137$ in QED. Also in QED the coupling is relevant for unitarity: by a proper redefinition of the photon field, the kinetic operator turns into $- F^2/(4 e^2)$ and an opposite choice of the initial condition for $e^2$, namely $e^2 \propto - 1/137$ would change the photon into a ghost. Notice that the nature of the photon would not change in perturbative quantum field theory, consistently with the Cutkosky rules. 

We do not have to repeat the same calculation above in pure nonlocal gravity, since the counterterms are still the same, $\propto R_{\mu\nu}R^{\mu\nu}$ and $R^2$ (up to topological and superficial terms). Since these terms do not appear with nonzero coefficients in the original classical Lagrangian $\cL$, in order to compensate the one-loop divergences with two independent counterterms it is sufficient to add to the classical Lagrangian the quadratic terms
\begin{equation}
{\cal L}_{\rm quad}=(\al-\tilde{\al})R_{\mu\nu}R^{\mu\nu}+(\sigma-\tilde{\sigma})R^{2}\,,
\end{equation}
which are subject to quantum renormalization. At the classical level, we choose $\al=\tilde{\al}$ and $\sigma=\tilde{\sigma}$ and then the Lagrangian ${\cal L}+{\cal L}_{\rm quad}$ reduces to the original unitary theory. At the quantum level, we face two possibilities. If the theory is finite and all the beta functions vanish (odd spacetime dimensions or even spacetime dimensions with killers), we have scale invariance and the above classical identification is valid also at the quantum level. Instead, if the theory has one-loop divergences (even spacetime dimensions without killers), then the parameters $\tilde{\al}$ and $\tilde{\sigma}$ are just the initial conditions for the RG equations of the running coupling constants. At the quantum level, the couplings $\al(\mu)$ and $\sigma(\mu)$ run with the renormalization energy scale $\mu$, and so does ${\cal L}_{\rm quad}$. However, these contributions can be absorbed in the finite part of the one-loop quantum effective action, which involves the same operators as in ${\cal L}_{\rm quad}$ with insertions of $\ln(-\Box/\mu^2)$ in between. This is a simple consequence of RG invariance, as shown above. A notable difference with respect to Stelle gravity is that the running equations \eq{RGab} and \eq{RGab2} are exact in nonlocal quantum gravity, so that the RG equations
\be
\al(\mu)=\al(\mu_0)
  + \frac{\beta_\alpha}{(4\pi)^2} \ln \frac{\mu}{\mu_0}\,,\qquad\sigma(\mu)=\sigma(\mu_0)
  + \frac{\beta_\sigma}{(4\pi)^2} \ln \frac{\mu}{\mu_0}\,,\label{RGab3}
\ee
are valid at all scales, not just in the UV.

Since the RG flow and its initial conditions can be formulated in a very general way on curved spacetimes and for almost any quantum field theory with any matter content, the same mechanism applies, \emph{mutatis mutandi}, to nonminimally coupled nonlocal quantum gravity with matter.

The ultimate message of this ``third solution'' to the multiplicative renormalizability problem is that what really matters is the number of counterterms rather than the number of operators in the classical action. If two theories have the same divergences, the same power counting, and a sufficiently high number of derivatives so that the beta functions do not depend on the running couplings, but one of them has less operators in the classical action, then their quantum behavior is the same. Hence, it is sufficient to select the classical couplings as initial conditions for the RG equations. 


\subsection{Finiteness and conformal invariance}\label{sec.ci}

It is well-known that \emph{any} classical gravitational theory, local or nonlocal, can be made conformally invariant \cite{Englert:1976ep,tHooft:2011aa}. Indeed, in a purely gravitational theory (no matter) one can make the following replacement of the metric $g_{\mu\nu}$, 
\be\label{phighat}
g_{\mu\nu} =g_{\mu\nu}(\phi,\hat{g}_{\mu\nu})= \phi^{\frac{4}{D-2}} \hat{g}_{\mu\nu}
\, , 
\ee
in which $\phi$ is a real scalar field called dilaton, and the action
\be
S[\hat g,\phi]\equiv S[g]\big|_{g=g(\phi,\hat{g})}
\ee
turns out to be trivially invariant under the conformal transformation
\be\label{resca}
\hat{g}^\prime_{\mu\nu} = \Omega^2 \, \hat{g}_{\mu\nu} \, , \qquad 
\phi^\prime = \Omega^{-\frac{D-2}{2}} \phi\,,
\ee
where $\Omega(x)$ is an arbitrary function of spacetime coordinates. Hence the tensor $g_{\mu\nu}$ is conformally invariant, $g_{\mu\nu}^\prime=g_{\mu\nu}$. The above replacement (\ref{phighat}) is sufficient for a purely gravitational theory, while in the presence of matter we must also introduce a proper power of the dilaton for each field in the action. Calling, as before, $\Phi_i$ the collection of all fields including the spacetime metric,
\be
\Phi_i \in \{ g_{\mu\nu}, \Phi, \Psi, A_\mu \} \, , 
\label{fields}
\ee
in order to make explicit the hidden conformal symmetry of the theory (\ref{action}) we can make the replacement
\be
\Phi_i = \phi^{\Delta_i} \, \hat{ \Phi}_i \,,
\label{MatterConfInv}
\ee
where $\Delta_i$ is the conformal weight of the field. For example, in $D=4$ dimensions, 
\be
g_{\mu\nu} = \phi^2 \, \hat{g}_{\mu\nu} \,, \quad 
 \Phi =
 \frac{\hat{\Phi}}{{\phi} } \, , \quad \Psi = 
 \frac{\hat{\Psi}}{\phi^{\frac{3}{2}}} \, , \quad A_\mu = \hat{A}_\mu \, .
\label{MatterConfInv1}
\ee
All the fields $\hat{g}_{\mu\nu}, \hat{\Phi}, \hat{\Psi}, \hat{A}_\mu$ are rescaled by $\phi$ in such a way that the combined terms take invariant forms under Weyl transformations when $\phi$ is treated like a scalar, as in \eq{resca}. Once the fields (\ref{MatterConfInv}) are replaced in the action (\ref{action}), all the mass scales disappear upon a redefinition of $\phi$, which acquires canonical dimension $[\phi] = (D-2)/2$. 

In many quantum field theories, conformal invariance is a symmetry valid at the tree level, but broken at the quantum level, since the coefficients of the operators appearing in the conformal anomaly are proportional to the nonzero $\beta$-functions of the theory. However, in finite theories where the $\beta$-functions all vanish, conformal symmetry is preserved at the quantum level.

Let us elaborate this feature for super-renormalizable or finite theories in general and, in particular, for those studied in this paper. For the sake of simplicity, we here consider a purely gravitational theory. In the presence of divergences, the quantum effective action consists of the classical (bare) action $S_{\rm cl}$, a divergent part, and a finite contribution $\Gamma_{\rm finite}$. Since the dimensional regularization scheme breaks conformal invariance, we have to split the divergent from the finite part in the quantum effective action. For this purpose, we evaluate all the operators in $D= 4 - 2 \varepsilon$ and, afterwards, take the limit $\varepsilon \rightarrow 0$. Therefore, the one-loop quantum effective action reads
\be
&& \Gamma^{(1)} = S_{\rm cl} + \frac{1}{\varepsilon}\sum_i\beta_i\int \rd^{4 -2 \varepsilon} x \, {\cal O}_i(g) + \Gamma_{\rm finite} \, , 
\qquad \frac{1}{\varepsilon}=\frac{2}{4-D}\equiv\ln\left(\frac{\Lambda_\textsc{uv}}{\mu}\right)^2, 
\label{QA}  \\
&&{\cal O}_i(g)\in\left\{\sqrt{|g|}\Box R\,,\,\sqrt{|g|} R_{\mu\nu\rho\sigma} R^{\mu\nu\rho\sigma},\, \sqrt{|g|} {R}^2\,,\,\sqrt{|g|}R_{\mu\nu}R^{\mu\nu},\, \sqrt{|g|}R\,,\, \sqrt{|g|}\right\}\Big|_{\substack{g=g(\phi,\hat{g}) \\ D = 4 - 2 \varepsilon}} \,  ,\nonumber\\
\label{counterT}
\ee
where $\Lambda_{\textsc{uv}}$ is the UV cutoff in the cutoff regularization scheme and $\mu$ is the renormalization scale. Each operator has a different $\beta$-function $\beta_i$ in front. Clearly, the replacement (\ref{phighat}) in (\ref{counterT}) produce several terms proportional to $\varepsilon$ that contribute to the finite part of the quantum effective action when replaced in the second term of $\Gamma^{(1)}$ in (\ref{QA}). Such terms are independent of $\varepsilon$. The extra $1/\varepsilon$ terms define the divergent part of the quantum effective action. As a particular example, let us consider the metric density $\sqrt{|g|}$, i.e., the last operator in (\ref{counterT}), which is also present in all the others. Call $\beta_\Lambda$ the coefficient associated to it in the divergent part of (\ref{QA}). When the metric (\ref{phighat}) with $D=4 - 2 \varepsilon$ is plugged into $\sqrt{|g|}$, the determinant of the metric produces a factor $\phi^{2\varepsilon}$:
\be
\left(\sqrt{|g|}\right)_{D=4-2\varepsilon}\simeq\phi^{4+2\varepsilon}\sqrt{|\hat{g}|} =\phi^{2\varepsilon}\underbrace{\phi^4\sqrt{|\hat{g}|}}_{\text{conf.\ inv.}} =\phi^{2\varepsilon} \sqrt{|g|}\,.
\ee
where we approximated for small $\varepsilon$. On the other hand, the above operator contributes to (\ref{QA}) linearly as $\propto\beta_\Lambda \varepsilon$, since
\be
\hspace{-.5cm}\frac{1}{\varepsilon}  \beta_\Lambda \phi^{2\varepsilon} \sqrt{|g|} = \frac{1}{\varepsilon} \beta_\Lambda [1+2\varepsilon\ln\phi + O(\varepsilon^2)]\sqrt{|g|}
= \frac{1}{\varepsilon} \beta_\Lambda  \sqrt{|g|}+ 2 \beta_\Lambda \ln\phi \,   \sqrt{|g|} + O(\varepsilon) \, .
\label{Lambda}
\ee
Hence, the divergent contribution is $\beta_\Lambda \sqrt{|g|}/\varepsilon$, while the second term is finite and violates conformal invariance. 
In general, one gets the following anomalous contributions to the action:
\be
\hspace{-.8cm}\lim_{\varepsilon\to 0}\frac{\beta_i}{\varepsilon} \phi^{2\varepsilon} {\cal O}_i(\phi^2 \hat{g}; \varepsilon)&=&\lim_{\varepsilon\to 0}\frac{\beta_i}{\varepsilon} (1+2\varepsilon\ln\phi) \, {\cal O}_i(\phi^2 \hat{g};\varepsilon)\nonumber\\
&=& \lim_{\varepsilon\to 0} \frac{\beta_i}{\varepsilon}\,{\cal O}_i(\phi^2\hat{g}) +\beta_i\tilde{\cal O}_i(\phi^2 \hat{g}) +2\beta_i\ln\phi\,{\cal O}_i(\phi^2 \hat{g}),\label{anomaly}
\ee
where $\tilde{\cal O}_i(\phi^2 \hat{g})$ is the finite contribution to $\lim_{\varepsilon\to 0}{\cal O}_i(\phi^2 \hat{g}; \varepsilon)/\varepsilon$. The first term can be eliminated by adding a counter\-term with opposite sign, while the last two contributions in (\ref{anomaly}) are finite (independent of $\varepsilon$) and explicitly violate conformal invariance. Therefore, a nonfinite gravitational theory can be conformally invariant at the classical level but not at the quantum level.

However, conformal invariance is preserved if the theory is finite, as is the case of minimally coupled nonlocal quantum gravity \cite{Modesto:2016max} and of the nonlocal theory presented here with killer operators, as explicitly shown in section \ref{finite}. All the operators in eqs.\ (\ref{counterT}) and (\ref{anomaly}) have vanishing coefficients $\beta_i=0$ from the start and we can take the limit $\varepsilon\rightarrow 0$ consistently with conformal invariance. This result is not a fine tuning but, actually, is one-loop exact because the theory has no divergences for $L>1$. The physical consequences of conformal invariance for the theory (\ref{action}) with the unit field-space metric \eq{id} are explored in \cite{Modesto:2022asj,Calcagni:2022tuz}.


\section{Conclusions}\label{concl}

In this paper, we have studied the power-counting renormalizability of the nonlocal field theory \eq{action}--\eq{FF2} where gravity and matter are nonminimally coupled. We found that the theory may be super-renormalizable in $D=4$ dimensions and that it can easily be rendered finite. More precisely, for the theory~\eq{action} with local action~\eq{aclocal} the number of UV divergences is infinite but the theory is strictly renormalizable if $n=(D-4)/2$ (eq.\ \eq{stric}), while it is finite and the theory is super-renormalizable if $n > (D-4)/2$ (eq.\ \eq{con1}), where $n+1$ is the degree of the polynomial $p(z)$ in the nonlocal form factor ${\rm H}$. Therefore, in four-dimensional spacetime, the model is power-counting renormalizable if $n=0$ and super-renormalizable if $n>0$. Finally, if the summation over the field indices is performed with the delta~\eq{de-field}, the UV action does not possess terms with an odd number of fields $\ph$ provided that the potential $V(\ph)$ is an even function. In this case, divergences with an odd number of $\ph$ are not generated.

We conclude that the nonlocal proposal \eq{action}, which is also unitary, could serve not just as a quantum gravity, but also as a ``theory of everything'' where all fields are quantized in the same way. Although there is no unification in the sense of supersymmetric multiplets or of particle modes arising from the vibrations of a string, and despite the fact that the choice of nonlocal form factors is restricted but not quite unique, the attractiveness of this proposal might lie in its simplicity: it is just a perturbative QFT of point particles, with no extra symmetries or extra dimensions.

In its finite version, however, conformal invariance does appear as an added symmetry and this has momentous consequences for cosmology. Indeed, the theory is conformally invariant thanks to finiteness and, because of that, one can dispense with inflation to solve the problems of the hot-big-bang scenario and to subsequently generate the primordial tensor and scalar spectra in a natural way \cite{Modesto:2022asj}. In particular, the primordial tensor spectrum is blue-tilted and there is a relatively large \emph{lower bound} on its amplitude. This prediction can be tested within the next five years, thus allowing us to either detect a signature of quantum gravity or to rule out the theory in the very near future \cite{Calcagni:2022tuz}. The results of the present work put this phenomenology on a firm theoretical ground.


\section*{Acknowledgments}

L.R.\ thanks the Department of Physics, UFJF, for the warm hospitality. G.C., L.M.\ and L.R.\ are supported by grant PID2020-118159GB-C41 funded by MCIN/AEI/10.13039/ 501100011033. L.M.\ is also supported by the Basic Research Program of the Science, Technology, and Innovation Commission of Shenzhen Municipality (grant no.\ JCYJ2018030\-2174206969).


\appendix

\section{Dimensional analysis of the theory}
\label{AppA}

Let us consider the general model~\eq{action} with fields $\Phi^i$ with dimensionality $[\Phi^i]$. 
Since the functional derivative is defined by
\be
\hspace{-1cm}S [\Phi \, + \de \Phi] -  S [\Phi] &=& 
\int \rd^D x \sqrt{|g (x)|} \, \frac{\de S}{\de \Phi^i (x)} \, \de \Phi^i (x)\nonumber \\
&& + \frac12 \iint \rd^D x \rd^D y \sqrt{|g(x) g(y)|} \de \Phi^i(x)  \frac{\de^2 S}{\de \Phi^j(y) \de \Phi^i (x)} \de \Phi^j(y)
+ \ldots\,,
\ee
the extremals $E_i$ (see~\eq{EiEq1}) of the underlying local model
have the dimension 
\be
\n{D-EOM}
[E_i] = D - [\Phi^i]\,
,
\ee
while for the components of the Hessian~\eq{HessianEq1}
we have
\be
[\De_{ij} (x,y) ] = 2D - [\Phi^i] - [\Phi^j]\,.
\ee
Notice that the components of the Hessian have different dimensions if the dimensions of the fields $\Phi^i$ are not the same.

It is useful to define the operator $\hat{\De}_{ij}$ through
\be
\De_{ij} (x,y) \eqqcolon \hat{\De}_{ij} \, \de^D(x,y).
\ee
Therefore,
\be
[\hat{\De}_{ij}] = D - [\Phi^i] - [\Phi^j]
\ee
and we can factorize the dimension of the covariant delta ($[\de^D(x,y)] = D$). The main reason for doing this is because the dimension of the delta is always compensated by an integration, when the operator is applied on a field.

Hence, from \eq{action} and \eq{D-EOM}, we have
\be
[F^{ij} (\hat{\De})] = [\Phi^i] + [\Phi^j] - D= - [\hat{\De}_{ij}]\,,
\ee
in agreement with formula~\eq{FF2}. 
In the definition of the form factor, we introduce the scale of nonlocality $\La_*$ with 
$
[\La_*]=1
$ 
and define \eq{DefStar}, so that
\be
[( \hat{\De}_{\La_*}) _{ij}] = 0\,.
\ee

In the generalized model with the contravariant field space metric $\mathscr{G}^{ij}$ (defined in eq.~\eq{de-field}), one can define the Hessian with mixed position of the indices $i$ and $j$:
\be
(\hat{\Delta}_{\Lambda_{\star}})^{i}{}_{j}=\mathscr{G}^{ik}(\hat{\Delta}_{\Lambda_{\star}})_{kj}\,.
\ee
Since $\mathscr{G}^{ij}$ is dimensionless in our convention, 
$[\mathscr{G}^{ij}]=0$, it follows that
\be
[(\hat{\Delta}_{\Lambda_{\star}})^{i}{}_{j}]=0\,.
\ee
Then, it is natural to consider arbitrary powers of this operator,
given explicitly by (with standard matrix multiplication and contraction
of dummy indices)
\be
(\hat{\Delta}_{\Lambda_{\star}}^{0})^{i}{}_{j}=\delta^{i}_{j}\,,\qquad (\hat{\Delta}_{\Lambda_{\star}}^{1})^{i}{}_{j}=(\hat{\Delta}_{\Lambda_{\star}})^{i}{}_{j}\,,
\ee
\be
(\hat{\Delta}_{\Lambda_{\star}}^{n})^{i}{}_{j}= (\hat{\Delta}_{\Lambda_{\star}})^{i}{}_{k_{1}}
(\hat{\Delta}_{\Lambda_{\star}})^{k_{1}}{}_{k_{2}}
\cdot\ldots\cdot
(\hat{\Delta}_{\Lambda_{\star}})^{k_{n-1}}{}_{j}\,,\qquad n\geqslant2\,.
\ee 
Thus, 
\be
[(\hat{\De}_{\La_*}^n)^{i} {}_{j}] = 0\,.
\ee

In the UV, according to eq.~\eq{UVlimit}, the function $e^{{\rm H}(z)}$ tends to a polynomial $p(z)$ of degree $n+1$ (with $n\geqslant 0$). 
Therefore, 
\be
F^{i} {}_j (\hat{\De})   \simeq    
\alpha \sum_{\ell=0}^{n} \frac{(\hat{\De}_{\La_*}^\ell)^{i}_{j}}{\left( \La_* \right)^{[\hat{\De}_{ij}]}}
,
\ee 
where $\al = e^{\ga_{\rm E}} e^{-{\rm H}(0)}/2$ and $[\al]=0$. Using the above formula, we obtain the UV action~\eq{UV-act}.
Notice that 
\be
\left[ \int \rd^D x \sqrt{|g|}  E_i  \frac{(\hat{\De}_{\La_*}^\ell)^{ij}}{\left( \La_* \right)^{[\hat{\De}_{ij}] }} E_j \right]  
& = & - D + [E_i] + [E_j] + [(\hat{\De}_{\La_*}^\ell)_{ij}] - [\hat{\De}_{ij}]
\nonumber
\\
& = & - D + 2D - [\Phi^i] - [\Phi^j] + 0  - (D - [\Phi^i] - [\Phi^j]) 
\nonumber
\\
& = & 0\,,
\ee
as it should be. 

In the explicit example of section \ref{model gravity-scalar}, i.e., in the model 
based in the local action~\eq{aclocal} with fields $\Phi^i = (g_{\mu\nu}, \ph)$, 
the dimensions of the fields are\footnote{The metric $g_{\mu\nu}$ can be kept dimensionless in arbitrary $D$ dimensions by changing 
the dimension of $\ka$.}
\be
[g_{\mu\nu}] = 0\,, \qquad 
[\ph]= \frac{D-2}{2}\,,
\ee
and for $n = 0$ we get 
\be
S_{\rm UV}^{n=0} = \int \rd^D x \sqrt{|g|} \, \left[  \cL_{\loc} 
+ \frac{\eta_1}{\La_*^D} E_{\mu\nu} E^{\mu\nu} 
+ \frac{\eta_2}{\La_*^D} E_{\mu}^\mu  E_\nu^\nu
+ \frac{\eta_3}{\La_*^{D-[\ph]}} E_{\mu}^\mu E_\ph
+ \frac{\eta_4}{\La_*^{D-2[\ph]}} E_\ph E_\ph   \right] ,\nonumber\\
\ee
where the coefficients $\eta_k$ depend on the contravariant field-space metric $\mathscr{G}^{ij}$, i.e., $\eta_k = \eta_k (\ga_i)\,$ with $[\eta_k] = 0$; see eq.~\eq{act-UV-gen}. In particular, for $D=4$ the above formula gives the result 
\be
S_{\rm UV}^{n=0} = \int \rd^4 x \sqrt{|g|} \, \left[  \cL_{\loc} +    \frac{\eta_1}{\La_*^4} E_{\mu\nu}^2 
+ \frac{\eta_2}{\La_*^4} E_{\mu}^\mu E_\nu^\nu  
+ \frac{\eta_3}{\La_*^{3}} E_{\mu}^\mu E_\ph
+ \frac{\eta_4}{\La_*^{2}} E_\ph E_\ph    \right] 
.
\ee


\section{Effect of a nonminimal term in \texorpdfstring{$\cL_{\loc}$}{Lloc}}
\label{AppB}

In section \ref{Sec.PCn=0}, we showed that even with a minimally coupled local subsidiary Lagrangian $\cL_{\loc}$, the resultant UV action~\eq{UV-act} of the nonlocal model can contain the nonminimal terms necessary for renormalization. Here we show that, on the other hand, if the local Lagrangian $\cL_{\loc}$ possesses a nonminimal (NM) term, additional structures are generated in the nonlocal action, which may cause the model to become nonrenormalizable. To this end, let us consider the nonminimal extension of the local action~\eqref{aclocal}
\be\label{nmac}
S_{\loc}^{\rm (NM)} = S_{\loc} - \frac{\xi}{2} \int \rd^D x \sqrt{|g|}  \ph^2 R\,,
\ee
where $S_{\loc}$ is given by~\eqref{aclocal}. The presence of the term $R \ph^2$ modifies the extremals \eq{EOMmunu} and \eq{EOMph} as
\be
E^{\mu\nu}_{\rm (NM)} & = &  E^{\mu\nu} + \frac{\xi \ph^2 }{2}  \left( R^{\mu \nu } - \frac{1}{2} g^{\mu \nu } R  \right) 
+  \xi g^{\mu\nu} \left[  \ph \Box \ph + (\na \ph)^2 \right] \nonumber\\
&& - \xi \left[ \ph (\na^\mu \na^\nu \ph) + (\na^\mu \ph)(\na^\nu \ph)  \right],\\
E_\ph^{\rm (NM)} & = &  E_\ph -  \xi  \ph R\,
.
\ee
Notice that, because of the nonminimal term, now the extremals have terms with maximal number of derivatives that depend on $\ph$, instead of $\nabla\ph$ only. Therefore, the extension of the action~\eqref{act-UV-gen} will also contain this type of terms. In fact, the action still has the form of~\eq{1x}, but instead of the coefficients~\eq{OsAs1}--\eq{OsCes3}, more coefficients become functions of $\ph$. For example, instead of~\eq{OsAs2} and~\eq{OsAs3} we now have
\be
a_2(\ph) & = & \al_1 \left( \frac{1}{\ka^2} - \frac{\xi \ph^2}{2} \right)^2,\label{a_2(ph)}\\
a_3(\ph) & = & \frac{1}{4} \left[  \al_1 (D-4)  + \al_2 (D-2)^2 \right] \left( \frac{1}{\ka^2} - \frac{\xi \ph^2}{2} \right)^2\nonumber\\
&&  -  \al_3 \left(\frac{D-2}{2} \right)  \left( \frac{1}{\ka^2} - \frac{\xi \ph^2}{2}  \right) \xi \ph + \al_4  \xi^2 \ph^2 .\label{a_3(ph)}
\ee

The nontrivial dependence of the above coefficients in $\ph$ introduces in the model new vertices with the maximal number of derivatives. Indeed, because of~\eq{a_2(ph)} and~\eq{a_3(ph)}, the list of terms that generate vertices with four derivatives is now augmented by\footnote{Beside these, there might be other terms with four derivatives that can contribute external scalar legs without derivatives, such as $ \sqrt{\vert g\vert} \ph^2 R ( \nabla \ph )^2 $,  $\sqrt{\vert g\vert} \ph^2 R^{\mu\nu}  (\nabla_\mu \ph) ( \nabla_\nu \ph )$,  $\sqrt{\vert g\vert} \ph^2 R  \Box \ph$, and $\sqrt{\vert g\vert} \ph^2 (\Box \ph)^2$.}
\be
\label{newver4} 
\sqrt{\vert g\vert} \ph^2 R^2_{\mu\nu}  , 
\quad \sqrt{\vert g\vert} \ph^4 R^2_{\mu\nu}  ,
\quad \sqrt{\vert g\vert} \ph R^2  , \quad  \sqrt{\vert g\vert} \ph^2 R^2 , \quad  \sqrt{\vert g\vert} \ph^3 R^2, \quad  \sqrt{\vert g\vert}  \ph^4 R^2
.
\ee
All these new vertices can yield an external scalar leg without derivative. Thus, the model based on a nonminimal local action has new diagrams formed by the substitution of an arbitrary number of the ``old'' four-derivative vertices originated from~\eq{1x} by the ``new'' vertices associated with~\eq{newver4} ---with up to four external scalar legs without derivatives, per vertex. These diagrams will have the same superficial degree of divergence of the original ones. Since there are logarithmically diverging diagrams with $V_4$ arbitrary, we can use the vertices from~{\eq{newver4}} to construct divergent diagrams with internal graviton lines and \emph{an arbitrary number of scalar external legs without derivatives}, which would call for counterterms in the form of eq.~\eq{1x} with the coefficients being of the type $\sum_{k=0}^\infty \la_k \ph^k$. Note that this result is independent of the potential $V(\ph)$ and of the choice of the metric in the space of fields. 

The considerations above can be easily extended to the more general case in which $n>0$ and $D>4$ in eq.~\eq{UV-act}, with a similar conclusion: The nonminimal term in $\cL_{\loc}$ will generate terms with maximal number of derivatives which depend not only on $\nabla\ph$, but also on $\ph$. Since vertices originated from these terms can occur in arbitrary number in logarithmically diverging diagrams (see the general discussion in section \ref{Sec.PC-gen}, and in section \ref{SubSecD=4} for the particular case $D=4$), the number of counterterms are infinite and the nonlocal theory with the nonminimal term in \eq{nmac} is not power-counting renormalizable in the usual sense.


\section{Hessian} 
\label{App3}

For the two-derivative local action~\eq{aclocal}, the Hessian is given by
\be
\label{HssianApp}
\De_{ij}  =
\begin{pmatrix}
 \frac{\de^2 S}{ \de g_{\al\beta} \de g_{\mu\nu}}  & \frac{\de^2 S}{  \de \ph \de g_{\mu\nu}} \\
\frac{\de^2 S}{ \de g_{\al\beta} \de \ph} &  \frac{\de^2 S}{  \de \ph \de \ph}
\end{pmatrix}
= \begin{pmatrix}
 \hat{\De}^{\mu\nu,\al\beta}_{11}  & \hat{\De}^{\mu\nu}_{12} \\
\hat{\De}^{\al\beta}_{21} & \hat{\De}_{22}
\end{pmatrix} \de^D (x,y)\,,
\ee
where
\be
\hat{\De}_{11}^{\mu\nu,\al\beta} &=& \,\,  
\frac{1}{\ka^2} \Bigg[ \frac{1}{2} \de^{\mu\nu,\al\beta} \Box 
- \frac12 g^{\mu\nu} g^{\al\beta} \Box
+ \frac12 ( g^{\mu\nu} \na^{\al} \na^{\beta}
+ g^{\al\beta} \na^{\mu} \na^{\nu} )\nonumber\\
&&
- \frac14 (
g^{\mu\al} \na^\nu \na^\beta +  g^{\nu\al} \na^\mu \na^\beta
+ g^{\mu\beta} \na^\nu \na^\al +  g^{\nu\beta} \na^\mu \na^\al)
\Bigg] + \Pi^{\mu\nu,\al\beta}\,,
\n{He11}\\
\hat{\De}_{12}^{\mu\nu} &=&  2 \mathscr{G}^{\mu\nu,\la\tau}_0 (\na_\la \ph) \na_\tau - \frac12 g^{\mu\nu} V'(\ph)\,,\n{He12}\\
\hat{\De}_{21}^{\al\beta} &=&  - 2 \mathscr{G}^{\al\beta,\la\tau}_0 [(\na_\la \ph) \na_\tau + (\na_\la \na_\tau \ph) ] - \frac12 g^{\al\beta} V'(\ph)\,,
\n{He21}\\
\hat{\De}_{22} &=&  \Box - V'' (\ph) \,,\n{He22}
\ee
where we defined the tensorial objects
\be
\mathscr{G}^{\mu\nu,\al\beta}_0 \coloneqq \frac12 (\de^{\mu\nu,\al\beta} - \tfrac12 g^{\mu\nu} g^{\al\beta} )
,
\ee
and 
\be
\Pi^{\mu\nu,\al\beta} &\coloneqq& \,\, \frac{1}{\ka^2} \left[ \frac14 \left(R^{\mu\al\nu\beta} + R^{\nu\al\mu\beta} + R^{\mu\beta\nu\al} + R^{\nu\beta\mu\al} \right)- \frac12 \left( g^{\mu\nu} R^{\al\beta} + g^{\al\beta} R^{\mu\nu} \right) \right.\nonumber\\
&& \left. + \frac14 \left(g^{\mu\al} R^{\nu\beta} + g^{\nu\al} R^{\mu\beta} + g^{\mu\beta} R^{\nu\al} + g^{\nu\beta} R^{\mu\al}   \right)
- \mathscr{G}_0^{\mu\nu,\al\beta} R \right]\nonumber\\
&&+\frac12 \mathscr{G}^{\mu\nu,\al\beta}_0 (\na \ph)^2 + \mathscr{G}^{\mu\nu,\al\beta}_0 V(\ph)
+ \frac14 \left[ g^{\mu\nu} (\na^\al  \ph) (\na^\beta  \ph) + g^{\al\beta} (\na^\mu  \ph) (\na^\nu  \ph) \right] \nonumber\\
&&- \frac14 \left[g^{\mu\al} (\na^\nu  \ph) (\na^\beta  \ph) + g^{\nu\al} (\na^\mu  \ph) (\na^\beta  \ph) + g^{\mu\beta} (\na^\nu  \ph) (\na^\al  \ph) + g^{\nu\beta} (\na^\mu  \ph) (\na^\al  \ph)   \right]\!.\nonumber\\
\ee
Note that the Hessian is self-adjoint, i.e., given a generic field $\Phi^i = (h_{\mu\nu}, \ph)$, where $h_{\mu\nu}$ and $\ph$ are, respectively, an arbitrary symmetric rank-2 tensor and a scalar field,  
the inner product 
\be
(\Phi, \hat{\De} \Phi ) \coloneqq \int \Phi^i \hat{\De}_{ij} \Phi^j ,
\ee
satisfies
\be
(\Phi, \hat{\De} \Phi ) = ( \hat{\De} \Phi, \Phi ).
\ee
Or, explicitly,
\be
\int \Phi^i \hat{\De}_{ij} \Phi^j = \int \Phi^j \hat{\De}_{ji} \Phi^i ,
\ee
where in the above formul\ae\ we omitted the integration measure and there is no summation on repeated indices. 
Indeed, using~\eq{He12} and~\eq{He21}, after integrating by parts, one can directly verify that, e.g.,
\be
\int \rd^D x \sqrt{|g|} \, h_{\mu\nu} \hat{\De}^{\mu\nu}_{12} \ph  =
\int \rd^D x \sqrt{|g|} \, \ph \hat{\De}^{\mu\nu}_{21} h_{\mu\nu} \,.
\ee




\begin{thebibliography}{99}
\bibitem{Krasnikov:1987yj} \au{N.V}{Krasnikov}, \tia{Nonlocal gauge theories} \doinn{10.1007/BF01017588}{Theor.\ Math.\ Phys.}{73}{1184}{1987} [\ndoinn{http://www.mathnet.ru/php/archive.phtml?wshow=paper&jrnid=tmf&paperid=5624&option_lang=eng}{Teor.\ Mat.\ Fiz.}{73}{235}{1987}].
\bibitem{Kuzmin:1989sp} \au{Yu.V}{Kuz'min}, \tia{The convergent nonlocal gravitation} Sov.\ J.\ Nucl.\ Phys.\ {\bf 50}, 1011 (1989) [Yad.\ Fiz.\ {\bf 50}, 1630 (1989)].
\bibitem{Modesto:2011kw} \au{L}{Modesto}, \tia{Super-renormalizable quantum gravity} \doin{10.1103/PhysRevD.86.044005}{Phys.\ Rev.}{D}{86}{044005}{2012} [\arX{1107.2403}].
\bibitem{Biswas:2011ar}  \au{T}{Biswas}, \au{E}{Gerwick}, \au{T}{Koivisto} and \au{A}{Mazumdar}, \tia{Towards singularity and ghost free theories of gravity} \doinn{10.1103/PhysRevLett.108.031101}{Phys.\ Rev.\ Lett.}{108}{031101}{2012} [\arX{1110.5249}].
\bibitem{Dona:2015tra} \au{P}{Don\`a}, \au{S}{Giaccari}, \au{L}{Modesto}, \au{L}{Rachwa\l} and \au{Y}{Zhu}, \tia{Scattering amplitudes in super-renormalizable gravity} \doij{10.1007/JHEP08(2015)038}{JHEP}{1508}{038}{2015} [\arX{1506.04589}].
\bibitem{BasiBeneito:2022wux} \au{A}{Bas i Beneito}, \au{G}{Calcagni} and \au{L}{Rachwa\l}, \tia{Classical and quantum nonlocal gravity}, 
to appear \procsinm{Handbook of Quantum Gravity}{\au{C}{Bambi}, \au{L}{Modesto} and \au{I.L}{Shapiro}}{Springer}{Switzerland}{2023?} [\arX{2211.05606}].
\bibitem{Buoninfante:2022ild} \au{L}{Buoninfante}, \au{B.L}{Giacchini} and \au{T}{de Paula Netto}, \tia{Black holes in non-local gravity} \arX{2211.03497}.
\bibitem {CANTATA:2021ktz} \au{G}{Calcagni}, \tia{\href{https://doi.org/10.1007/978-3-030-83715-0_9}{\cob Non-local gravity}} \procsinm{Modified Gravity and Cosmology}{\au{E.N}{Saridakis}, \au{R}{Lazkoz}, \au{V}{Salzano}, \au{P}{Vargas Moniz}, \au{S}{Capozziello}, \au{J}{Beltrán Jiménez}, \au{M}{De Laurentis} and \au{G.J}{Olmo}}{Springer}{Switzerland}{2021} [\arX{2105.12582}].
\bibitem{Koshelev:2023elc} \au{A.S}{Koshelev}, \au{K.S}{Kumar} and \au{A.A}{Starobinsky}, \tia{Cosmology in nonlocal gravity} \arX{2305.18716}.
\bibitem{Modesto:2021ief} \au{L}{Modesto}, \tia{Nonlocal spacetime-matter} \arX{2103.04936}.
\bibitem{Modesto:2021okr} \au{L}{Modesto}, \tia{The Higgs mechanism in nonlocal field theory} \doij{10.1007/JHEP06(2021)049}{JHEP}{2106}{049}{2021} [\arX{2103.05536}].
\bibitem{Modesto:2021soh} \au{L}{Modesto} and \au{G}{Calcagni}, \tia{Tree-level scattering amplitudes in nonlocal field theories} \doij{10.1007/JHEP10(2021)169}{JHEP}{2110}{169}{2021} [\arX{2107.04558}].
\bibitem{Modesto:2022asj} \au{L}{Modesto} and \au{G}{Calcagni}, \tia{Early universe in quantum gravity} \arX{2206.06384}.
\bibitem{Calcagni:2022tuz} \au{G}{Calcagni} and \au{L}{Modesto}, \tia{Testing quantum gravity with primordial gravitational waves} \arX{2206.07066}.
\bibitem{Kaku:1993ym} \au{M}{Kaku}, \book{Quantum field theory}{Oxford University Press}{Oxford}{UK}{1993}, p.\ 216.
\bibitem{PeSc}  \au{M.E}{Peskin} and \au{D.V}{Schroeder}, \book{An Introduction to Quantum Field Theory}{Perseus}{Reading}{MA}{1995}, p.\ 321.
\bibitem{Buchbinder:2021wzv} \au{I.L}{Buchbinder} and \au{I}{Shapiro}, \book{Introduction to Quantum Field Theory with Applications to Quantum Gravity}{Oxford University Press}{Oxford}{UK}{2021}.
\bibitem{NIST} \au{F.W.J}{Olver} {et al.} (Eds.) \book{NIST Handbook of Mathematical Functions}{Cambridge University Press}{Cambridge}{UK}{2010}.
\bibitem{Briscese:2018oyx} \au{F}{Briscese} and \au{L}{Modesto}, \tia{Cutkosky rules and perturbative unitarity in Euclidean nonlocal quantum field theories} \doin{10.1103/PhysRevD.99.104043}{Phys.\ Rev.}{D}{99}{104043}{2019} [\arX{1803.08827}].
\bibitem{complexGhosts} \au{J}{Liu}, \au{L}{Modesto} and \au{G}{Calcagni}, \tia{Quantum field theory with ghost pairs} \doij{10.1007/JHEP02(2023)140}{JHEP}{2302}{140}{2023} [\arX{2208.13536}].
\bibitem{Eran:1998pga} \au{E}{Marcus}, \emph{Higher-Derivative Gauge and Gravitational Theories}, Ph.D.\ thesis (UCLA, CA, 1998).
\bibitem{Ohta:2020bsc} \au{N}{Ohta} and \au{L}{Rachwa\l}, \tia{Effective action from the functional renormalization group} \doin{10.1140/epjc/s10052-020-8325-8}{Eur.\ Phys.\ J.}{C}{80}{877}{2020} [\arX{2002.10839}].
\bibitem{Stelle:1976gc} \au{K.S}{Stelle}, \tia{Renormalization of higher-derivative quantum gravity} \doin{10.1103/PhysRevD.16.953}{Phys.\ Rev.}{D}{16}{953}{1977}.
\bibitem{Asorey:1996hz} \au{M}{Asorey}, \au{J.L}{L\'opez} and \au{I.L}{Shapiro}, \tia{Some remarks on high derivative quantum gravity} \doin{10.1142/S0217751X97002991}{Int.\ J.\ Mod.\ Phys.}{A}{12}{5711}{1997} [\oarX{hep-th/9610006}].
\bibitem{Elizalde:1994nz} \au{E}{Elizalde}, \au{A.G}{Zheksenaev}, \au{S.D}{Odintsov} and \au{I.L}{Shapiro}, \tia{A four-dimensional theory for quantum gravity with conformal and nonconformal explicit solutions} \doinn{10.1088/0264-9381/12/6/006}{Classical Quantum Gravity}{12}{1385}{1995} [\oarX{hep-th/9412061}].
\bibitem{Lavrov:2019nuz} \au{P.M}{Lavrov} and \au{I.L}{Shapiro}, \tia{Gauge invariant renormalizability of quantum gravity} \doin{10.1103/PhysRevD.100.026018}{Phys.\ Rev.}{D}{100}{026018}{2019} [\arX{1902.04687}].
\bibitem{tHooft:1974toh}  \au{G}{'t Hooft} and \au{M.J.G}{Veltman}, \tia{One-loop divergencies in the theory of gravitation} \ndoin{http://www.numdam.org/item?id=AIHPA_1974__20_1_69_0}{Ann.\ Poincar\'e Phys.\ Theor.}{A}{20}{69}{1974}.
\bibitem{Deser:1974cz} \au{S}{Deser} and \au{P}{van Nieuwenhuizen}, \tia{One-loop divergences of quantized Einstein--Maxwell fields} \doin{10.1103/PhysRevD.10.401}{Phys.\ Rev.}{D}{10}{401}{1974}.
\bibitem{gosa} \au{M.H}{Goroff} and \au{A}{Sagnotti}, \tia{The ultraviolet behavior of Einstein gravity} \doin{10.1016/0550-3213(86)90193-8}{Nucl.\ Phys.}{B}{266}{709}{1986}.
\bibitem{vandeVen:1991gw} \au{A.E.M}{van de Ven}, \tia{Two loop quantum gravity} \doin{10.1016/0550-3213(92)90011-Y}{Nucl.\ Phys.}{B}{378}{309}{1992}.
\bibitem{Tomboulis:1997gg} \au{E.T}{Tomboulis}, \tia{Superrenormalizable gauge and gravitational theories} \oarX{hep-th/9702146}.
\bibitem{Starobinsky:1980te} \au{A.A}{Starobinsky}, \tia{A new type of isotropic cosmological models without singularity} \doin{10.1016/0370-2693(80)90670-X}{Phys.\ Lett.}{B}{91}{99}{1980}.
\bibitem{Vilenkin:1985md} \au{A}{Vilenkin}, \tia{Classical and quantum cosmology of the Starobinsky inflationary model} \doin{10.1103/PhysRevD.32.2511}{Phys.\ Rev.}{D}{32}{2511}{1985}.
\bibitem{Maeda:1987xf} \au{K.-i}{Maeda}, \tia{Inflation as a transient attractor in $R^2$ cosmology} \doin{10.1103/PhysRevD.37.858}{Phys.\ Rev.}{D}{37}{858}{1988}.
\bibitem{Modesto:2014lga} \au{L}{Modesto} and \au{L}{Rachwa\l}, \tia{Super-renormalizable and finite gravitational theories} \doin{10.1016/j.nuclphysb.2014.10.015}{Nucl.\ Phys.}{B}{889}{228}{2014} [\arX{1407.8036}].
\bibitem{Modesto:2015lna} \au{L}{Modesto} and \au{L}{Rachwa\l}, \tia{Universally finite gravitational and gauge theories} \doin{10.1016/j.nuclphysb.2015.09.006}{Nucl.\ Phys.}{B}{900}{147}{2015} [\arX{1503.00261}].
\bibitem{Modesto:2017sdr} \au{L}{Modesto} and \au{L}{Rachwa\l}, \tia{Nonlocal quantum gravity: a review} \doin{10.1142/S0218271817300208}{Int.\ J.\ Mod.\ Phys.}{D}{26}{1730020}{2017}.
\bibitem{Julve:1978xn}    \au{J}{Julve} and \au{M}{Tonin}, \tia{Quantum gravity with higher derivative terms} \doin{10.1007/BF02748637}{Nuovo Cimento}{B}{46}{137}{1978}.
\bibitem{Barvinsky:1985an} \au{A.O}{Barvinsky} and \au{G.A}{Vilkovisky}, \tia{The generalized Schwinger--DeWitt technique in gauge theories and quantum gravity} \doinn{10.1016/0370-1573(85)90148-6}{Phys.\ Rep.}{119}{1}{1985}.
\bibitem{Avramidi:1985ki} \au{I.G}{Avramidi} and \au{A.O}{Barvinsky}, \tia{Asymptotic freedom in higher-derivative quantum gravity} \doin{10.1016/0370-2693(85)90248-5}{Phys.\ Lett.}{B}{159}{269}{1985}.
\bibitem{Shapiro:2009dh} \au{I.L}{Shapiro} and \au{J}{Sola}, \tia{On the possible running of the cosmological `constant'} \doin{10.1016/j.physletb.2009.10.073}{Phys.\ Lett.}{B}{682}{105}{2009} [\arX{0910.4925}].
\bibitem{Fradkin:1981iu} \au{E.S}{Fradkin} and \au{A.A}{Tseytlin}, \tia{Renormalizable asymptotically free quantum theory of gravity} \doin{10.1016/0550-3213(82)90444-8}{Nucl.\ Phys.}{B}{201}{469}{1982}.
\bibitem{Eichhorn:2017muy} \au{A}{Eichhorn}, \au{A}{Held} and \au{C}{Wetterich}, \tia{Quantum-gravity predictions for the fine-structure constant} \doin{10.1016/j.physletb.2018.05.016}{Phys.\ Lett.}{B}{782}{198}{2018} [\arX{1711.02949}].
\bibitem{Englert:1976ep} \au{F}{Englert}, \au{C}{Truffin} and \au{R}{Gastmans}, \tia{Conformal invariance in quantum gravity} \doin{10.1016/0550-3213(76)90406-5}{Nucl.\ Phys.}{B}{117}{407}{1976}.
\bibitem{tHooft:2011aa} \au{G}{'t Hooft}, \tia{A class of elementary particle models without any adjustable real parameters} \doinn{10.1007/s10701-011-9586-8}{Found.\ Phys.}{41}{1829}{2011} [\arX{1104.4543}].
\bibitem{Modesto:2016max} \au{L}{Modesto} and \au{L}{Rachwa\l}, \tia{Finite conformal quantum gravity and nonsingular spacetimes} \arX{1605.04173}.
\end{thebibliography}
\end{document}